\documentclass[twoside,12pt]{article}
\usepackage{epsfig}
\usepackage{graphicx}
\def\Journal#1#2#3#4{{#1} {#2} (#4) #3 }

\def\NAT{\em Nature}
\def\NATAS{\em Nature Astronomy}
\def\NP{\em Nucl. Phys.}
\def\NPA{{\em Nucl. Phys.} A}

\def\PRO{{\em Prog. Theor. Phys.}}
\def\PTPS{{\em Prog. Theor. Phys.} Suppl.}
\def\PTEP{\em Prog. Theor. Exp. Phys.}

\def\PLB{{\em Phys. Lett.} B}

\def\PRL{\em Phys. Rev. Lett.}
\def\PREV{\em Phys. Rev.}
\def\PREP{\em Phys. Rep.}
\def\PRA{{\em Phys. Rev.} A}
\def\PRD{{\em Phys. Rev.} D}
\def\PRC{{\em Phys. Rev.} C}

\def\ZP{\em Z. Phys.}

\def\RMP{{\em Rev. Mod. Phys.}}

\def\INT{{\em Int. J. Mod. Phys.} E}
\def\ApJ{\em Astrophysical J.}
\def\ApJS{\em Astrophysical J., Suppl.}
\def\ApJL{\em Astrophysical J. Letters}
\def\ADNDT{\em At. Data Nucl. Data Tables}
\def\ARNPS{\em Ann. Rev. Nucl. Part. Sci.}
\def\ASPRv{\em Sov. Sci. Rev. E Astrophys. Space Phys.}
\def\PASJ{\em Pub. Astr. Soc. Japan}
\def\PASA{\em Pub. Astr. Soc. Australia}
  
\def\EPJ{{\em Eur. Phys. J.} A}
\def\EPJWC{\em EPJ Web of Conf.}
\def\JPC{\em J. Phys. Conf. Ser}
\def\JPG{{\em J. Phys.} G} 
\def\CPC{\em Comput. Phys. Commun.}
\def\PPNP{\em Prog. Part. Nucl. Phys.}
\def\RPPh{\em Rep. Prog. Phys.}
\def\NIMA{{\em Nucl. Instrum. Methods} A}
\def\HYP{\em Hyperfine Interactions}
\def\JINST{\em J. Instrum.}
\def\MNRAS{\em Monthly Notices Roy. Astron. Soc.}
\def\AA{\em Astronomy and Astrophysics}
\def\SCI{\em Science}
\def\EPSL{\em Earth Planet Sci. Lett.}
\def\CG{\em Chem. Geol.}
\def\SJNP{\em Sov. J. Nucl. Phys.}

\def\JPSC{\em JPS Conf. Proc.} 
\def\JCAP{\em J. Cosmol. Astropart. Phys.}
\def\NJP{\em New J. Phys.}
\def\ADV{\em Adv. in Phys.}

\newcommand{\be}{\begin{equation}}
\newcommand{\ee}{\end{equation}}
\newcommand{\bea}{\begin{eqnarray}}
\newcommand{\eea}{\end{eqnarray}}

\begin{document}

\title{ \vspace{1cm} Nuclear Weak Rates and Nuclear Weak Processes in Stars}
\author{Toshio\ Suzuki, \\ Department of Physics, College of Humanities and Sciences,\ Nihon University\\
Sakurajosui 3-25-40, Setagaya-ku, Tokyo 156-8550, JAPAN. \\}
%$^{*}$
%and\\
%National Astronomical Observatory of Japan\\
%$^{*}$
%Mitaka, Tokyo 181-8588, \ JAPAN.\\ }
\maketitle
\begin{abstract} Nuclear weak rates in stellar environments are 
%updated 
obtained by taking into account recent advances in shell-model studies of  spin-dependent excitation modes in nuclei including Gamow-Teller (GT) and spin-dipole transitions.   
They are applied to nuclear weak processes in stars such as cooling and heating of the cores of stars and nucleosynthesis in supernovae.  
The important roles of accurate weak rates for the study of astrophysical processes are pointed out in the following cases.
(1) The electron-capture (e-capture) and $\beta$-decay rates in $sd$-shell are evaluated with the USDB Hamiltonian and used to study the evolution of O-Ne-Mg cores in stars with 8-10 M$_{\odot}$. 
The important roles of the $A$ =23 and 25 pairs of nuclei for the cooling of the cores by nuclear Urca processes are investigated.
(2) They are also used to study heating of the O-Ne-Mg core by double e-captures on $^{20}$Ne in later stages of the evolution.
Especially, the e-capture rates for a second-forbidden transition in $^{20}$Ne are evaluated with 
the multipole expansion method by Walecka 
as well as the method of  Behrens-B$\ddot{\mbox{u}}$hring.
Possible important roles of the transition in heating the O-Ne-Mg cores and implications on the final fate of the cores (core-collapse or thermonuclear explosion) are discussed.
(3) The weak rates in $pf$-shell nuclei are evaluated with a new Hamiltonian, GXPF1J, and applied to nucleosynthesis of iron-group elements in Type Ia supernova explosions.
The over-production problem of neutron-rich iron isotopes compared with the solar abundances, which remained for the rates according to Fuller, Fowler and Newman, is much improved, and the over-production is now reduced to be within a factor of two. 
(4) The weak rates for nuclei with two-major shells are evaluated.
For $sd$-$pf$ shell in the island of inversion, the weak rates for the $A$=31 pair of nuclei, which are important for nuclear Urca processes in neutron-star crusts, are evaluated with the effective interaction obtained by the extended Kuo-Krenciglowa (EKK) method.   
Neutron-rich nuclei with and near neutron number ($N$) of 50 are important for core-collapse processes in supernova explosions.
The transition strengths and e-capture rates in $^{78}$Ni are evaluated with a new shell-model Hamiltonian for the $pf$-$sdg$ shell, and compared with those obtained by the random-phase-approximation (RPA) and an effective rate formula.  
(5) $\beta$-decay rates and half-lives of N =126 isotones, the waiting point nuclei for r-process nucleosynthesis, are evaluated by shell-model calculations with both the GT and first-forbidden transitions. The important roles of the forbidden transitions are pointed out for the isotones with larger proton number ($Z$).
The half-lives are found to be shorter than those obtained by standard models such as the finite-range droplet model (FRDM) by M$\ddot{\mbox{o}}$ller. 
(6) Neutrino-nucleus reaction cross sections on $^{13}$C, $^{16}$O and $^{40}$Ar are obtained with new shell-model Hamiltonians. 
Implications on nucleosynthesis, neutrino detection, neutrino oscillations and neutrino mass hierarchy are discussed.    
\end{abstract}


%\eject
%\tableofcontents
\section{Introduction}
Due to the recent progress in the studies of both experimental and theoretical aspects of nuclear physics, more accurate evaluations of nuclear weak rates have become feasible, and transition rates important for astrophysical processes have been updated and accumulated. A number of new exotic nuclei have been produced at radio-active beam (RIB) factories in the world, and the number of nuclides amounted up to more than three thousands until now.    
The shell structure is found to change toward driplines, with disappearance of traditional magic numbers and appearance of new magic numbers. These shell evolutions have been studied based on monopole components of nucleon-nucleon interactions \cite{RMP2020,LMRMP}. 
The important roles of various parts of nuclear forces, such as central, tensor, spin-orbit and three-nucleon interactions, for the shell evolution have been also clarified. 

Electron-capture (e-capture) and $\beta$-decay rates in stellar environments as well as neutrino-induced reaction cross sections have been updated with the use of new shell-model Hamiltonians constructed on the basis of the developments mentioned above.
Spin-dependent transitions play dominant roles in the weak rates and cross sections.  The leading contributions come from the Gamow-Teller (GT) and spin-dipole (SD) transitions. 
In the present review, we discuss recent progress in the refinement of the weak rates and cross sections, and 
%highlight 
its important implications for astrophysical processes.

%First, 
We discuss the precise evaluations of e-capture and $\beta$-decay rates and their applications to the stellar evolution and nucleosynthesis in stars.
Up to now, many studies have been done to obtain the weak rates in various regions of nuclides since the pioneering work by Fuller, Fowler and Newman \cite{FFN},
%\cite{Fuller1980,Fuller1982a,Fuller1982b}, 
where allowed Fermi and GT transitions were taken into account.
The weak rates were then refined and tabulated by using shell-model calculations combined with available experimental data for $sd$-shell \cite{Oda} and $pf$-shell nuclei \cite{LM}. 
For heavier nuclei in the $pfg$/$sdg$-shell (A = 65 -112), the rates were obtained by the shell model Monte Carlo (SMMC) approach combined with random phase approximation (RPA) including both allowed and forbidden transitions \cite{HL}. 

Here, we further refine the weak rates for $sd$-shell and $pf$-shell nuclei induced by GT transitions by using new shell-model Hamiltonians.
The obtained rates for $sd$-shell nuclei are used to study 
the evolution of O-Ne-Mg cores in stars with 8-10 $M_{\odot}$.
Cooling of the core by nuclear Urca processes and heating of the core by double e-capture processes in the evolution are investigated.
In particular, we discuss the weak rates 
induced by a second-forbidden transition in $^{20}$Ne, which can be important for heating the core in the late stage of the evolution and its final fate, either core-collapse or thermonuclear explosion.

The rates for $pf$-shell nuclei are refined and applied to nucleosynthesis of iron-group elements in Type Ia supernova (SN) explosions.
The over-production of neutron-rich isotopes compared with the solar abundances is shown to be suppressed when using the improved rates. 

We then extend our study to nuclides where 
two-major shells are concerned.  
The weak rates for $sd$-$pf$-shell nuclei in the island of inversion relevant to nuclear Urca processes in neutron star crusts will be evaluated with a new effective interaction derived from an extended G-matrix method applicable to two-major shells.
The weak rates in neutron-rich $pf$-$g$-shell nuclei with neutron magicity at $N$=50, which are important for gravitational core-collapse processes, will also be evaluated by shell-model calculations at $Z$=28 with full $pf$-$sdg$ shell model space.
Spin-dipole transition strengths and e-capture rates in $^{78}$Ni are investigated.
Improvements in the method of calculation and extension of the model space are shown to be important for accurate evaluations of the rates.

The $\beta$-decay rates and half-lives of waiting-point nuclei at $N$=126, important for r-process nucleosynthesis, are studied by including both GT and first-forbidden (spin-dipole) transitions.
Half-lives consistent with recent experimental data but short compared to those obtained by the standard FRDM method are obtained. 
%The r-process nucleosynthesis near and beyond the third-peak region is investigated for possible r-process sites, in core-collapse supernova (CCSN) explosions and binary neutron star mergers.   

As the treatment of forbidden transitions is rather complex and not easy to access for primers, formulae for the first and second-forbidden transitions by the multipole expansion method of Walecka as well as of Behrens-B$\ddot{\mbox{u}}$hring are explained.

Finally, we discuss neutrino-induced reactions on carbon isotopes, as well as on $^{16}$O and $^{40}$Ar, at reactor, solar and supernova neutrino energies. 
The cross sections refined by recent shell-model calculations are important for neutrino detection by recent carbon-based scintillators, water Cerenkov detectors and liquid argon time projection chambers, as well as for nucleosynthesis in SN and study of neutrino properties such as their mass hierarchy.

Astrophysical topics treated here are not inclusive, and more or less 
related to the weak rates 
%updated 
discussed in this work. 
%biased on those related to the work of the author.
More 
%focusses are 
emphasis is put on the role of nuclear physics in 
%updating 
obtaining the weak rates.    

In Sect. 2, we discuss e-capture and $\beta$-decay rates for $sd$-shell nuclei and evolution of high-density O-Ne-Mg cores. 
The weak rates for $pf$-shell nuclei and nucleosynthesis of iron-group elements in Type Ia SN explosions are discussed in Sect. 3. 
In Sect. 4, the weak rates for cross-shell nuclei in $sd$-$pf$ and $pf$-$g$ shells are studied. 
$\beta$-decay rates for isotones with $N$ = 126 
%and r-process nucleosynthesis 
are investigated in Sect. 5. 
In Sect. 6, we discuss neutrino-nucleus reactions relevant to neutrino detection, nucleosynthesis and study of neutrino mass hierarchy. 
Summary is given in Sect. 7. 

\section{Weak Rates for $sd$-Shell Nuclei and Evolution of High-Density O-Ne-Mg Cores}
\subsection{\it Evolution of 8-10 Solar Mass Stars \label{sec:evolstar}}
The evolution and final fate of stars depend on their initial masses $M_I$.
Stars with $M_I$ = 0.5-8 $M_{\odot}$ form electron degenerate C-O cores after helium burning and end as C-O white dwarfs.
Stars more massive than 8 $M_{\odot}$ form O-Ne-Mg cores after carbon burning.
The O-Ne-Mg core is mostly composed of $^{16}$O and $^{20}$Ne, with minor amounts of $^{23}$Na, $^{24}$Mg, $^{25}$Mg and $^{27}$Al.
Stars with $M_{I} >$ 10 $M_{\odot}$ form Fe cores and later explode as core-collapse supernovae (CCSN). 
Stars with $M_I$ = 8-10 $M_{\odot}$ can end up in various ways such as
(1) O-Ne-Mg white dwarfs, (2) e-capture SN explosion with neutron star (NS) remnants 
\cite{Miyaji,Nomoto1984,Nomoto1988,Gutierrez1996},
or (3) thermonuclear explosion with O-Ne-Fe white dwarf remnants \cite{Jones2016}.

In cases (2) and (3), as the O-Ne-Mg cores evolve, the density and temperature in the central region increase and the chemical potential (Fermi energy) of electrons reaches the threshold energy of e-capture on various nuclei in stars.
The e-capture process, on one hand, leads to the contraction of the core due to the loss of the pressure of degenerate electrons. 
The energy production associated with the e-capture, on the other hand, increases the temperature and induces explosive oxygen burning. 
During the evolution, the densities and temperatures of the core are of the order 10$^{8-10}$ g cm$^{-3}$ and 10$^{7-9}$ K, respectively.   

As the density increases, the e-capture process is favored because of larger chemical potential of electrons, while the $\beta$-decay process is hindered because of smaller phase space of the decaying electrons.
For a particular nuclear pair, X and Y, a condition becomes fulfilled at a certain density such that both e-capture and $\beta$-decay    
\begin{eqnarray}
^{A}_{Z}X + e^{-} \rightarrow ^{A}_{Z-1}Y + \nu_{e} \nonumber\\
^{A}_{Z-1}Y \rightarrow ^{A}_{Z}X + e^{-} + \bar{\nu_e}
\end{eqnarray}
occur simultaneously.
In such a case, both the emitted neutrinos and anti-neutrinos take away energy from the star, which leads to an efficient cooling of the core.
This cooling mechanism is called the nuclear Urca process. 
The fate of the stars with $M_I \sim$ 8-10 $M_{\odot}$ is determined by the competing processes of contraction and cooling or heating induced by e-capture and $\beta$-decay processes.
Theoretical predictions of the fate also depend on the treatment of convection and Coulomb effects \cite{Isern1991,Canal1992,Gutierrez1996,Gutierrez2005}.

As the O-Ne-Mg core evolves and the density of the core increases, e-captures on nuclei are triggered in order of their $Q$-values.
For the even mass number components of the core, as energies of even-even nuclei are lowered by pairing effects, e-captures on even-even nuclei have larger magnitude of $Q$-values than odd mass number cases in general.
The e-captures on odd mass number nuclei, therefore, take place first and the cooling of the core by nuclear Urca processes occurs.
The e-captures on even mass number components of the core are triggered later at higher densities. 
Successive e-captures on the odd-odd daughter nuclei occur immediately after the first e-captures because of small magnitude of $Q$-values due to the pairing effects, thus leading to double e-capture processes on even mass number components.
The core is heated by $\gamma$ emission from excited states of daughter nuclei in these processes.

\begin{figure}[tb]
\begin{center}
\begin{minipage}[t]{16.5 cm}
\hspace*{-0.5cm}
\epsfig{file=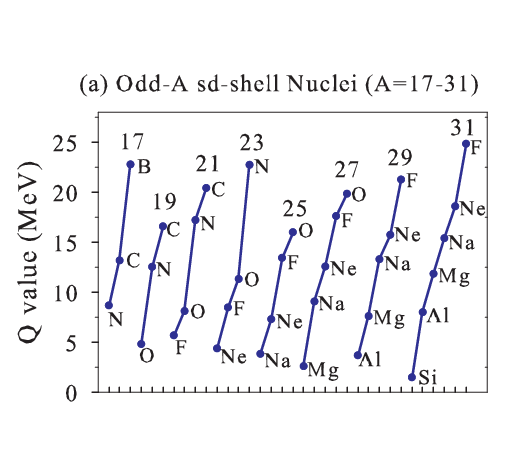,scale=1.0}
\epsfig{file=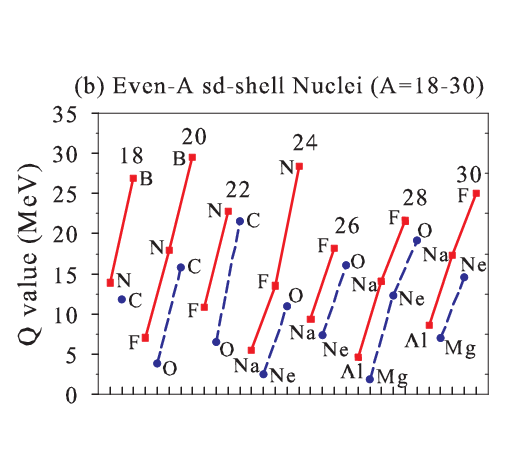,scale=1.0}
\end{minipage}
\begin{minipage}[t]{16.5 cm}
\caption{$\beta$-decay $Q$-values for $sd$-shell nuclei. (a) Odd and (b) even mass number cases are shown separately. \label{fig1}}
\end{minipage}
\end{center}
\end{figure}

The $Q$-value of the weak process determines the density at which the process is triggered.
$\beta$-decay $Q$-values for (a) odd mass number ($A$ = 17-31) and (b) even mass number ($A$ =18-30) $sd$-shell nuclei are shown in Fig.~\ref{fig1}. 
Solid and dashed lines in Fig.~\ref{fig1}(b) denote odd-odd and even-even nuclei, respectively. 

We see that the $Q$-values are small especially for odd mass number nuclear pairs with $A$ = 23, 25 and 27, that is, for (X, Y) = ($^{23}$Na, $^{23}$Ne), ($^{25}$Mg, $^{25}$Na) and ($^{27}$Al, $^{27}$Mg).
The $Q$-values are 4.376 MeV, 3.835 MeV and 2.610 MeV, respectively, for the $A$ = 23, 25 and 27 pairs.
Among even mass number nuclei, e-captures on $^{20}$Ne and $^{24}$Mg, which are components of the O-Ne-Mg core, are important. The $Q$-values of the pairs of $A$ =24 and 20, for (X, Y) = ($^{24}$Mg, $^{24}$Na) and ($^{20}$Ne, $^{20}$F), are 5.516 MeV and 7.024 MeV, respectively.  
$Q$-values of the successive e-captures on $^{24}$Na and $^{20}$F are -2.470 MeV and -3.814 MeV, respectively. 

The electron chemical potential, $\mu_e$, at high densities $\rho Y_e$ with $\rho$ the baryon density and $Y_e$ the proton fraction, 
and high temperatures, $T$, is determined by,
\begin{eqnarray}
\rho Y_e &=& \frac{1}{\pi^2 N_A}\bigl(\frac{m_e c}{\hbar}\bigr)^3 \int_{0}^{\infty} (S_e -S_p) p^2 dp  \nonumber\\
S_{\ell} &=& \frac{1}{exp(\frac{E_\ell -\mu_\ell}{kT})+1}
\end{eqnarray}
where $\mu_p$ = $-\mu_e$. 
The electron chemical potentials at $\rho Y_e$ = 10$^{7}$ - 10$^{11}$ g$\cdot$cm$^{-3}$ and $T_{9}$ = 0.01-10 with $T$ = $T_{9}\times$ 10$^{9}$ K are shown in Table 1.
In the electron degenerate limit, namely at $T$ =0,
\begin{equation}
\mu_e = m_e c^2 \{ \Bigl(\frac{3\rho Y_e}{[\frac{(m_e c)^3}{\pi^2 \hbar^3 N_A}]}\Bigr)^{2/3} +1 \}^{1/2}
\end{equation}
with $\rho Y_e$ in g$\cdot$cm$^{-3}$ and $\frac{(m_e c)^3}{\pi^2 \hbar^3 N_A}$ = 2.9218$\times$10$^{6}$ g$\cdot$cm$^{-3}$.   
It can be expressed approximately as $\mu_e$ $\approx$ 5.2$(\rho Y_e/10^9)^{1/3}$. 
As the temperature increases, $\mu_e$ decreases gradually from its value at $T\approx$ 10$^{9}$ K.
In the limit of $T$ = $\infty$, $\mu_e$ approaches 0. 
We find that $\mu_e$ becomes as large as 4-5 MeV at $\rho Y_e$ $\sim$ 10$^{9}$ g$\cdot$cm$^{-3}$. 

Here, we define 'Urca density' as that density where both the e-capture and $\beta$-decay processes take place simultaneously almost independent of the temperature. 
The cooling timescale, which depends on the e-capture and $\beta$-decay rates, needs to be shorter than the crossing timescale, which is related to the time the core evolves in the density range where the Urca process is active \cite{SMN2021}.  
The Urca density is estimated to be log$_{10}$($\rho Y_e$) $\sim$8.8 -9.0 for the $A$=23 and 25 pairs. 
In case of the $A$=27 pair, 
the transitions between the ground states are forbidden ones, and the Urca cooling effect is negligible \cite{SMN2021}.
Nuclear Urca processes occur for the $^{25}$Mg-$^{25}$Na pair and then for the $^{23}$Na-$^{23}$Ne pair.  
The cooling of the core by the Urca process for the pairs with $A$=23 and 25 will be discussed in Sect. 2.3.

\begin{table}
\begin{center}
\begin{minipage}[t]{16.5 cm}
\caption{Electron chemical potential $\mu_e$ (in units of MeV) at high densities, $\rho Y_e$ = 10$^{7}$ -10$^{11}$ g$\cdot$cm$^{-3}$, and high temperatures, $T$ =$T_{9}\times$10$^{9}$ K.
%\protect{\cite{}
}
\label{tab:cheme}
\end{minipage}
\begin{tabular}{r|rrrrrrrr}
\hline
&&&&&&&&\\[-2mm]
&\multicolumn{2}{c}{$T_{9}$}\\
$\rho Y_e$ & 0.01 & 0.1 & 1 & 2 & 3 & 5 & 7 & 10\\
&&&&&&&&\\[-2mm]
\hline
&&&&&&&&\\[-2mm]
10$^{7}$ & 1.223 & 1.222 & 1.200 & 1.133 & 1.021 & 0.698 & 0.404 & 0.196 \\[2mm]
10$^{8}$ & 2.447 & 2.447 & 2.437 & 2.406 & 2.355 & 2.192 & 1.952 & 1.493 \\[2mm]
10$^{9}$ & 5.180 & 5.180 & 5.176 & 5.161 & 5.138 & 5.062 & 4.948 & 4.708 \\[2mm]
10$^{10}$ & 11.118 & 11.118 & 11.116 & 11.109 & 11.098 & 11.063 & 11.011 & 10.898 \\[2mm]
10$^{11}$ & 23.934 & 23.934 & 23.933 & 23.930 & 23.924 & 23.908 & 23.884 & 23.832 \\[2mm]
&&&&&&&&\\[-2mm]\hline
\end{tabular}
%noalign{\smallskip\hrule}\cr}
\begin{minipage}[t]{16.5 cm}
\vskip 0.5cm
\noindent
\end{minipage}
\end{center}
\end{table}

Electron-captures on $^{24}$Mg and $^{20}$Ne are triggered at higher densities given by log$_{10}$($\rho Y_e$) $\approx$ 9.3 and 9.5, respectively.
In the later stage of the evolution of the O-Ne-Mg core, double e-capture reactions on $^{24}$Mg and $^{20}$Ne, namely
$^{24}$Mg (e$^{-}$, $\nu_e$) $^{24}$Na (e$^{-}$, $\nu_e$) $^{24}$Ne and
$^{20}$Ne (e$^{-}$, $\nu_e$) $^{20}$F (e$^{-}$, $\nu_e$) $^{20}$O, 
become important for the heating of the core. 

The e-capture reaction, $^{20}$Ne (0$^{+}$, g.s.) (e$^{-}$, $\nu_e$) $^{20}$F (2$^{+}$, g.s.) is a second-forbidden transition. 
The transition was pointed out to be rather important at densities of log$_{10}$ ($\rho Y_e$) = 9.2-9.6 and temperatures of log$_{10}$ ($T$) $\leq$ 8.8 \cite{Pinedo}.  
Recently, it was argued that the heating of the O-Ne-Mg core might lead to thermonuclear expansion of the star instead of e-capture SN explosion, because of the contributions from the second-forbidden transition \cite{Jones2016}.     
We will discuss this issue in Sect. 2.5. 

\subsection{Electron-Capture and $\beta$-Decay Rates in $sd$-Shell \label{sec:sdrate}}
In this subsection, electron-capture and $\beta$-decay rates for $sd$-shell nuclei in  stellar environments are updated by shell-model calculations with the use of USDB Hamiltonian \cite{USDB}. 
The weak rates for $sd$-shell nuclei ($A$ = 17 -39) obtained with the USD Hamiltonian \cite{USD,BW} were tabulated in Ref.~\cite{Oda}.
The USDB is an updated version of the USD improved by taking into account recent data of neutron-rich nuclei. While neutron-rich oxygen and fluorine isotopes were overbound for the USD, the new version is free from this problem.

The e-capture rates at high densities and temperatures are evaluated as \cite{FFN,LM,Suzu2011}
\begin{eqnarray}
\lambda &=& \frac{ln 2}{6146 (s)} \sum_{i}W_i \sum_{j}(B_{ij}(GT)+B_{ij}(F))\Phi^{ec}_{ij} \nonumber\\
\Phi^{ec}_{ij} &=& \int_{\omega_{min}}^{\infty} \omega p (Q_{ij}+\omega)^2 F(Z, \omega) S_e(\omega) d\omega \nonumber\\
Q_{ij} &=& (M_p c^2 -M_d c^2 +E_i -E_f )/m_e c^2 \nonumber\\
W_{i} &=& (2J_i +1) e^{-E_i/kT} /\sum_{i} (2J_i +1) e^{-E_i/kT}, 
\end{eqnarray}
where $\omega$ ($p$) is electron energy (momentum) in units of $m_e c^2$ ($m_e c$), $M_p$ and $M_d$ are nuclear masses of parent and daughter nuclei, respectively, and $E_i$ ($E_f$) is the excitation energy of initial (final) state.
Here, $B$(GT) and $B$(F) are the GT and Fermi transition strengths, respectively, given by
\begin{eqnarray}
B_{ij}(\mbox{GT}) &=& (g_A/g_V)^2 \frac{1}{2J_i +1} |<f||\sum_{k} \sigma^k t_{+}^k ||i>|^2 \nonumber\\ 
B_{ij}(\mbox{F}) &=& \frac{1}{2J_i +1} |<f||\sum_{k} t_{+}^k ||i>|^2, 
\end{eqnarray} 
where $J_i$ is the total spin of initial state and $t_{+}|p> =|n>$.
$F(Z, \omega)$ is the Fermi function and $S_e(\omega)$ is the Fermi-Dirac distribution for electrons, where the chemical potential is determined from the density $\rho Y_e$ as shown in Eq. (2) and discussed in Sect. 2.1.
   
In the case of $\beta$-decay, $t_{+}$ is replaced by $t_{-}$, where $t_{-}|n> =|p>$, and $\Phi^{ec}_{ij}$ in Eq. (4) is replaced by 
\begin{equation}
\Phi^{\beta}_{ij} (Q_{ij}) = \int_{1}^{Q_{ij}} \omega p (Q_{ij}-\omega)^2 F(Z+1, \omega) (1 - S_e(\omega)) d\omega.
\end{equation}

Transitions from the excited states of the parent nucleus are taken into account  by the partition function $W_i$, as they can become important for high temperatures and low excitation energies. 
Because of the factors $S_e(\omega)$ and $1-S_e(\omega)$ in the integrals of $\Phi^{ec}_{ij}$ and $\Phi^{\beta}_{ij}$, respectively, the e-capture ($\beta$-decay) rates increase (decrease) as the density and the electron chemical potential increase. 
The neutrino-energy-loss rates and $\gamma$-ray heating rates are also evaluated. 
The rates are evaluated for log$_{10}(\rho Y_e)$ = 8.0 -11.0 in fine steps of 0.02 and log$_{10}T$ = 8.0 -9.65 (7.0 -8.0) in steps of 0.05 (0.20).

\begin{figure}[tb]
\begin{center}
\begin{minipage}[t]{16.5 cm}
\hspace*{-1.0cm}
\epsfig{file=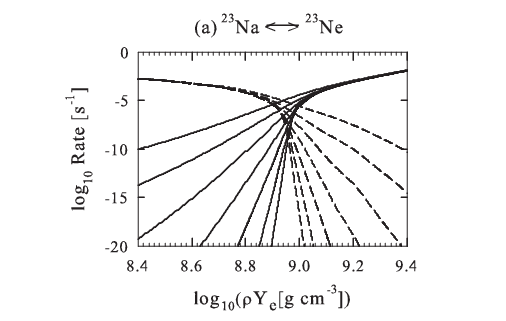,scale=1.1}
%urcanane23scr.eps, scale=1.1}
\hspace*{-1.5cm}
\epsfig{file=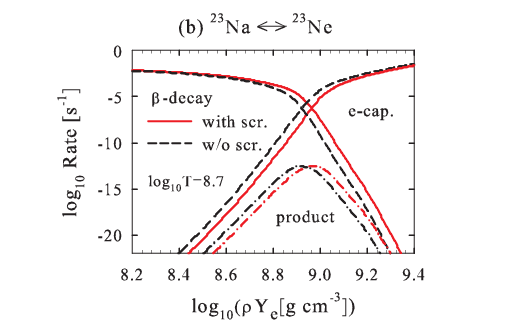,scale=1.1}
%product-nane23.eps, scale=1.1} 
\end{minipage}
\begin{minipage}[t]{16.5 cm}
\caption{(a) Electron-capture and $\beta$-decay rates for the $A$=23 Urca nuclear pair, ($^{23}$Na, $^{23}$Ne), including the Coulomb effects, are shown as functions of density log$_{10}$($\rho Y_e$) for temperatures log$_{10}T$ = 8.0-9.2 in steps of 0.2.
Electron-capture rates (solid curves) increase with density, while $\beta$-decay rates (dashed curves) decrease with density.
(b) Comparison of the cases with and without the Coulomb effects at log$_{10}T$ =8.7.
\label{fig2}}
\end{minipage}
\end{center}
\end{figure}  

\begin{figure}[tb]
\begin{center}
\begin{minipage}[t]{16.5 cm}
\epsfig{file=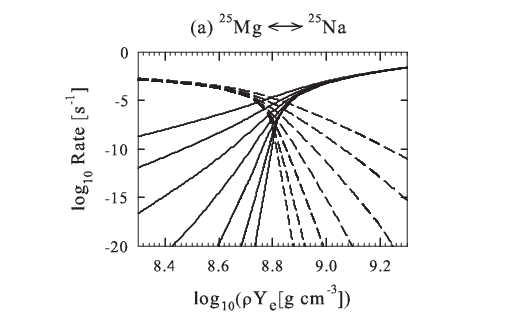,scale=1.1}
%urcamgna25scr.eps, scale=1.1}
\hspace*{-1.5 cm}
\epsfig{file=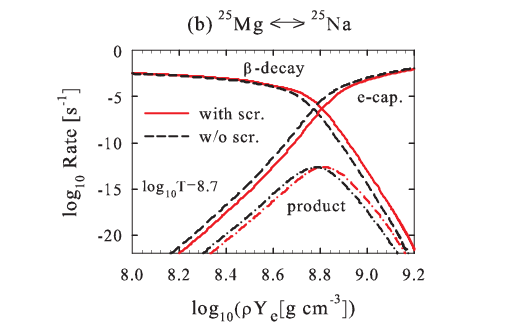,scale=1.1}
%product-mgna25.eps, scale=1.1}
\end{minipage}
\begin{minipage}[t]{16.5 cm}
\caption{Same as in Fig.~\ref{fig2} for the $A$=25 Urca nuclear pair ($^{25}$Mg, $^{25}$Na). \label{fig3}}
\end{minipage}
\end{center}
\end{figure}

The e-capture and $\beta$-decay rates for the ($^{23}$Na, $^{23}$Ne) and ($^{25}$Mg, $^{25}$Na) 
pairs are shown in Figs.~\ref{fig2} and ~\ref{fig3}, respectively. 
Here the Coulomb corrections are taken into account \cite{Juod,Toki,Suzu2016}.
The Coulomb corrections affect the thermodynamic properties of a high density plasma.
The interaction of ions in a uniform electron background leads to  corrections to the equation of state of matter, and modifies the chemical potential of the ions. 
The corrections to the chemical potential have significant effects on the abundance distributions of nuclei in nuclear statistical equilibrium (NSE), as well as on the conductive thermonuclear flames and the neutralization rate of matter in NSE in massive white dwarfs near the Chandrasekhar point \cite{Bravo}.

The Coulomb effects on the weak rates are mainly caused by the modification of the threshold energy,
\begin{equation}
\Delta Q_C = \mu_C (Z-1) -\mu_C (Z),
\end{equation}
where $\mu_C$ ($Z$) is the Coulomb chemical potential of the nucleus with charge number $Z$ \cite{Slattery,Ichimaru,Yakov}.
The threshold energy is enhanced for e-capture processes, and the e-capture ($\beta$-decay) rates are reduced (enhanced) by the Coulomb effects. 
  
Another correction to the rates comes from the reduction of the electron chemical potential. 
The amount of the reduction is evaluated by using the dielectric function obtained by relativistic random phase approximation (RPA) \cite{Itoh}. 
This correction also leads to a slight reduction (enhancement) of e-capture ($\beta$-decay) rates.

Effects of the Coulomb effects on the rates are  shown in Figs.~\ref{fig2}(b) and ~\ref{fig3}(b) for the $A$=23 and $A$=25 pairs, respectively, at log$_{10}T$ = 8.7 in comparison with the case without the Coulomb effects.
Here, the quenching of the axial-vector coupling constant is taken to be $g_A^{eff}/g_A$ = 0.764 \cite{Richter}. 
Transitions from the states with excitation energies up to $E_x$ = 2 MeV are taken into account. 
For the $A$=23 pair ($^{23}$Na, $^{23}$Ne), transitions from 3/2$_{g.s.}^{+}$, 5/2$^{+}$, 7/2$^{+}$ and 1/2$^{+}$ states of $^{23}$Na are included for the e-capture reactions, while the 5/2$_{g.s.}^{+}$ and 1/2$^{+}$ states of $^{23}$Ne are taken into account for the $\beta$-decays.
The Urca density is found at log$_{10}$($\rho Y_e$) = 8.96 and 8.92 for the case with and without the Coulomb effects, respectively. 
The dependence on the temperature is quite small. 
The Urca density is shifted upwards by $\Delta$log$_{10}$($\rho Y_e$) = 0.04 by the Coulomb effects. 
  
For the ($^{25}$Mg, $^{25}$Na) pair, the Urca density is also found at log$_{10}$($\rho Y_e$) = 8.81 (8.77) for the case with (without) the Coulomb effects. 
Transitions from 5/2$_{g.s.}^{+}$, 1/2$^{+}$, 3/2$^{+}$, 7/2$^{+}$ and 5/2$_{2}^{+}$ states in $^{25}$Mg and 5/2$_{g.s.}^{+}$, 3/2$^{+}$ and 1/2$^{+}$ states of $^{25}$Na are included. 
  
In case of the $A$=27 pair ($^{27}$Al, $^{27}$Mg), the GT transition does not occur between the ground states, 
as the ground states of $^{27}$Al and $^{27}$Mg are 5/2$^{+}$ and 1/2$^{+}$, respectively.     
An evaluation of the weak rates including the second-forbidden transition has been done in Ref.~\cite{SMN2021}.  

Now, we comment on the difference of the GT strengths between the USDB and USD cases.
The GT strengths in $^{23}$Na and $^{25}$Mg are more spread for USDB compared to USD, and larger strengths remain more in the higher excitation energy region for the USDB case\cite{Suzu2016}. 
However, when available experimental data of $B$(GT) and energies are taken into account as in Ref.~\cite{Oda}, the differences in the calculated rates become quite small. 

Besides the ($^{23}$Na, $^{23}$Ne) and ($^{25}$Mg, $^{25}$Na) pairs, Urca processes can occur for ($^{24}$Mg, $^{24}$Na), ($^{21}$Ne, $^{21}$F), ($^{25}$Na, $^{25}$Ne), ($^{23}$Ne, $^{23}$F), and ($^{27}$Mg, $^{27}$Na) pairs at log$_{10}$($\rho Y_e$) $\approx$ 9.3, 9.5, 9.6, 9.8 and 9.9, respectively, for the case without the Coulomb effects.

The e-capture and $\beta$-decay rates, neutrino-energy-loss rates, and $\gamma$-ray heating rates for $sd$-shell nuclei with $A$ =17-28, evaluated with the Coulomb effects by the USDB Hamiltonian, have been tabulated in Ref.~\cite{Suzu2016} for densities log$_{10}$($\rho Y_e$) = 8.0-11.0 in steps of 0.02 and temperatures log$_{10} T$ = 8.0 -9.65 (7.0-8.0) in steps of 0.05 (0.20). % with fine meshes. 
Experimental $B$(GT) and excitation energies available \cite{BW,NNDC,Endt,Tilley} are taken into account here.
Note that the rates in the table of Ref.~\cite{Oda} were evaluated without the Coulomb effects, and were given only at log$_{10}$($\rho Y_e$) = 7, 8, 9, 10 and 11.
On the other hand, as the rates in Ref.~\cite{Oda} have been obtained with larger number of excited nuclear states than in the present calculation, they may be more  
%more accurate for 
suitable for use at higher densities and temperatures where Si burning occurs.

Here, we comment on our choice of fine meshes with steps of 0.02 in log$_{10}$($\rho Y_e$) and 0.05 in log$_{10} T$.
It is not possible to get an accurate rate by interpolation procedures with a sparse grid of densities.  
It is true that a procedure using effective log $ft$ values proposed by Ref.~\cite{Fuller} works well for certain cases where the change of the rates by orders of magnitude comes mainly from the phase space factor while the remaining parts, including the nuclear transition strength, do not change drastically.
When the transitions between the ground states are forbidden or transitions from excited states give essential contributions, this method becomes invalid, for example for the pairs ($^{27}$Al, $^{27}$Mg) and ($^{20}$Ne, $^{20}$F). 
As will be shown in the next subsection, the use of fine grids works well for the calculation of the cooling of the O-Ne-Mg core.
Instead of using tabulated rates with fine grids of density and temperature, an alternative way is to use analytic expressions for the rates as in Ref.~\cite{Pinedo}.
Such a 'on the fly' approach was recently implemented in the stellar evaluation code MESA \cite{Paxton2015} including an extension to forbidden transitions \cite{SMN2021}.
   
\subsection{Cooling of the O-Ne-Mg Core by Nuclear Urca Processes \label{sec:urca}}      
\begin{figure}
\begin{center}
%\begin{minipage}[t]{16.5 cm}
%\vspace*{-5cm}
%\hspace*{-1.5cm}
%\epsfig{file=fig6_ppnp.eps,scale=0.8}
\epsfig{file=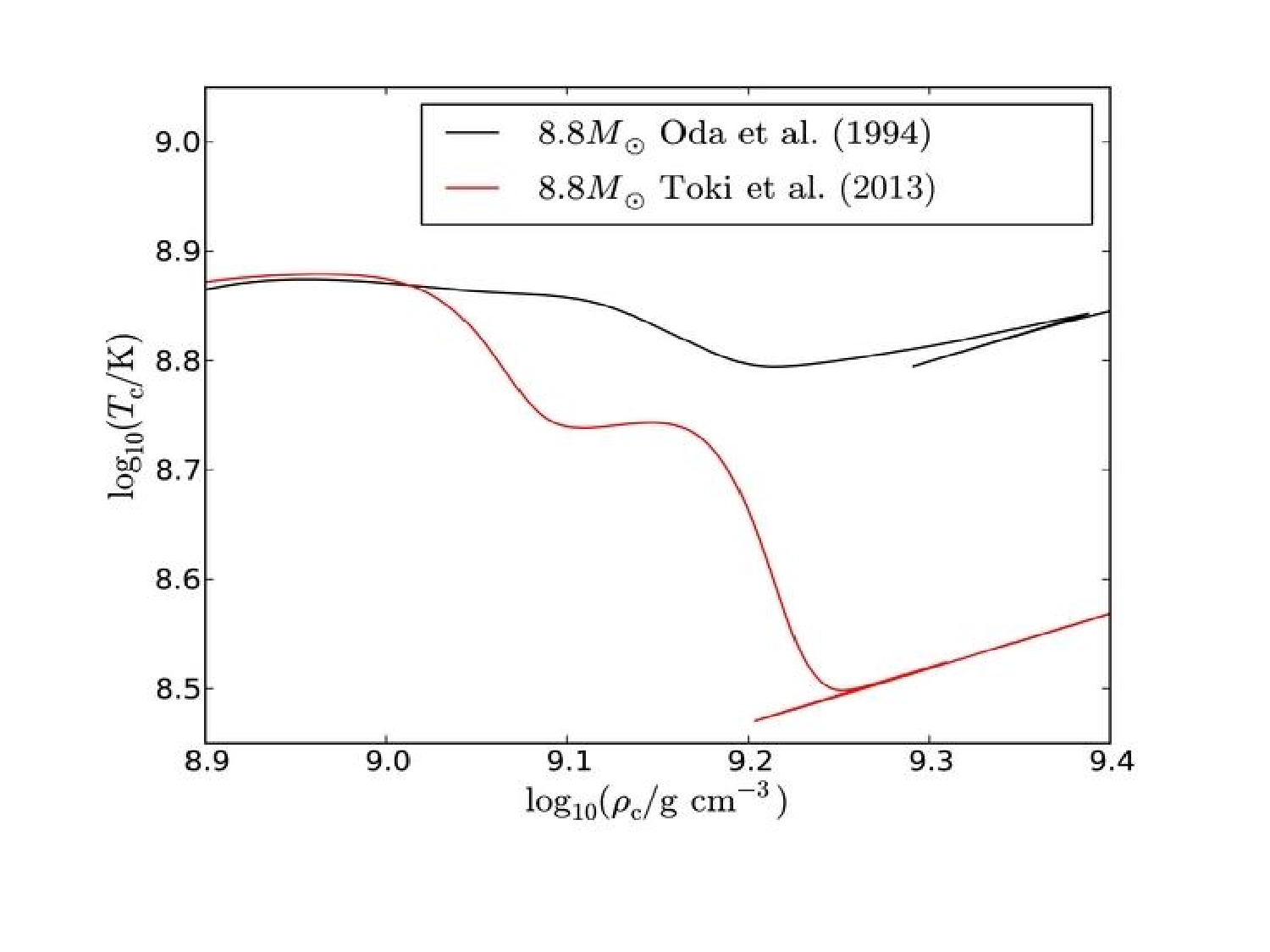, scale =0.5}
%\includegraphics[width=5cm,pagebox=cropbox,clip]{apj476841f12.pdf}
%\epsfig{file=apj476841f12.pdf, scale=1.0}
%\end{minipage}
\begin{minipage}[t]{16.5 cm}
%\vspace*{-3.5cm}
\vspace*{-1.2cm}
\caption{The evolution of the central temperature $T_c$ as a function of central density log$_{10}$($\rho_c$/g cm$^{-3}$).
Cooling of the O-Ne-Mg core of the 8.8$M_{\odot}$ star by the nuclear Urca processes of the pairs ($^{25}$Mg, $^{25}$Na) and ($^{23}$Na, $^{23}$Ne) is shown by the lower curve. 
The upper curve shows the case with the rates of Ref.~\cite{Oda} without using the fine meshes for densities and temperatures.
From Ref.~\cite{Jones}.
%time in units of years. 
\label{fig4}}
\end{minipage}
\end{center}
\end{figure}

Now we show how the cooling of the O-Ne-Mg core is realized by the Urca processes in the nuclear pairs with $A$=23 ($^{23}$Na, $^{23}$Ne) and $A$=25 ($^{25}$Mg, $^{25}$Na).
The time evolution of the central temperature in the 8.8$M_{\odot}$ star is shown in Fig.~\ref{fig4}. 

The star forms an electron-degenerate O-Ne-Mg core after C burning in the central region and, during t = 0 to 4 yr, its central density increases from log$_{10}$ $\rho_c$ = 9.0 to 9.4.
Two distinct drops of the temperature due to $A$=25 Urca cooling up to around log$_{10}\rho_c$ = 9.1,
%t = 1 yr 
and due to $A$=23 Urca cooling, between log$_{10}\rho_c$ =9.15 and 9.25, can be seen.
%t = 2 yr and T = 2.5 yr.
When the temperature drops take place, the abundances of $^{25}$Mg and $^{23}$Na also drop due to the e-captures, eventually dominating over $\beta$-decays. 
%The sharp drops around t = 3 yr and t = 3.7 yr are due to sudden expansion caused by neon shell flash in the outer layers. 
%As a result of the large temperature drop due to Urca cooling, 

After the Urca cooling, double e-captures on $^{24}$Mg and $^{20}$Ne occur inside the 8.8$M_{\odot}$ star, which leads to heating of the core and ignition of oxygen deflagration, resulting in an e-capture SN (ECSN).
If the contraction of the core is fast enough, it will collapse, otherwise thermonuclear explosion may occur.
The final fate, collapse or explosion, is determined by the competition between the contraction of the core due to e-captures on post-deflagration material and the energy release by the propagation of the deflagration flames.
This subject is discussed in Sect.2.5.

Cases for progenitor masses of 8.2$M_{\odot}$, 8.7$M_{\odot}$, 8.75$M_{\odot}$ and 9.5$M_{\odot}$ have also been investigated up to the ignition of the oxygen deflagration in Ref.~\cite{Jones}. 
The 8.2$M_{\odot}$ star is found to end up as O-Ne WD. 
The 8.75$M_{\odot}$ and 8.7$M_{\odot}$ stars evolve toward ECSN due to double e-capture processes.
The 9.5$M_{\odot}$ star evolves to Fe-CCSN.
More detailed discussion can be found in Ref. ~\cite{Jones}.

\begin{figure}[tb]
\begin{center}
\begin{minipage}[t]{16.5 cm}
\hspace*{-1.0 cm}
\epsfig{file=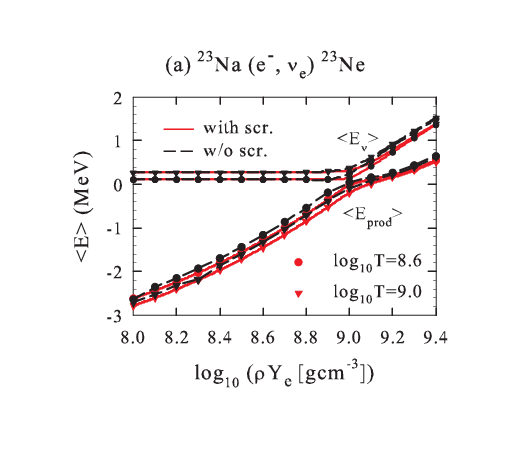,scale=1.0}
%fig7_a23a.eps,scale=1.1}
%energtnane23.eps, scale=1.1}
\hspace*{-1.2 cm}
\epsfig{file=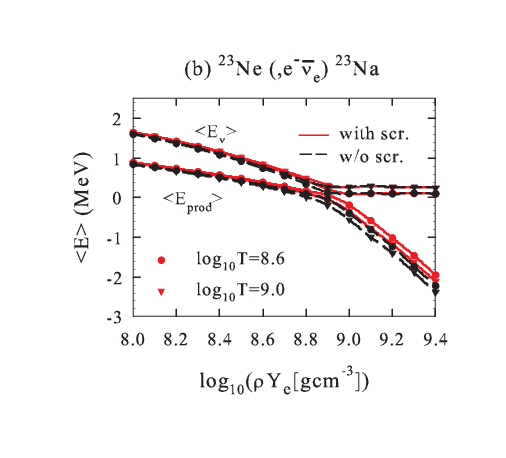,scale=1.1}
%fig7_a23b.eps,scale=1.1}
%energtnena23.eps, scale=1.1}
\end{minipage}
\begin{minipage}[t]{16.5 cm}
\vspace*{-1cm}
\caption{Averaged neutrino energy $< E_{\nu} >$ and averaged energy production $< E_{prod} >$ in (a) e-capture reactions on $^{23}$Na and (b) $\beta$-decay transitions from $^{23}$Ne for temperatures log$_{10} T$ =8.6 and 9.0 as functions of density log$_{10}$($\rho Y_e$). 
Cases with and without the Coulomb effects are denoted by solid and dashed curves, respectively. \label{fig5}}
\end{minipage}
\end{center}
\end{figure}

Next, we discuss the energy loss by neutrino emissions, and the heating by $\gamma$ emissions during the e-capture and $\beta$-decay processes.
The averaged energy production for e-capture processes is defined as
\begin{equation}
< E_{prod} > = \mu_e -Q_{nucl}^{ec} -< E_{\nu} > 
\end{equation}
where $< E_{\nu} >$ is the averaged energy loss due to neutrino emissions, and $Q_{nucl}^{ec}$ = M$_{d}$c$^{2}$ -M$_{p}$c$^{2}$ is the energy difference between the ground states of daughter and parent nuclei.
The averaged energy production for $\beta$-decay processes is defined as
\begin{equation}
< E_{prod} > = Q_{nucl}^{ec} -\mu_e -< E_{\nu} >.
\end{equation}
$< E_{prod} >$ and $< E_{\nu} >$ are shown in Fig.~\ref{fig5} for the Urca process in the $^{23}$Na-$^{23}$Ne pair. 
In the case of the e-capture reaction, $< E_{\nu} >$ increases above the Urca density and the increase of $< E_{prod} >$ starts to be suppressed just at the density, where the energy production becomes positive. 
In the case of the $\beta$-decay transition, $< E_{prod} >$ is suppressed by neutrino emissions below the Urca density. When it becomes negative at the Urca density, the energy loss begins to increase monotonically with increasing density.
In both cases, the energy production is suppressed by neutrino emissions when it becomes positive.

\subsection{Weak Rates for the Forbidden Transitions $^{20}$Ne (0$_{g.s.}^{+}$) $\leftrightarrow$ $^{20}$F (2$_{g.s.}^{+}$) \label{sec:ne20fb}}

The weak rates for $A$=20 pairs, ($^{20}$Ne, $^{20}$F) and ($^{20}$F, $^{20}$O), were evaluated in Refs.~\cite{Suzu2016,Takahara,Pinedo}.
In the case of the ($^{20}$Ne, $^{20}$F) pair, the transitions between the ground states are forbidden and the main GT contributions come from transitions between $^{20}$Ne (0$_{g.s.}^{+}$) and $^{20}$F (1$^{+}$, 1.057 MeV) and those between $^{20}$Ne (2$^{+}$, 1.634 MeV) and $^{20}$F (2$_{g.s.}^{+}$).
The rates were obtained in Ref.~\cite{Takahara} by assuming only GT transitions. 
The effects from the second forbidden transition between the ground states were also estimated in Refs.~\cite{Pinedo,Suzu2016} by assuming that the transition is an allowed GT one, with the strength determined to reproduce log {\it ft} = 10.5, which was the experimental lower limit for the $\beta$-decay, $^{20}$F (2$_{g.s.}^{+}$) $\rightarrow$ $^{20}$Ne (0$_{g.s.}^{+}$) \cite{NNDC}.  
However, the strengths for forbidden transitions generally depend on the lepton energies, contrary to the case of allowed transitions.
Recently, a new log {\it ft} value for the $\beta$-decay was measured: log {\it} = 10.89$\pm$0.11 \cite{Kirsebom}.

Here, we evaluate the weak rates for forbidden transitions in proper ways by using the multipole expansion method of Walecka \cite{Walecka} as well as the method of Behrens-B$\ddot{\mbox{u}}$hring \cite{bb1971}.
Electrons are treated as plane waves in the method of Walecka, while 
in the method of Behrens-B$\ddot{\mbox{u}}$hring electrons are treated as distorted waves in a Coulomb potential, and coupling terms between the transition operators and the Coulomb wave functions are taken into account.    
The latter method is more accurate, but its formulae are rather complex for primers.
We start from the method of Walecka, which is easier to handle, and compare it with that of Behrens-B$\ddot{\mbox{u}}$hring, clarifying their differences.
%The difference in the calculated rates between the two methods is will be shown to be small in Sect. 2.4 for the present case with $Z$=10 as far as the conserved-vector-current (CVC) relation for the electric quadrupole (E2) operator is fulfilled.}

The e-capture rates for finite density and temperature are given as
\cite{Walecka,Ocon,Paar,Vretenar},
\begin{eqnarray}
\lambda^{ecap}(T) &=& \frac{V_{ud}^2 g_V^2 c}{\pi^2 (\hbar c)^3}
\int_{E_{th}}^{\infty} \sigma(E_e,T) E_e p_e c S_e (E_e) dE_e \nonumber\\
\sigma(E_e,T) &=& \sum_{i} \frac{(2J_i +1)e^{-E_i/kT}}{G(Z,A,T)}
 \sum_{f}\sigma_{f,i}(E_e)\nonumber\\
G(Z,A,T) &=& \sum_{i} (2J_i +1) e^{-E_i/kT}, 
\end{eqnarray}
where $V_{ud} = \cos \theta_C$ is the up-down element in the Cabibbo-Kobayashi-Maskawa quark mixing matrix with $\theta_C$ the Cabibbo angle, $g_V = 1$ is the weak vector coupling constant, $E_e$ and $p_e$ are electron energy and momentum, respectively, $E_{th}$ is the threshold energy for the electron capture, and
$S_e (E_e)$ is the Fermi-Dirac distribution for the electron.
%The electron chemical potential is determined from $\rho Y_e$ with $\rho$ the baryon density and $Y_e$ is the proton fraction.
%Here, $i$ denotes the initial state with excitation energy $E_{i}$ and angular momentum $J_i$, and $f$ specifies the final state.  
The cross section $\sigma_{f,i}(E_e)$ from an initial state with $E_i$ and spin $J_i$ to a final state with excitation energy $E_f$ and spin $J_f$ is evaluated with the multipole expansion method \cite{Walecka,Ocon} as follows: 
\begin{eqnarray}
\sigma_{f,i}(E_e) 
%\int (\frac{d\sigma}{d\Omega})_{f,i} d\Omega \nonumber\\
%(\frac{d\sigma}{d\Omega})_{f,i} 
&=& \frac{G_F^2}{2\pi} F(Z,E_e) W(E_{\nu}) C_{f,i}(E_e)\nonumber\\ 
C_{f,i}(E_e) &=& \frac{1}{2J_i +1} \int d\Omega \Bigl(\sum_{J\geq1}\{[1-(\hat{\vec{\nu}}\cdot\hat{\vec{q}})(\vec{\beta}\cdot\hat{\vec{q}})]  
[|\langle J_f || T_J^{mag} || J_i\rangle|^2 +|\langle J_f || T_J^{elec} || J_i\rangle|^2]\nonumber\\
&-&2\hat{\vec{q}}\cdot(\hat{\vec{\nu}}-\vec{\beta}) Re \langle J_f || T_J^{mag} || J_i\rangle \langle J_f || T_J^{elec} ||J_i\rangle^{*}\} \nonumber\\ 
&+& \sum_{J\geq0}\{[1-\hat{\vec{\nu}}\cdot\vec{\beta}+2(\hat{\vec{\nu}}\cdot\hat{\vec{q}})(\vec{\beta}\cdot\hat{\vec{q}})]
|\langle J_F || L_J || J_i\rangle|^2 + (1+\hat{\vec{\nu}}\cdot\vec{\beta})|\langle J_f || M_J || J_i \rangle|^2 \nonumber\\
&-&2\hat{\vec{q}}\cdot(\hat{\vec{\nu}}+\vec{\beta})Re\langle J_f||L_J ||J_i\rangle \langle J_f || M_J|| J_i\rangle^{*}\}\Bigr) , 
\end{eqnarray}
where $\vec{q} = \vec{\nu} -\vec{k}$ is the momentum transfer with $\vec{\nu}$ and $\vec{k}$ the neutrino and electron momentum, respectively, $\hat{\vec{q}}$ and $\hat{\vec{\nu}}$ are the corresponding unit vectors and $\vec{\beta}$ =$\vec{k}/E_e$.   
$G_F$ is te Fermi coupling constant, $F(Z, E_e)$ is the Fermi function, and
$W(E_{\nu})$ is the neutrino phase space factor given by 
\begin{equation}
W(E_{\nu}) = \frac{E_{\nu}^2}{1+E_{\nu}/M_{T}}, 
\end{equation}
where $E_{\nu} = E_e +Q +E_i -E_f$ is the neutrino energy with $E_i$ and $E_f$ the excitation energies of initial and final nuclear states, respectively, and $M_T$ is the target mass. The $Q$ value is determined from $Q = M_i -M_f$, where $M_i$ and $M_f$ are the masses of parent and daughter nuclei, respectively.
%$E_i$ and $E_f$ are excitation energies of initial and final states, respectively.
The Coulomb, longitudinal, transverse magnetic and electric multipole operators with multipolarity $J$ are denoted as $M_J$, $L_J$, $T_J^{mag}$ and $T_J^{elec}$, respectively.

In the multipole expansion formula, transition matrix elements of the Coulomb multipole with $J$=0 corresponds to allowed Fermi transition.
%$q=|\vec{q}| \rightarrow$ 0,
$M_0(q)=F_{1}^{V}(q^2) \sum_{k} j_0(qr_{k}) Y^{0}(\Omega_{k}) t_{+}^{k}$, where $F_1^{V}(q^2)$ is the nucleon Dirac form factor and the sum over each nucleon, denoted by k, is taken.
In the limit of low momentum transfer, $q=|\vec{q}| \rightarrow$ 0, this simplifies to $M_{0}(q)$ = $\sum_{k} \frac{1}{\sqrt{4\pi}} t_{+}^{k}$, and then 
\begin{eqnarray}
C_{f,i}(E_e) = \frac{1}{2J_i +1}\int (1+\hat{\vec{\nu}}\cdot\vec{\beta}) \langle J_f|| M_0 ||J_i \rangle|^2 d\Omega 
&=& \frac{1}{2J_i +1} \int (1+\beta cos\theta) \frac{1}{4\pi} |\langle J_f|| \sum_{k} t_{+}^{k} ||J_i \rangle |^2 2\pi sin\theta d\theta\nonumber\\
= \frac{1}{2J_i +1} \int_{-1}^{+1} (1+\beta t) dt \frac{1}{2} | \langle J_f|| \sum_{k} t_{+}^{k} || J_i \rangle |^2
&=& \frac{1}{2J_i +1}|\langle J_f|| \sum_{k} t_{+}^{k} ||J_i \rangle|^2
\end{eqnarray}
where $t =\hat{\vec{\nu}}\cdot\hat{\vec{\beta}} =cos\theta$ and $\beta$ =$|\vec{\beta}|$.
Note that energy transfer to the isobaric analog state ($-q_0$) is zero, and the longitudinal multipole $L_0$ does not contribute.  
The transition matrix elements of the axial electric dipole operator, $T_{1}^{elec,5}$ $\approx F_{A}(q^2) \sum_{k} \sqrt{\frac{2}{3}} j_0(qr_{k})Y^{0}(\Omega_{k}) \vec{\sigma}_{k}t_{+}^{k}$, and the axial longitudinal dipole operator, $L_{1}^{5}$ $\approx F_A(q^2) \sum_{k} \sqrt{\frac{1}{3}} j_0(qr_{k})Y^{0}(\Omega_{k}) \vec{\sigma}_{k}t_{+}^{k}$, where $F_A(q^2)$ is the nucleon axial-vector form factor and $F_A(0)$ = $g_A$, correspond to allowed GT transitions.
In the limit of $q \rightarrow$ 0,
% and vanishing lepton mass , 
\begin{eqnarray}
C_{f,i}(E_e) &=& \frac{1}{2J_i +1} \int [\{(1-(\hat{\vec{\nu}}\cdot\hat{\vec{q}})(\vec{\beta}\cdot\hat{\vec{q}})\}  
|\langle J_f || T_{1}^{elec, 5} || J_i\rangle|^2 \nonumber\\
&+& \{(1-\hat{\vec{\nu}}\cdot\vec{\beta})+2(\hat{\vec{\nu}}\cdot\hat{\vec{q}})(\vec{\beta}\cdot\hat{\vec{q}})\}
|\langle J_f || L_{1}^{5} || J_i\rangle|^2 ] d\Omega \nonumber\\
&=& \frac{1}{2J_i +1}  \int[\{1-(\hat{\vec{\nu}}\cdot\hat{\vec{q}})(\vec{\beta}\cdot\hat{\vec{q}})\}
\frac{1}{3} |\langle J_f || g_A \sum_{k} \vec{\sigma}_{k} t_{+}^{k} ||J_i \rangle|^2 \nonumber\\
&+&
\{(1-\hat{\vec{\nu}}\cdot\vec{\beta})+2(\hat{\vec{\nu}}\cdot\hat{\vec{q}})(\vec{\beta}\cdot\hat{\vec{q}})\} |\frac{1}{6} |\langle J_f || g_A \sum_{k} \vec{\sigma}_{k} t_{+}^{k} ||J_i \rangle|^2] dt \nonumber\\
%= \int_{1-}^{+1} [(1-t + \frac{1+t}{2}\frac{q^2-q_0^2}{q^2}) \frac{2}{3} \frac{1}{4\pi} |\langle J_f || \sigma t_{+} || J_i\rangle |^2 
%&+& ((1+t)\frac{q^2-q_0^2}{q^2} \frac{1}{3}\frac{1}{4\pi} |\langle J_f || \sigma t_{+} || J_i \rangle|^2) 2\pi dt\nonumber\\
&=& \frac{1}{2J_i +1} \int_{-1}^{+1} [\frac{1}{3} + \frac{1}{6}(1-\beta t)] |\langle J_f || g_A \sum_{k} \vec{\sigma}_{k} t_{+}^{k} || J_i\rangle |^2 dt \nonumber\\ 
&=& \frac{1}{2J_i +1} |\langle J_f|| g_A \sum_{k} \vec{\sigma}_{k} t_{+}^{k} || J_i\rangle|^2. 
\end{eqnarray}

In the case of first-forbidden transitions, the axial Coulomb and longitudinal multipoles contribute for 0$^{-}$ and 2$^{-}$, and axial electric and vector magnetic quadrupoles additionally contribute for 2$^{-}$. 
For 1$^{-}$, there are contributions from the Coulomb, longitudinal and electric dipoles from the weak vector current, and the axial magnetic dipole from the weak axial-vector current.

For second-forbidden transitions, 0$^{+}$ $\leftrightarrow$ 2$^{+}$, the transition matrix elements of the Coulomb, longitudinal and electric transverse operators from the weak vector current, as well as the axial magnetic operator from the weak axial-vector current with multipolarity $J=2$ contribute to the rates. 
\begin{eqnarray}
M_2(q) &=& \sum_{k} F_1^{V}(q^2) j_2(qr_{k})Y^{2}(\Omega_{k}) t_{\pm}^{k}
\nonumber\\
L_2(q) &=& \sum_{k} \frac{1}{M} F_1^{V}(q^2) (\sqrt{\frac{2}{5}}j_1(qr_{k})[Y^1(\Omega_{k}) \times \vec{\nabla}_{k}]^2 +\sqrt{\frac{3}{5}}j_3(qr_{k})[Y^3(\Omega_{k}) \times \vec{\nabla}_{k}]^2) t_{\pm}^{k}  \nonumber\\ 
T_2^{elec}(q) &=& \sum_{k} \frac{q}{M}F_1^{V}(q^2) (\sqrt{\frac{3}{5}}j_1(qr_{k})[Y^1(\Omega_{k}) \times\frac{\vec{\nabla}_{k}}{q}]^2 -\sqrt{\frac{2}{5}}j_3(qr_{k})[Y^3(\Omega_{k}) \times\frac{\vec{\nabla}_{k}}{q}]^2) t_{\pm}^{k}\nonumber\\
  & &+ \sum_{k} \frac{q}{2M}\mu_{V}(q^2) j_2(qr_{k})[Y^2(\Omega_{k}) \times\vec{\sigma}_{k}]^2 t_{\pm}^{k} \nonumber\\
T_2^{mag,5} &=& \sum_{k} F_A(q^2) j_2(qr_{k})[Y^2(\Omega_{k}) \times\vec{\sigma}_{k}]^2 t_{\pm}^{k}, 
\end{eqnarray}
where $M$ is nucleon mass and  
%$F_1^{V}$, 
$\mu^{V}$ 
%and $F_A$ are
is the nucleon magnetic form factor \cite{Kura}. 
%Here, the relation $L_2(q)$=$\frac{q_0}{q}M_2(q)$ from conservation of vector current is used.  
In the low momentum transfer limit, using the following definitions of matrix elements \cite{Schopper}
\begin{eqnarray}
x &=& \frac{1}{\sqrt{2J_i +1}} \langle J_{f}|| \sum_{k}r_{k}^2 C^{2}(\Omega_{k}) ||J_{i}\rangle \nonumber\\
y &=& \frac{1}{\sqrt{2J_i +1}} \langle J_{f}|| \sum_{k} r_{k}[C^{1}(\Omega_{k})\times \frac{\vec{\nabla}_{k}}{M}]^{2} ||J_{i}\rangle \nonumber\\
u &=& \frac{1}{\sqrt{2J_i +1}} g_A \langle J_{f}|| \sum_{k} r_{k}^2 [C^{2}(\Omega_{k})\times \vec{\sigma}_{k}]^{2} ||J_{i}\rangle
\end{eqnarray}
with $C^{\lambda}$ = $\sqrt{\frac{4\pi}{2\lambda+1}} Y^{\lambda}$, 
one can express $C_{f,i}(E_e) \equiv C(k, \nu)$, where $k=|\vec{k}|=\sqrt{E_e^2-m_e^2}$ and $\nu$ =$E_{\nu}$, in the following way:
\begin{eqnarray}
C(k, \nu) &=& \frac{1}{45}x^2 (k^4 -\frac{4}{3}\beta k^3\nu +\frac{10}{3}k^2\nu^2 -\frac{4}{3}\beta k\nu^3 +\nu^4) 
+\frac{2}{15}y^2 (k^2-2\beta k\nu+\nu^2) \nonumber\\
&+& \frac{2}{45}\sqrt{6}xy (\beta k^3 -\frac{5}{3}k^2\nu +\frac{5}{3}\beta k\nu^2 -\nu^3) \nonumber\\ 
%\frac{1}{45}x^2 \frac{8}{3}k^2\nu^2 \nonumber\\
&+&\frac{1}{5}y^2 (k^2+\nu^2+\frac{4}{3}\beta k\nu) + \frac{1}{45}u^2 (k^4+2\beta k^3\nu+\frac{10}{3}k^2\nu^2+2\beta k\nu^3+\nu^4)\nonumber\\
&-& \frac{2}{15}yu (\beta k^3+\frac{5}{3}k^2\nu+\frac{5}{3}\beta k\nu^2+\nu^3). 
\end{eqnarray}
%with $q_0$=$E_{\nu}-E_e$. 
The first, second and third terms in Eq. (17) correspond to the Coulomb, the longitudinal and the interference of the Coulomb and the longitudinal form factors, respectively.
The next terms proportional to $y^2$, $u^2$ and $yu$ denote the transverse electric, the axial magnetic form factors, and their interference form factors, respectively.   
%In the limit of $m_e$=0.0, $\beta$=1 and the first three terms become $\frac{1}{45}x^2\frac{8}{3}k^2\nu^2$. 
%with $k=|\vec{k}|=\sqrt{E_e^2 -m_e^2}$ and $\nu$ = $E_\nu$.

The matrix elements $y$ defined above are related to $x$ via the conservation of the vector current $V_{\pm}$ (CVC).
The longitudinal and the transverse E2 operator in the long-wavelength limit can also be expressed as \cite{Forest}
\begin{eqnarray}
L_2(q) &=& -\frac{i}{q} \sum_{k} \frac{q^2 r_{k}^2}{15} 
%j_2(qr_{k}) 
\vec{\nabla}\cdot\vec{V}_{\pm,k} Y^2(\Omega_{k})  \nonumber\\
T_2^{elec}(q) &=& -\frac{i}{q} \sqrt{\frac{3}{2}} \sum_{k} \frac{q^2 r_{k}^2}{15}
%j_2(qr_{k}) 
(\vec{\nabla}\cdot\vec{V}_{\pm,k} + \frac{q^2}{3}\frac{\mu_{V}(q^2)}{2M}\vec{\nabla}\cdot[\vec{r}_{k}\times\vec{\sigma}_{k}]) Y^2(\Omega_{k})
\end{eqnarray}
The CVC relation can be expressed as 
%we get
\begin{equation}
\vec{\nabla}\cdot\vec{V}_{\pm} = -\frac{\partial\rho_{\pm}}{\partial t} = -i[H, \rho_{\pm}]
\end{equation}
where $\rho_{\pm}= F_1^{V}(q^2) \sum_{k} \vec{\delta}(\vec{r}-\vec{r}_{k}) t_{\pm}$ is the time component of $V_{\pm}$ and H is the total Hamiltonian of the nucleus.
From Eqs. (15), (18) and (19), one can show that the relations $\langle J_f || L_2(q) ||J_i\rangle = - \frac{E_f -E_i}{q} \langle J_f || M_2(q) ||J_i \rangle$, with $E_i$ and $E_f$ the energies of the initial and final states, respectively, and 
$\langle J_f || T_2(q) || J_i \rangle = -\sqrt{\frac{3}{2}}\frac{E_f -E_i}{q} \langle J_f || M_2(q) ||J_i\rangle$, are fulfilled in the low momentum transfer limit. 
These relations are equivalent to 
\begin{equation}
y = -\frac{E_f -E_i}{\sqrt{6}\hbar}x.
\end{equation}
When the electromagnetic interaction is added to the CVC relation \cite{Eichler,BlinStoyle},
the energy difference is modified to include the isovector part of the electromagnetic interaction rotated into the $\pm$ direction in isospin space, that is, the Coulomb energy difference, and the neutron-proton mass difference \cite{Fujita1,Fujita2},
\begin{equation}
\Delta E \equiv E_f -E_i \pm V_C \mp (m_n -m_p).
\end{equation}
%\cite{Fujita1,Fujita2,BlinStoyle}.  
For the transition $^{20}$Ne (0$^{+}$, g.s.) $\rightarrow$ $^{20}$F (2$^{+}$, g.s.), this is just the excitation energy of $^{20}$Ne (2$^{+}$, T=1. 10.274 MeV), the analog state of $^{20}$F (2$^{+}$, T=1, g.s.).

\begin{figure}[tb]
\begin{center}
\begin{minipage}[t]{16.5 cm}
\hspace*{-1.0 cm}
\epsfig{file=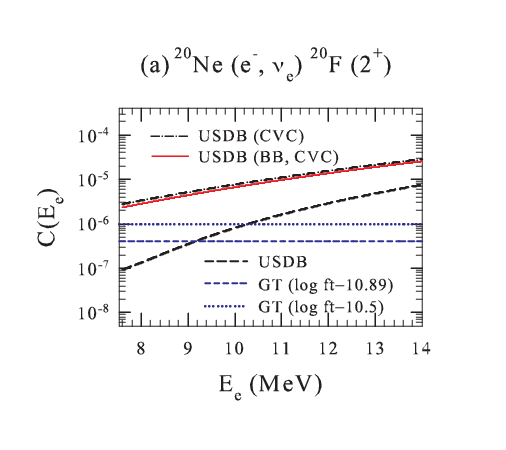, scale=1.15}
\hspace*{-1.0 cm}
\epsfig{file=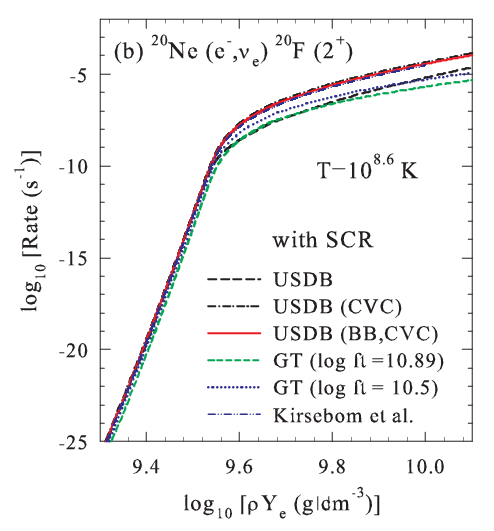, scale=1.0}
%ratenef20walbb.eps, scale=1.05}
\end{minipage}
\begin{minipage}[t]{16.5 cm}
%:q
\vspace*{-1cm}
\caption{(a) Shape factors as functions of electron energy, and  (b) e-capture rates for the second-forbidden transition, $^{20}$Ne (0$_{g.s.s}^{+}$) (e$^{-}$, $\nu_e$) $^{20}$F (2$_{g.s.}^{+}$) as functions of density log$_{10}$($\rho Y_e$). 
The rates are evaluated with the Coulomb (screening) effects at log$_{10} T$ =8.6. 
The dashed and dash-dotted curves are obtained by shell-model calculations with the USDB Hamiltonian by using the multipole expansion method of Walecka without and with the CVC for the evaluation of the transverse E2 transition matrix elements, respectively.
The solid curve is obtained by the method of Behrens-B$\ddot{\mbox{u}}$hring with the CVC for the transverse E2 matrix elements.
The short-dashed and dotted curves are results of GT prescription, in which the transition is treated as an allowed GT one with the strength determined to reproduce log $ft$ = 10.89 and 10.5, respectively, for the $\beta$-decay.         
The dashed-two-dotted curve is from \cite{Kirsebom}.
 \label{fig6}}
\end{minipage}
\end{center}
\end{figure}

Here, we evaluate the electron-capture rates for the forbidden transition $^{20}$Ne (0$_{g.s.}^{+}$) $\rightarrow$ $^{20}$F (2$_{g.s.}^{+}$) by using the USDB shell-model Hamiltonian \cite{USDB} within $sd$-shell. 
The calculated shape factors and e-capture rates for the forbidden transition 
%obtained with the USDB 
are shown in Fig.~\ref{fig6}.
Here, the quenching factor for the axial-vector coupling constant $g_A$ is taken to be $q =0.764$ \cite{Richter}, and 
harmonic oscillator wave functions with a size parameter $b=1.85$ fm are used.

We now use the CVC relation, $y=-\frac{\Delta E}{\sqrt{6}\hbar}x$, to evaluate the longitudinal and transverse electric form factors for $\Delta E$ = 10.274 MeV. 
We note that, when $y$ is evaluated within the $sd$-shell with harmonic oscillator wave functions instead of using the CVC relation, we obtain $y$ = $(E_{sd} -E_{sd}) x$/$\sqrt{6}\hbar$ = 0, as the energy of initial and final states are the same for the single-particle states in the $sd$-shell: $E_{sd}$ = $\frac{7}{2}\hbar\omega$.
The e-capture rates evaluated with the CVC relation are denoted as USDB (CVC),  while those obtained without the CVC relation for the transverse E2 form factor are denoted as USDB in Fig.~\ref{fig6}.
The longitudinal form factor is always evaluated with the CVC relation, as usually done in the method of Walecka. 
  
The calculated rates obtained with the assumption of an allowed GT transition, with a $B$(GT) value corresponding to $\log ft = 10.89$ \cite{Kirsebom}, that is, $B$(GT) $= 0.396 \times 10^{-6}$,  are also shown in Fig.~\ref{fig6}.   
We refer to this method as $^\prime$GT prescription$^\prime$ hereafter.     
A sizeable difference is found between this method and the other two methods explained above. 

In the multipole expansion method of Ref.~\cite{Walecka}, leptons are treated as plane waves 
and effects of Coulomb interaction between the electron and the nucleus are taken into account by the Fermi function. 
However, in forbidden transitions, the Coulomb distortion of electron wave functions needs a more careful treatment.   
This has been done with explicit inclusion of Coulomb wave functions \cite{bb1971,buhring,behrens}.
The shape factor for the e-capture rates for the second-forbidden transition in $^{20}$Ne is given as
\begin{eqnarray}
C(k, \nu) &=& \frac{\nu^2}{3} \{ [y +\sqrt{\frac{2}{3}}x (\frac{E_e}{3}-\frac{\nu}{5}) -u(\frac{E_e}{3}+\frac{\nu}{5}) +\frac{1}{3}3\xi(\sqrt{\frac{2}{3}}x'_1 -u'_1)]^2 +\frac{1}{9}(\sqrt{\frac{2}{3}}x-u)^2 \}\nonumber\\
&+& \frac{k^2}{3} \{[y +\sqrt{\frac{2}{3}}x (\frac{E_e}{5}-\frac{\nu}{3}) -u(\frac{E_e}{5}+\frac{\nu}{3}) + \frac{3\xi}{5}(\sqrt{\frac{2}{3}}x'_2-u'_2)]^2 +\frac{1}{25}(\sqrt{\frac{2}{3}}-u)^2\}\nonumber\\
&+& \frac{\nu^4}{50}(\sqrt{\frac{2}{3}}x-\frac{2}{3}u)^2 + \frac{k^2\nu^2}{27}\frac{2}{3}x^2 + \frac{k^4}{50}(\sqrt{\frac{2}{3}}x+\frac{2}{3}u)^2,
\end{eqnarray}
with     
$\xi$=$\frac{\alpha Z}{2R}$.
Here, $R$ is the radius of a uniformly charged sphere approximating the nuclear charge distribution, and $\alpha$ is the fine structure constant.
The electron radial wave functions are solved in a potential of a uniformly charged sphere whose radius is the nuclear radius, leading to the modifications of the matrix elements, $x$ and $u$.   
The modified matrix elements, $x'_1, x'_2, u'_1$ and $u'_2$, are given as
\begin{eqnarray}
x'_1 &=& \frac{1}{\sqrt{2J_i +1}} \langle f|| r^2C^2(\Omega) I(1,1,1,1;r) ||i\rangle \nonumber\\
x'_2 &=& \frac{1}{\sqrt{2J_i +1}} \langle f|| r^2C^2(\Omega) I(2,1,1,1;r) ||i\rangle \nonumber\\
u'_1 &=& \frac{1}{\sqrt{2J_i +1}} g_A \langle f|| r^2[C^2(\Omega)\times \vec{\sigma}]^2 I(1,1,1,1;r) ||i \rangle \nonumber\\
u'_2 &=& \frac{1}{\sqrt{2J_i +1}} g_A \langle f|| r^2[C^2(\Omega)\times \vec{\sigma}]^2 I(2,1,1,1;r) ||i \rangle
\nonumber\\
%\end{eqnarray} 
%where $I(1,1,1,1;r)$ and $I(2,1,1,1;r)$ are $\frac{2}{3}$ times of those defined in \cite{bb1971};  
I(k,1,1,1;r) &=& \left\{
\begin{array}{ll}
\displaystyle{\frac{2}{3}\{\frac{3}{2}-\frac{2k+1}{2(2k+3)}(\frac{r}{R})^2\}} & \qquad r\leq R \\
\displaystyle{\frac{2}{3}\{\frac{2k+1}{2k}\frac{R}{r}-\frac{3}{2k(2k+3)}(\frac{R}{r})^{2k+1}\}} & \qquad r\geq R
\end{array} \right. 
\end{eqnarray}
where $I(k,1,1,1;r)$'s ($k$=1, 2) are $\frac{2}{3}$ times  of those defined in Ref.~\cite{bb1971}.  
%$I(k,1,1,1;r)$ = $\frac{2}{3}\{\frac{3}{2}-\frac{2k+1}{2(2k+3)}(\frac{r}{R})^2\}$ for $r\leq R$,
%and =$\frac{2}{3}\{\frac{2k+1}{2k}\frac{R}{r}-\frac{3}{2k(2k+3)}(\frac{R}{r})^{2k+1}\}$ for $r\geq R$.   
%I(k,1,1,1;r) = 2/3*{I(k,1,1,1;r) in \cite{bb1971}}
The matrix elements $x'_1$ and $u'_1$ ($x'_2$ and $u'_2$) are reduced about by 22-23$\%$ (25-27$\%$) compared with $x$ and $u$. 
  
%We adopt the CVC relation $y=\frac{q_0}{\sqrt{6}}x$ for the evaluation of the weak rates.   
When the terms with $\xi$ are neglected, Eq. (22) becomes equal to Eq. (17) 
in the limit of $m_e$=0.0,  except for a term 
$(\frac{\nu^2}{27}+\frac{k^2}{75}) (\sqrt{\frac{2}{3}}x-u)^2$, 
which is of higher order in $(\frac{m_e c^2}{\hbar c})^2$ compared to the other terms and negligibly small.  
We adopt here the CVC relation $y=-\frac{\Delta E}{\sqrt{6}}x$ for the evaluation of the weak rates.  
The $\xi$ terms represent coupling between nuclear operators and electron wave functions. 
The shape factor and e-capture rates on $^{20}$Ne are also shown for this method (referred as BB) in Fig.~\ref{fig6}. 

As we see from Fig.~\ref{fig6}(a), the shape factors obtained by the multipole expansion method depend on the electron energies, while those of the GT prescription are energy independent.
When the CVC relation is used for the evaluation of the transverse E2 matrix elements, the shape factor is enhanced especially in the low electron energy region. 
The difference between the Behrens-B$\ddot{\mbox{u}}$hring (BB) and the Walecka methods is insignificant.
The e-capture rates obtained with the CVC are also found to be enhanced compared to those without the CVC relation by an order of magnitude at log$_{10}(\rho Y_e) \ge$ 9.6.
The difference between the BB and the Walecka methods with the CVC relation is rather small.  
The rates obtained with the GT prescription with log $ft$ =10.5, which was adopted in Ref.~\cite{Pinedo}, 
are close to the rates obtained with the CVC at log$_{10}\rho Y_e<$ 9.6, while they become smaller beyond log$_{10}\rho Y_e$ =9.6.
Note that the optimum log {\it ft} value for a constant shape factor determined from likelihood fit to the experimental $\beta$-decay spectrum is equal to 10.46 (see Table III of Ref.~\cite{Kirsebom2}). 
The e-capture rates have been also evaluated in Refs.~\cite{Kirsebom,Kirsebom2} by the method of Behrens-B$\ddot{\mbox{u}}$hring (BB) with the use of the CVC relation for the transverse E2 matrix element. The calculated rates at log$_{10}T$ =8.6 are shown in Fig.~\ref{fig6}(b). 
In Ref.~\cite{Kirsebom2}, the experimental strength of $B$(E2) for the transition $^{20}$Ne (0$^{+}_{g.s.}$) $\rightarrow$ $^{20}$Ne (2$^{+}$, 10.273 MeV) has been used for the evaluation of the matrix element $x$ in Eq. (16). 
This results in a reduction of $x$ by about 27$\%$ compared with the calculated value.
Except for this point, the rates obtained in Refs.~\cite{Kirsebom,Kirsebom2} are essentially the same as those denoted by USDB (BB, CVC) in Fig.~\ref{fig6}.
They are close to each other as we see from Fig.~\ref{fig6}(b).
The rates in Ref.~\cite{Suzuki2019} correspond to those denoted by USDB obtained without the CVC relation.

\begin{figure}
%[tb]
\begin{center}
\begin{minipage}[t]{16.5 cm}
\hspace*{-1.0 cm}
\epsfig{file=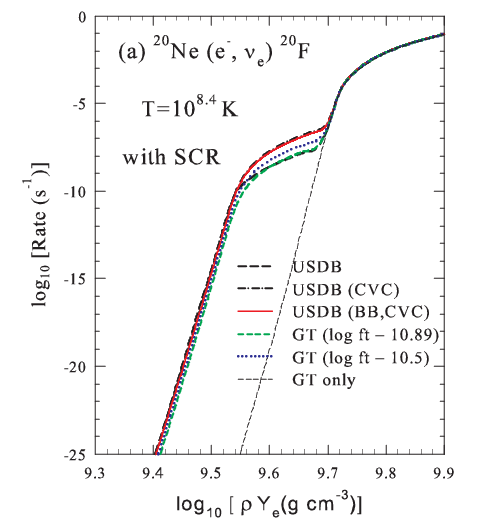, scale=1.0}
%ratenef20walbb84.eps, scale=1.0}
\hspace*{-0.5 cm}
\epsfig{file=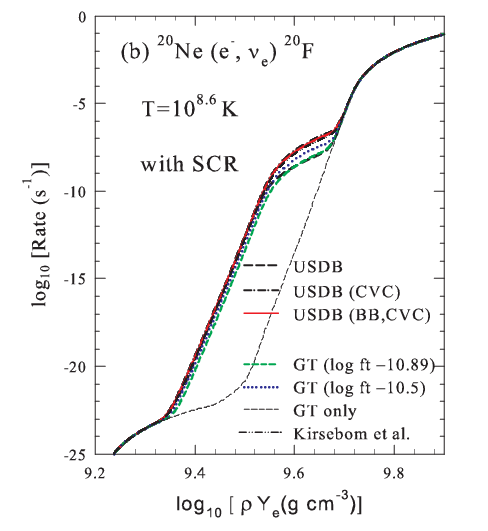, scale=1.0}
%ratenef20walbb86.eps, scale=1.0}
\hspace*{-0.5cm}
\epsfig{file=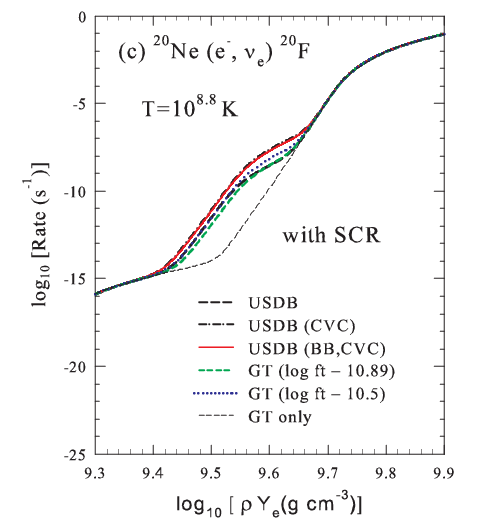,scale=1.0}
%ratenef20walbb88.eps, scale=1.0}
\hspace*{0.3 cm}
\epsfig{file=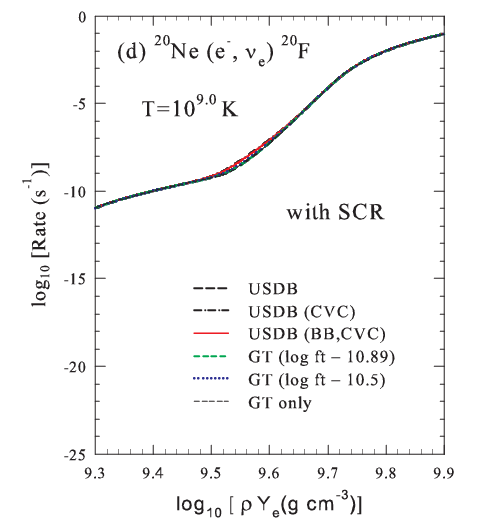,scale=1.0}
%ratenef20walbb90.eps, scale=1.0}
%\vspace*{-2.2cm}
%\hspace*{-3.0cm}
%\epsfig{file=fig9e-a.eps,scale=0.35}
\end{minipage}
\begin{minipage}[t]{16.5 cm}
\vspace*{1.8cm}
\caption{Total e-capture rates on $^{20}$Ne with the Coulomb (screening) effects at log$_{10} T$ =(a) 8.4, (b) 8.6, (c) 8.8 and (d) 9.0 as functions of density log$_{10}$($\rho Y_e$). The curves denote the same cases as in Fig.~\ref{fig6}.
%(e) Electron-capture rates at log$_{10} T$ = 8.6 for the main transitions denoted in the figure. From Ref.~\cite{Kirsebom}.  
 \label{fig7}}
\end{minipage}
\end{center}
\end{figure}

The total e-capture rates on $^{20}$Ne with the contributions from both the GT and the second-forbidden transitions are shown in Fig.~\ref{fig7} for the cases of log$_{10}T$ =8.4, 8.6, 8.8 and 9.0.
The effects of the second-forbidden transition are found to be sizeable at log$_{10}\rho Y_e$ $\approx$ 9.4-9.7 for log$_{10}T\leq$8.8. 
The total e-capture rates evaluated in Refs.~\cite{Kirsebom,Kirsebom2} 
are also shown for log$_{10} T$ =8.6 in Fig.~\ref{fig7}(b).
The rates obtained with the CVC relation, denoted by USDB (CVC) and USDB (BB, CVC), are found to be enhanced by an order of magnitude compared with those without the CVC relation (USDB) as well as those denoted by GT (log $ft$ =10.89) at log$_{10}$($\rho Y_e$) $\approx$9.6 at log$_{10}T$ = 8.4-8.6.

Now we discuss $\beta$-decay rates for the forbidden transition, $^{20}$F (2$_{g.s.}^{+}$) $\rightarrow$ $^{20}$Ne (0$_{g.s.}^{+}$). 
The $\beta$-decay rate for finite density and temperature is given as \cite{Walecka,Ocon},
\begin{eqnarray}
\lambda^{\beta}(T) &=& \frac{V_{ud}^2 g_V^2 c}{\pi^2 (\hbar c)^3}
\int_{m_{e}c^2}^{Q} S(E_e,T) E_e p_e c (Q-E_e)^2 (1 -f(E_e)) dE_e \nonumber\\
S(E_e,T) &=& \sum_{i} \frac{(2J_i +1)e^{-E_i/kT}}{G(Z,A,T)}
 \sum_{f} \frac{G_F^2}{2\pi} F(Z+1, E_e) C_{f,i}(E_e)\nonumber\\
C_{f,i}(E_e) &=& \int \frac{1}{4\pi} d\Omega_{\nu} \int d\Omega_k
\frac{1}{2J_i+1}
\Bigl(\sum_{J\geq1}\{[1-(\hat{\vec{\nu}}\cdot\hat{\vec{q}})(\vec{\beta}\cdot\hat{\vec{q}})]  
[|\langle J_f || T_J^{mag} || J_i\rangle|^2\nonumber\\ 
&+&|\langle J_f || T_J^{elec} || J_i\rangle|^2]
+2\hat{\vec{q}}\cdot(\hat{\vec{\nu}}-\vec{\beta}) Re \langle J_f || T_J^{mag} || J_i\rangle \langle J_f || T_J^{elec} ||J_i\rangle^{*}\} \nonumber\\ 
&+& \sum_{J\geq0}\{[1-\hat{\vec{\nu}}\cdot\vec{\beta}+2(\hat{\vec{\nu}}\cdot\hat{\vec{q}})(\vec{\beta}\cdot\hat{\vec{q}})]
|\langle J_F || L_J || J_i\rangle|^2 + (1+\hat{\vec{\nu}}\cdot\vec{\beta})|\langle J_f || M_J || J_i \rangle|^2 \nonumber\\
&-&2\hat{\vec{q}}\cdot(\hat{\vec{\nu}}+\vec{\beta})Re\langle J_f||L_J ||J_i\rangle \langle J_f || M_J|| J_i\rangle^{*}\}\Bigr),  
\end{eqnarray}
where $\vec{q}$ = $\vec{k}+\vec{\nu}$, and the factor $1-f(E_e)$ denotes the blocking of the decay by electrons in high density matter.
Note that $\Delta E$ is negative ($\Delta E$ = -10.274 MeV), i.e., the sign 
is opposite to the case of the inverse reaction (the e-capture reaction).  

For the second-forbidden $\beta$-decay transition, $^{20}$F (2$_{g.s.}^{+}$) $\rightarrow$ $^{20}$Ne (0$_{g.s.}^{+}$), the shape factors $C_{f,i}(E_e)$  
%$\equiv C(k, \nu)$ 
are obtained by changing signs of $\nu$, $y$=-$\frac{\Delta E}{\sqrt{6}\hbar} x$ and $u$ in $C(k, \nu)$ of Eqs. (17) and (22) for the method of Refs.~\cite{Walecka} and \cite{bb1971}, respectively.

\begin{figure}[tb]
\begin{center}
\begin{minipage}[t]{16.5 cm}
\hspace*{-1.0 cm}
\epsfig{file=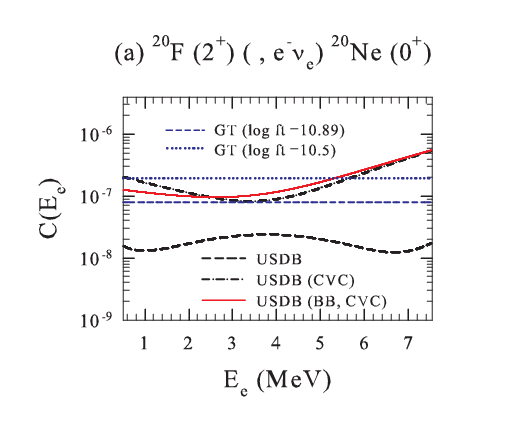,scale=1.05}
%shapebeta.eps, scale=1.05}
\hspace*{-1.0 cm}
\epsfig{file=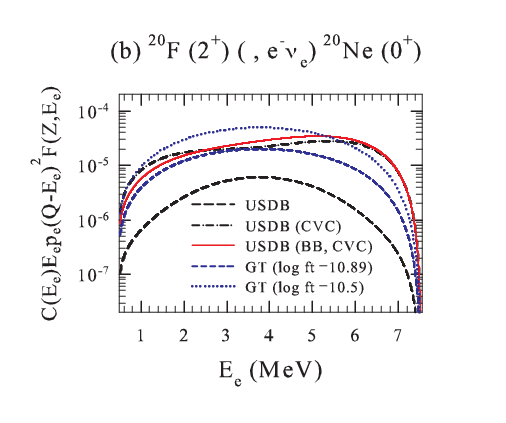,scale=1.10}
%shapebetaphs.eps, scale=1.10}
%ratenef20walbb.eps, scale=1.05}
\end{minipage}
\begin{minipage}[t]{16.5 cm}
%\vspace*{-1cm}
\caption{(a) Shape factors and (b) those multiplied by the phase space factor and Fermi function as functions of electron energy for the second-forbidden transition, $^{20}$F (2$_{g.s.}^{+}$) ( ,e$^{-}$ $\nu_e$) $^{20}$Ne (0$_{g.s.}^{+}$) as functions of density log$_{10}$($\rho Y_e$). 
The curves denote the same cases as in Fig.~\ref{fig6}.
 \label{fig8}}
\end{minipage}
\end{center}
\end{figure}

The shape factors obtained by the multipole expansion method with the USDB depend on the lepton energies as shown in Fig.~\ref{fig8} (a), while those obtained with the GT prescription are energy independent.
The shape factors with the CVC are enhanced compared with those without the CVC about one order of magnitude.
% near $E_e$ =7.5 MeV. 
The shape factors multiplied by the phase space factors also show similar characteristics (see Fig.~\ref{fig8}(b)). A large enhancement is seen for the case with the CVC at $E_e >$ 5 MeV.
The effects of the coupling between electron wave functions and operators are found to be important at lower $E_e$ regions, as indicated by the difference between the results of USDB (BB, CVC) and USDB (CVC).   

The log {\it ft} value for a $\beta$-decay transition is given as \cite{Oda,LM} 
\begin{eqnarray}
ft &=& {\rm ln} 2 \frac{I}{\lambda^{\beta}} \nonumber\\ 
I &=& \int_{m_e c^2}^{Q} E_e p_e c (Q-E_e)^2 F(Z+1, E_e)(1 -f(E_e)) dE_e. 
\end{eqnarray}
Here, $\lambda^{\beta}$ is the $\beta$-decay rate for the transition, and $I$ is the phase space integral. 
In the case of $\beta$-decay in vacuum at $T$ = 0, or in low-density matter at low temperature, the term $(1 -f(E_e))$ can be replaced by 1. 
The log $ft$ value for the $\beta$-decay from $^{20}$F (2$_{g.s.}^{+}$) is calculated to be log $ft$ =10.70 (10.65)
with the USDB for the Walecka (Behrens-B$\ddot{\mbox{u}}$hring) method, when 
the analog state energy is used for $\Delta E$ in $y$=-$\frac{\Delta E}{\sqrt{6}\hbar}x$.  
This value is close to the experimental value: log $ft$ =10.89$\pm$0.11 \cite{Kirsebom}. 
In Ref.~\cite{Kirsebom2}, log {\it ft} =10.86 is obtained with $x$ constrained by the experimental E2 strength in $^{20}$Ne.   
It becomes log $ft$ =11.49 when 
the transverse E2 form factor is calculated within the $sd$-shell using harmonic oscillator wave functions without the CVC relation.

\subsection{Heating of the O-Ne-Mg Core by Double Electron-Capture Processes and Evolution toward Electron-capture Supernovae\label{sec:heat}} 

In this subsection, we discuss the effects of the forbidden transition in the e-capture processes on $^{20}$Ne on the evolution of the final stages of the high-density electron-degenerate O-Ne-Mg cores.
When the core is compressed and the core mass becomes close to the Chandrasekhar mass, the core undergoes exothermic electron captures on $^{24}$Mg and $^{20}$Ne, that release enough energy to cause thermonuclear ignition of oxygen fusion and an oxygen-burning deflagration. 
The final fate of the core, whether collapse or explosion, is determined by the competition between the energy release by the propagating oxygen deflagration wave and the reduction of the degeneracy pressure due to electron captures on the post-deflagration material in nuclear statistical equilibrium (NSE).
As the energy release by double electron captures in $A$=20 and 24 nuclei is about 3 MeV and 0.5 MeV per a capture, respectively, heating of the core due to $\gamma$ emissions succeeding the reactions, $^{20}$Ne (e$^{-}$, $\nu_e$) $^{20}$F (e$^{-}$, $\nu_e$) $^{20}$O, is important in the final stage of the evolution of the core.

\begin{figure}[tb]
\begin{center}
\begin{minipage}[t]{16.5 cm}
\hspace*{-1.0 cm}
\epsfig{file=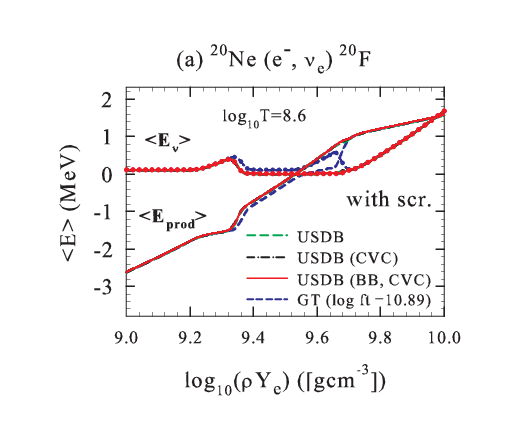,scale=1.0}
%0.72}
%energtnane23.eps, scale=1.1}
\hspace*{-1.0 cm}
\epsfig{file=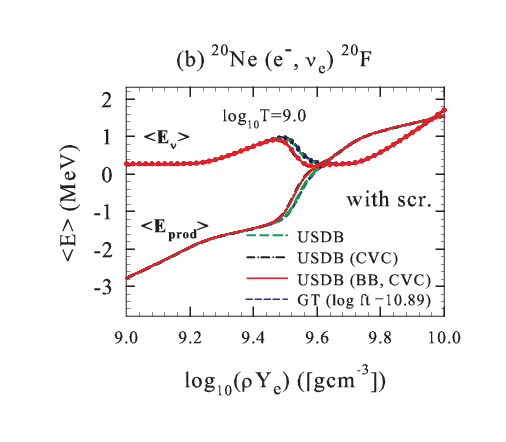,scale=1.0}
%0.72}
%energtnena23.eps, scale=1.1}
\hspace*{-1.0cm}
\epsfig{file=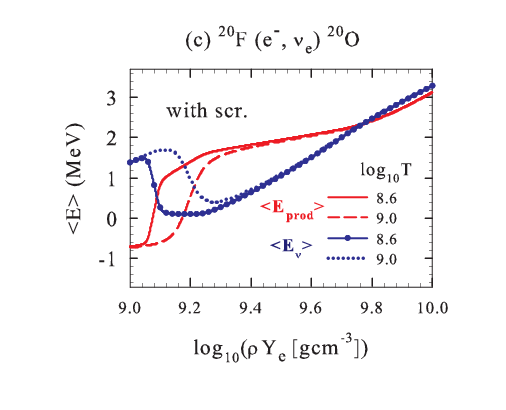,scale=1.0}
%0.76}
\end{minipage}
\begin{minipage}[t]{16.5 cm}
%\vspace*{-0.5cm}
\caption{Averaged neutrino energy $< E_{\nu} >$ and averaged energy production $< E_{prod} >$ in e-capture reactions on (a) $^{20}$Ne at log$_{10} T$=8.6, (b) $^{20}$Ne at log$_{10} T$=9.0, and (c) $^{20}$F at log$_{10} T$ =8.6 and 9.0 as functions of density log$_{10}$($\rho Y_e$) for the case with the Coulomb (screening) effects. \label{fig9}}
\end{minipage}
\end{center}
\end{figure}

Here we discuss the heating of O-Ne-Mg core by double e-capture reactions on $^{20}$Ne in late stages of star evolution. 
The averaged energy production and averaged energy loss by neutrino emissions for e-capture reactions on $^{20}$Ne and subsequent e-capture processes on $^{20}$F are shown in Fig.~\ref{fig9} for temperatures log$_{10} T$ =8.6 and 9.0. 
The energy production for e-capture on $^{20}$Ne is negative up to log$_{10}\rho Y_e$ =9.5 (9.6) at log$_{10} T$ = 8.6 (9.0), while it is positive on $^{20}$F beyond log$_{10}\rho Y_e$ $\approx$9.1 (9.2) at log$_{10} T$ =8.6 (9.0). 
For log$_{10} T$ =8.6, the contributions from the second-forbidden transition become important at log$_{10}\rho Y_e$ =9.3-9.6 and enhance the energy production (see Fig. 6 of Ref. \cite{Pinedo} also for the case without the forbidden transition).
Their effects on the averaged energy production and energy loss by $\nu$ emissions are quite similar in amount among the cases USDB with CVC, USDB without CVC, and BB with CVC, as shown in Fig.~\ref{fig9} (a).   
The net energy production for the double e-captures on $^{20}$Ne and $^{20}$F becomes positive at log$_{10}\rho Y_e$ =9.3 (9.4) at log$_{10} T$ =8.6 (9.0).

The oxygen ignition occurs in the central region of the core, within $\sim$100 km from the center, initiated by heating due to e-capture on $^{20}$Ne. 
Here, ignition is defined as the stage where the nuclear energy generation exceeds the thermal neutrino losses. 
The central density at the oxygen ignition is denoted as $\rho_{c,ign}$.
Subsequent oxygen burning grows into the thermonuclear runaway, that is, oxygen deflagration, when the time scale of temperature rise gets shorter than the dynamical time scale.    
The central density when the oxygen deflagration starts is denoted as $\rho_{c,def}$.
Note that at the oxygen ignition, the heating timescale by local oxygen burning is estimated to be $\sim$10$^{7-8}$ s, which is 8-9 orders of magnitude larger than the dynamical timescale \cite{Zha2019}, and the thermonuclear runaway of the local oxygen burning does not take place yet.     
Thus $\rho_{c,def}$ is not the same as $\rho_{c,ign}$, usually it is higher than $\rho_{c,ign}$.
% thanks to the convection thanks to the convection.  :q

Further evolution of the core depends on the competition between the nuclear energy release by the oxygen deflagration and the reduction of the degeneracy pressure by e-capture in the NSE ash \cite{NomotoKondo1991,Timmes1992,Jones2016,LeungNomoto2019a}.
Recent multidimensional simulations of the oxygen deflagration show that the competition depends sensitively on the value of $\rho_{c,def}$.
If $\rho_{c,def}$ is higher than a certain critical density $\rho_{cr}$, the core collapses to form a neutron star (NS) due to e-capture \cite{Fryer1999,Kitaura2006,Radice2017}, while if $\rho_{c,def}$ $< \rho_{cr}$ thermonuclear energy release dominates to induce partial explosion of the core \cite{Jones2016}. 

For $\rho_{cr}$, the values log$_{10}$($\rho_{cr}$/g cm$^{-3}$) = 9.95-10.3 and 9.90-9.95 have been obtained by the two-dimensional (2D; \cite{NomotoLeung2017a,LeungNomoto2019a,Leung2019a}) and three-dimensional (3D; \cite{Jones2016}) hydrodynamical simulations, respectively.
There exists a large uncertainty in the treatment of the propagation of the oxygen deflagration \cite{Timmes1992} as well as the e-capture rates \cite{Seitenzahl2009}.
The value of $\rho_{c,def}$ is also subject to uncertainties involved in the calculation of the final stage of the core evolution.
The evaluated value of log$_{10}$($\rho_{c,def}$/g cm$^{-3}$) is currently in the range of 9.9-10.2 depending on the treatment of convection \cite{Schwab2015,Schwab2017a,Takahashi2019}. 
Oxygen burning forms a convectively unstable region, which will develop above the oxygen-burning region. 
A smaller value for $\rho_{c,def}$ $\approx 10^{9.95}$ \cite{Schwab2017a} was obtained without convection.  
%Note that at the oxygen ignition, the heating timescale by local oxygen burning is estimated to be $\sim$10$^{7-8}$ s, which is 8-9 orders of magnitude larger than the dynamical timescale \cite{Zha2019}, and the thermonuclear runaway of local oxygen burning does not take place yet.     
  
The evolution of the 8.4 M$_{\odot}$ star from the main sequence until the oxygen ignition in the degenerate O-Ne-Mg core have been studied \cite{Zha2019} using the MESA code \cite{SchwabRocha2019}.
%MESA}.
The weak rates of Ref. \cite{Suzuki2019} including the second-forbidden transition for the e-capture on $^{20}$Ne (denoted as USDB in Figs.~\ref{fig6} and \ref{fig7}) have been used. 
The core evolves through complicated processes of mass accretion, heating by e-capture, cooling by Urca processes, and $Y_e$ change.
%Convective and semiconvective regions are formed. 
%Both the Ledoux and Schwarzschild criteria for the convective stability have been applied because of uncertainties in the semiconvective mixing. 
%We obtained the following findings for the core evolution \cite{Zha2019}.
It has been investigated how the location of the oxygen ignition (center or off-center) and the $Y_e$ distribution depend on the input physics and the treatment of the semiconvection and convection.
There are two extreme criteria for the convective stability, the Schwarzschild criterion and the Ledoux criterion \cite{MiyajiNomoto1987,Kippenhahn2012}.
The Schwarzschild criterion is given as
\begin{eqnarray}
\nabla_{rad} &<& \nabla_{ad} \nonumber\\
\nabla_{rad(ad)} &=& \Bigl(\frac{\partial ln T}{\partial ln P}\Bigr)_{rad(ad)}
\end{eqnarray}
where $\nabla_{rad}$ ($\nabla_{ad}$) is the radiative (adiabatic) temperature gradient.
The Ledoux criterion is given as
\begin{eqnarray}
\nabla_{rad} &<& \nabla_{ad} + (\chi_{Y_e}/\chi_{T})\nabla_{Y_e} \nonumber\\
\nabla_{Y_e} &=& -\frac{\mathrm{d} ln Y_e}{\mathrm{d} ln P} \nonumber\\
\chi_{Y_e} &=& \Bigl(\frac{\partial ln P}{\partial ln Y_e}\Bigr)_{T}, 
\hspace*{0.7cm} \chi_{T} = \Bigl(\frac{\partial ln P}{\partial ln T}\Bigr)_{Y_e} 
\end{eqnarray}
where the $\nabla_{Ye}$ term works to enhance the stability.  
In a region with homogeneous chemical composition, this term vanishes and the Ledoux criterion becomes identical to the Schwartzschild criterion.
The semiconvective region is treated as convectively unstable (stable) when using the Schwartzschild (Ledoux) criterion.   
   
When the Schwarzschild criterion for the convective stability 
%($\grad_{rad} < \grad_{ad}$ where $\grad_{rad}$ and $\grad_{ad}$ are the radiative and adiabatic temperature gradients, respectively) 
is applied, the oxygen ignition takes place at the center.  The convective energy transport delays the oxygen ignition until log$_{10}$($\rho_{c,ign}$/ g cm$^{-3}$) $\sim$10.0 is reached, and the convective mixing makes $Y_e$ in the convective region as high as 0.49.
When the Ledoux criterion for the convective stability is applied, the second-forbidden transition is so slow that it does not ignite oxygen burning at the related threshold density, but decreases the central $Y_e$ to $\sim$0.46 during the core contraction. 
The oxygen ignition takes place when the central density reaches log$_{10}$($\rho_{c,ign}$/g cm$^{-3}$) =9.96-9.97. 
The location of the oxygen ignition, center or off-center at $r_{ign}\sim$ 30-60 km, depends on the $^{12}$C ($\alpha$, $\gamma$) $^{16}$O reaction rate,
which affects the mass fraction of $^{20}$Ne in the core after carbon burning.
Larger (smaller) mass fraction of $^{20}$Ne favors oxygen ignition near (away from) the center.
Even with the Ledoux criterion, the oxygen ignition creates the convectively unstable region, and the convective mixing forms an extended region with $Y_{e}\sim$0.49 above the oxygen ignited shell. 
For both convective stability criteria, the convective energy transport would slow down the temperature increase, and the thermonuclear runaway to form a deflagration wave is estimated to occur at log$_{10}$($\rho_{c,def}$/ g cm$^{-3}$)$>$ 10.10. 
This estimate is consistent with log$_{10}$($\rho_{c,def}$/g cm$^{-3}$) $\approx$10.2 obtained with the semiconvective mixing \cite{Takahashi2019}.

\begin{figure}[tb]
\begin{center}
\begin{minipage}[t]{16.5 cm}
%\hspace*{-1.0 cm}
\epsfig{file=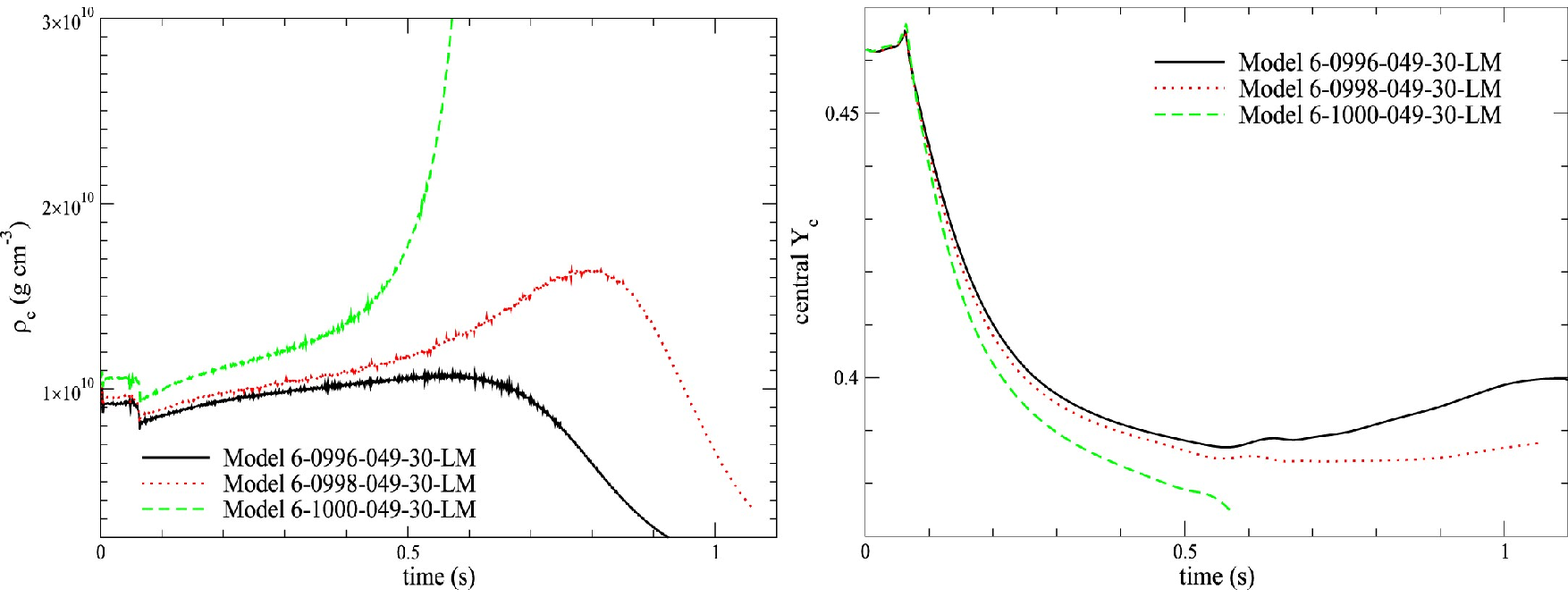,scale=0.55}
\end{minipage}
\begin{minipage}[t]{16.5 cm}
\vspace*{-2.0cm}
\caption{The $\rho_{c,def}$-dependence of the Ledoux models for mass accretion rate $\dot{M} =10^{-6}M_{\odot}$ yr$^{-1}$. 
(Left panel) Central density evolution of models with log$_{10}$($\rho_{c,def}$/g cm$^{-3}$) =9.96, 9.98 and 10.00. Initial value of $Y_e$ =0.49 and oxygen ignition takes place at 30 km from the center. 
The time lapse of $\sim$0.1 s is the time for the flame to arrive at the center to trigger the first expansion. The collapsing model shows a monotonic increase of the central density after the early expansion, while the other two exploding models show a turning point after which the star expands due to the energy input by oxygen deflagration. 
(Right panel) The same as the left panel, but for the central $Y_e$. 
From Ref.~\cite{Zha2019}. 
\label{fig10}}
\end{minipage}
\end{center}
\end{figure}

\begin{figure}[tbh]
\begin{center}
\begin{minipage}[tbh]{16.5 cm}
%\hspace*{-1.0 cm}
\epsfig{file=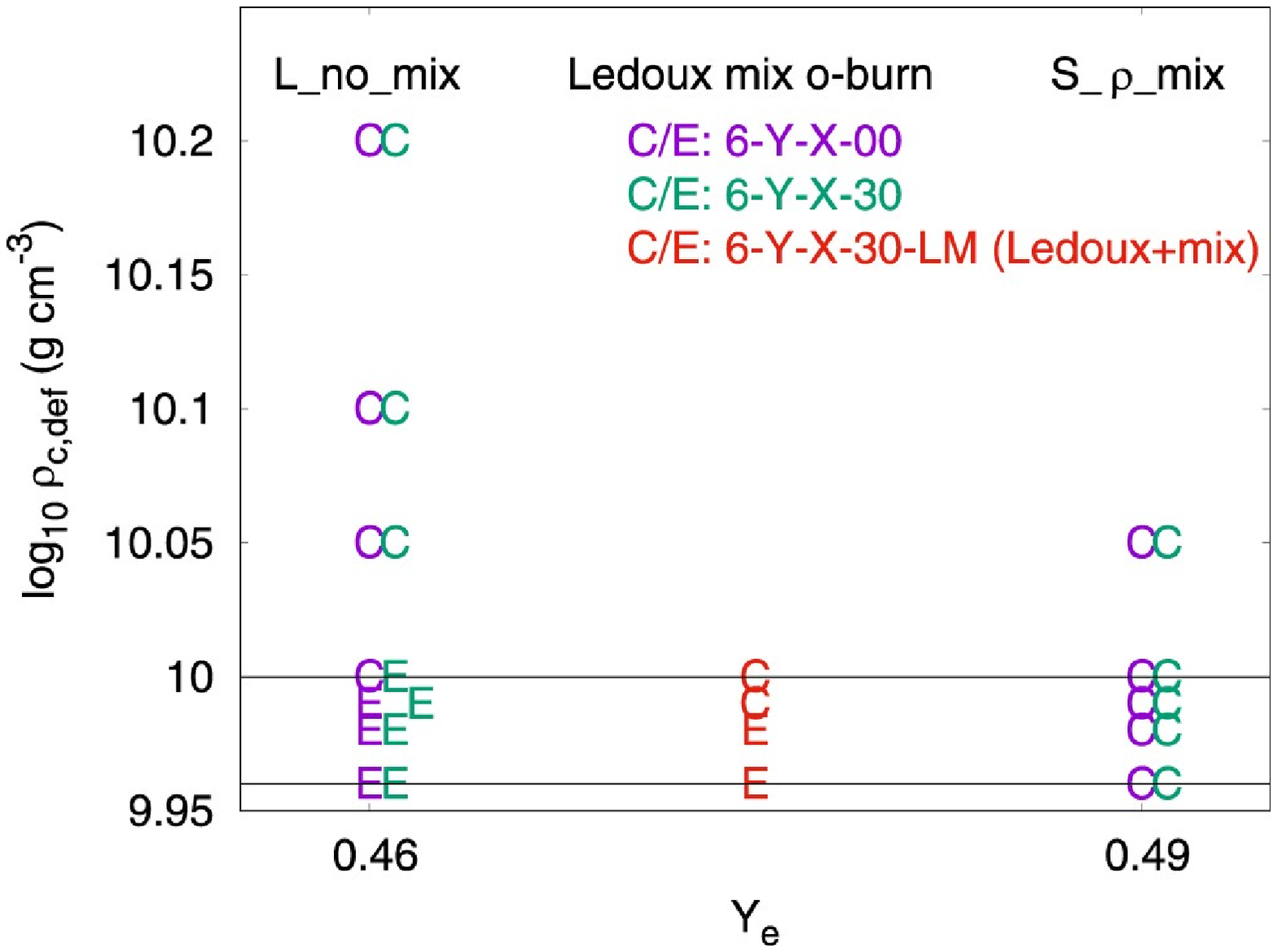,scale=0.275}
\hspace*{-0.5cm}
\epsfig{file=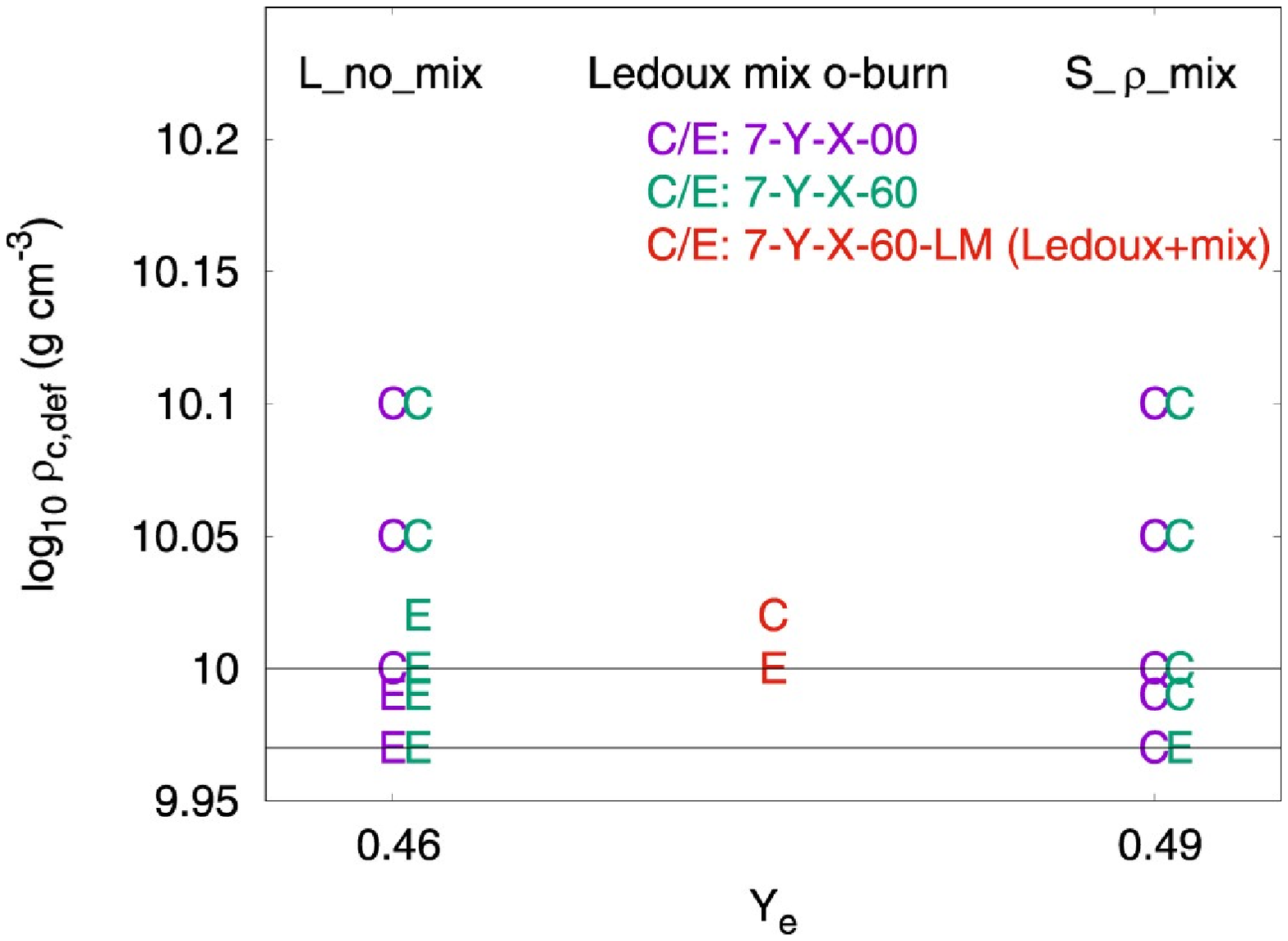,scale=0.275}
\end{minipage}
\begin{minipage}[tbh]{16.5 cm}
\vspace*{0.5cm}
\caption{Explosion-collapse bifurcation diagram as a function of $\rho_{c,def}$ and the initial $Y_e$ distribution for two Ledoux models and one Schwartzschild model. Here "E" and "C" stand for "explosion" and "collapse", respectively.
Left (right) symbols ("E" or "C") for L$_{-}$no$_{-}$mix and S$_{-}\rho_{-}$mix cases correspond to the oxygen ignition at the center (off-center at 30 km from the center).
Left (right) panel corresponds to the case of mass accretion rate of $10^{-6}$ ($10^{-7}$) $M_{\odot}$ yr$^{-1}$.
From Ref.~\cite{Zha2019}.  \label{fig11}}
\end{minipage}
\end{center}
\end{figure}

Then, to examine the final fate of the O-Ne-Mg core, 2D hydrodynamical simulation for the propagation of the oxygen deflagration wave has been performed based on the above simulation of the evolution of the star until the oxygen ignition. 
Three cases of $Y_e$ distributions (Schwartzschild, Ledoux and Ledoux with mixed region above the oxygen-ignited shell), three locations of the oxygen ignition (center, off-center at $r_{ign}$ = 30 km and 60 km), and various central densities at log$_{10}$($\rho_{c,def}$/g cm$^{-3}$) =9.96-10.2 are used for the initial configurations at the initiation of the deflagration (see Ref. \cite{Zha2019} for the details).
The explosion-collapse bifurcation analysis is shown for some initial configurations of the deflagration in Fig.~\ref{fig10},   
where the evolutions of the central density and $Y_e$ as functions of time are shown.     
The explosion-collapse bifurcation diagram is shown in Fig.~\ref{fig11} for the accretion mass rate of $\dot{M}$ = $10^{-6} M_{\odot}$ and $10^{-7} M_{\odot}$. 
The critical density for the explosion-collapse bifurcation is found to be at log$_{10}$($\rho_{crt}$/g cm$^{-3}$) =10.01.
The deflagration starting from log$_{10}$($\rho_{c,def}$/g cm$^{-3}$) $>10.01$ ($<10.01$) leads to a collapse ( a thermonuclear explosion). 
Since $\rho_{c,def}$ is estimated to well exceed this critical value, the O-Ne-Mg core is likely to collapse to form a NS irrespective of the central $Y_e$ and ignition position, although further studies of the convection and semiconvection before the deflagration are needed in future by improving the stellar evolution modeling. 
It would be interesting to see if the present conclusion remains valid for 
the rates calculated with the CVC relation discussed in Sect. 2.4, namely the USDB (CVC) and USDB (BB, CVC) cases, which are enhanced 
%by an order of magnitude 
compared to the rates used here around log$_{10}(\rho Y_e)$ =9.6 at log$_{10}T <$ 8.8. 

%When the rates with the CVC for the  transverse E2 transition, denoted as USDB (CVC) and USDB (BB, CVC), are used for the second-forbidden transition in $^{20}$Ne, the location of the oxygen ignition is pushed away from the center by $\sim$30 km \cite{Zha2020}.  
%This is in a direction in favor of explosion, but the effects on the critical central density for the bifurcation is expected to be as small as log$_{10}$($\rho_{cr}$/g cm$^{-3}$) $\sim$0.02 and the conclusion would remain the same. 
Jones et al. \cite{Jones2016}, using the rates GT (log ft=10.50), and Kirsebom et al. \cite{Kirsebom2} with the rates from USDB (BB, CVC), on the other hand, obtained the opposite conclusion in favor of thermonuclear explosion by assuming that the effects of convection and semiconvection would be small. 
Investigations whether the convective energy transport is efficient enough to delay the ignition and the start of the oxygen deflagration wave to densities above the critical density for collapse were left for the future.     

In case of thermonuclear explosions, the oxygen deflagration results in a partial disruption of the O-Ne-Mg core with an O-Ne-Fe WD left behind \cite{Jones2016}.
The turbulent mixing by the flame allows the ejecta to consist of both Fe-peak elements and O-Ne-rich fuel.
Ejecta can be rich in neutron-rich isotopes such as $^{48}$Ca, $^{50}$Ti, $^{54}$Cr, $^{60}$Fe and $^{66}$Zn, which are overproduced relative to their solar abundances \cite{Jones2019}.
%The enrichment of $^{54}$Cr and $^{50}$Ti is found to be consistent with the most extreme grain among the presolar oxide grains from the Orgueil CI meteorite \cite{Nittler2018}.
A substantial enrichment was reported for $^{54}$Cr and $^{50}$Ti with $^{54}$Cr/$^{52}$Cr and $^{50}$Ti/$^{48}$Ti ratios ranging from 1.2 to 56 and from 1.1 to 455 times the solar values, respectively, in the presolar oxide grains from the Orguil CI meteorite \cite{Nittler2018}.
The enrichment of $^{54}$Cr and $^{50}$Ti obtained in ejecta in the thermonuclear explosion simulation \cite{Jones2016} is found to be consistent with the most extreme grain, $2_{37}$, among the enriched grains.  

Using the solar abundance distribution as constraint, an upper limit of the frequency of thermonuclear e-capture SN explosion (ECSNe) has been estimated to be $\sim$1-3 $\%$ of the frequency of core-collapse SN explosion for the metalicity range $Z$ =0.004-0.00 \cite{Jones2019}.
This probability is 
%a bit or 1 dex 
similar to or one order of magnitude smaller than the following estimates for ECSNe.
%on par or 1 dex lower than the estimates for ECSNe. 
The ECSNe rate was predicted to be 3-21$\%$ \cite{Poela2007} (see \cite{Doherty2015} also) and $\sim$2-5$\%$ \cite{Doherty2015} for $Z$ =0.02-0.0001 from stellar evolution simulations.
The difference at lower $Z$ comes from the metalicity scaling of the mass-loss rate taken in \cite{Poela2007}.   
A narrow initial mass range for ECSNe, at most 0.2 $M_{\odot}$, obtained in \cite{Doherty2015} leads to a lower ECSNe rate, $\sim$2-5$\%$.
The ECSNe rate was predicted to be $\sim$4-20$\%$ at $Z$ =0.02 (solar abundance), where uncertainties in the third dredge-up efficiency and AGB (Asymptotic Giant Branch) mass-loss rate lead to a large span for the rate \cite{Poela2008}.  
The ECSNe rate was estimated to be $\sim$4-10$\%$ of all stellar core-collapse events from nucleosynthesis analysis of elements from Zn to Zr, $^{48}$Ca and $^{60}$Fe \cite{Wanajo2011,Wanajo2013a}.  
%54Cr  50Ti rich   in meteorite    
%Clear observational evidence for thermonuclear ECSN or collapsing ECSN is scarse.
It is not easy to find clear evidence for thermonuclear ECSN (tECSN) or collapsing ECSN (cECSN). 
In tECSN, O-Ne-Fe WD is expected to be formed as a remnant. Information on its mass-radius relation could assign O-Ne-Fe WD \cite{Jones2019}. 
The progenitor of SN2018zd, which proved to eject relatively small amount of $^{56}$Ni and faint X-ray radiation,  
%whose light curve behaved like a low-luminosity SN IIP, 
has been suggested as a massive AGB star that collapsed by ECSN \cite{Zhang2020}.
In Ref. ~\cite{Hira2021}, SN2018zd is shown to have strong evidence for or consistent with six indicators of e-capture supernovae, that is, progenitor identification, circumstellar material, chemical composition, explosion energy, light curve and nucleosynthesis. 
Theoretically, we need to understand more clearly the evolution from the oxygen ignition (at the end of the MESA calculations)  till the beginning of the deflagration by taking into account the semiconvection and convection.

\section{Electron-Capture Rates for $pf$-Shell Nuclei and Nucleosynthesis of Iron-Group Elements}

\subsection{Type Ia Supernova Explosion\label{sec:type1}}

In this section, we discuss important roles played by e-capture rates in $pf$-shell nuclei for the nucleosynthesis of iron-group elements in Type Ia supernova explosions (SNe).
Type Ia supernovae are thought to result from accreting C-O white dwarfs (WDs) in close binaries.

When the WD reaches a certain critical condition, thermonuclear burning initiated in the electron-degenerate core results in a violent explosion of the whole star. 
The subsequent nucleosynthesis results in an abundance of Fe-peak elements and intermediate-mass elements such as Ca, S, Si, Mg and O. 
The ejection of these elements into the interstellar medium (ISM) contributes to the galactic chemical enrichment.
Electron-captures reduce the electron mole fraction $Y_e$, and enhance the abundance of neutron-rich Fe-peak elements. 
The detailed abundance ratios with respect to $^{56}$Fe (or $^{56}$Ni) depend on the central densities of the WDs and the nature of flame propagation triggered by thermonuclear burning. 

There are two types of models for Type Ia SNe. The first is the typical case of accretion from a non-degenerate companion star, where the WD mass approaches the Chandrasekhar mass, inducing a SN Ia. This is known as the single-degenerate progenitor model. 
The other case is the double-degenerate model, where two WDs merge to produce a SN Ia.
The central densities ($\rho_c$) of the exploding WDs are different in the two cases: $\rho_c > 10^{9}$ g cm$^{-3}$ in the single-degenerate model, while $\rho_c \leq 10^{8}$ g cm$^{-3}$ in the double-degenerate model.   
In case of the single-degenerate model with high central densities, a significant amount of Fe-group elements are synthesized as a result of e-capture reactions.
On the other hand, in case of the double-degenerate model with lower central densities, less amount of stable Ni isotopes is produced due to little e-capture processes. 
It is thus important to accurately evaluate the e-capture rates relevant for nucleosynthesis in Type Ia SNe to constrain the explosion conditions and the explosion models.  

\subsection{Electron-Capture Rates in $pf$-Shell\label{sec:ratepf}}

In Ref. \cite{FFN} (hereafter referred as FFN), e-capture rates were obtained based on simple shell-model calculations as well as using available experimental GT strengths. 
It was noticed that an overproduction problem of iron-group elements relative to the solar abundances in Type Ia SNe occurs when the FFN rates are used \cite{Iwa}.
Evaluations of the rates have been improved by large-scale shell-model calculations (LSSM). It was found that the FFN e-capture rates overestimate the rates of the LSSM calculations in many cases \cite{LanM}. 
The e-capture rates in $pf$-shell nuclei are obtained by LSSM calculations with the use of KB3 Hamiltonians \cite{LMRMP}, and they are tabulated for a wide range of nuclei \cite{LM} for the KBF Hamiltonian \cite{KBF}.
The KBF rates are generally used as standards for nucleosynthesis calculations. 

Here, we use a new shell-model Hamiltonian for $pf$-shell, GXPF1J \cite{GX1J}, for the evaluation of the weak rates. 
The GXPF1J is a modified version of the original GXPF1 Hamiltonian \cite{GX1}.
New experimental data of neutron-rich Ca, Ti, and Cr isotopes are taken into account, and the peak position of the magnetic dipole (M1) strength in $^{48}$Ca is reproduced. 
The experimental $B$(M1) strength in $^{48}$Ca is reproduced by reducing the calculated one with the quenching of the spin $g$ factor, $g_{s}^{eff}/g_s$ =0.62$\pm$0.02.     
The KBF and KB3G \cite{KB3G}, the most recent version of KB3's, give energies for the 1$^{+}$ state in $^{48}$Ca about 1 MeV below the experimental one. The M1 strength is split into two states for KB3G.
The M1 transition strengths in $^{50}$Ti, $^{52}$Cr, and $^{54}$Fe are reproduced with $g_{s}^{eff}/g_s$ = 0.75$\pm$0.02 for the GXPF1J \cite{GX1J}. 
The GT$_{-}$ strength in $^{58}$Ni is also found to be well reproduced \cite{GX1J} with the universal quenching for the axial-vector coupling constant, $f_q$ = $g_{A}^{eff}/g_A$ = 0.74 \cite{Caurier}.  
The quenching factor $f_q$ =0.74 will be used for the GT strengths in $pf$-shell nuclei. 
The GT$_{-}$ strengths in Ni and Fe isotopes obtained with the GXPF1J are found to be more fragmented with remaining tails at high excitation energies, compared to those of the KB3G \cite{PTPS,Suzuki2013}. 
This is true also for the GT$_{+}$ strengths in $^{58}$Ni, $^{60}$Ni and $^{62}$Ni \cite{Suzu2011}. 
     
\begin{figure}[tb]
\begin{center}
\begin{minipage}[t]{16.5 cm}
\hspace*{-0.5 cm}
\epsfig{file=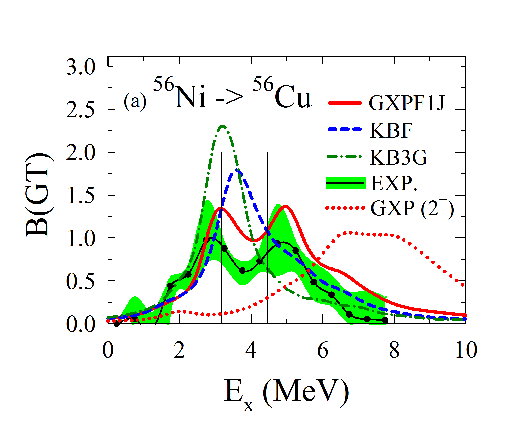,scale=1.0}
%energtnane23.eps, scale=1.1}
\hspace*{-0.3 cm}
\epsfig{file=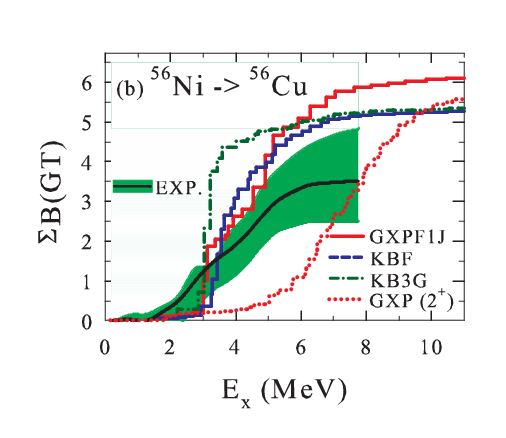,scale=1.0}
%energtnena23.eps, scale=1.1}
%\hspace*{-1.0cm}
%\epsfig{file=fig11_a20c.eps,scale=0.76}
\end{minipage}
\begin{minipage}[t]{16.5 cm}
%\vspace*{-0.5cm}
\caption{GT strengths obtained by shell-model calculations with GXPF1J, KBF and KB3G. (a) $B$(GT) as a function of excitation energy of $^{56}$Cu, and (b) accumulated sum of the $B$(GT) up to a certain excitation energy of $^{56}$Cu, are shown. Experimental data ~\cite{Sasano} are denoted by shaded areas. From Ref.~\cite{Mori2016}. \label{fig12}}
\end{minipage}
\end{center}
\end{figure}

Now we discuss the GT strength in $^{56}$Ni, as experimental data from ($p, n$) reactions are available and $^{56}$Ni is produced in large amounts in the inner part of the WDs. 
The e-capture rates on $^{56}$Ni affects the production yield of $^{58}$Ni, which was known to be over-produced by several times the solar abundance, if the FFN rates are used. 
Suppression of the rates, which leads to less neutron-rich environment with higher $Y_e$, can fix this over-production problem. 
Optical light curves in Type Ia SNe are dominated by photons 
produced in the radiative decay of $^{56}$Ni through $^{56}$Co to $^{56}$Fe.
The maximal luminosity is determined by the amount of $^{56}$Ni.   
Photon emissions from $^{56}$Co and $^{56}$Fe are important for the light curve at later times. 
More accurate evaluation of the e-capture rates on $^{56}$Ni is thus important to fix these issues.
Moreover, one of the most noticeable differences in the strength distribution among shell-model Hamiltonians is seen in the case of $^{56}$Ni.      
Values for $B$(GT) obtained by shell-model calculations with GXPF1J, KBF and KB3G, as well as the experimental data, are shown in Fig.~\ref{fig12} (a). 
 
The GT strength has two peaks for the GXPF1J, consistent with the experimental data. However, this structure is not seen for the KBF and KB3G. 
The strength from the 2$^{+}$ state at 2.70 MeV is also shown for the GXPF1J case.
The integrated GT strengths up to a certain excitation energy ($E_x$) of $^{56}$Cu are compared in Fig.~\ref{fig12}(b).    
Their difference becomes noticeable at $E_x >$ 3 MeV, while it is small in the region corresponding to electron chemical potential less than 1.5 MeV.

\begin{figure}[tb]
\begin{center}
\begin{minipage}[t]{16.5 cm}
\hspace*{-0.5 cm}
\epsfig{file=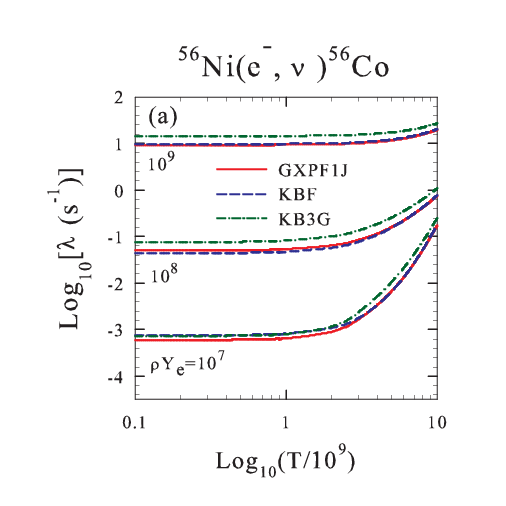,scale=1.0}
%energtnane23.eps, scale=1.1}
\hspace*{-0.5 cm}
\epsfig{file=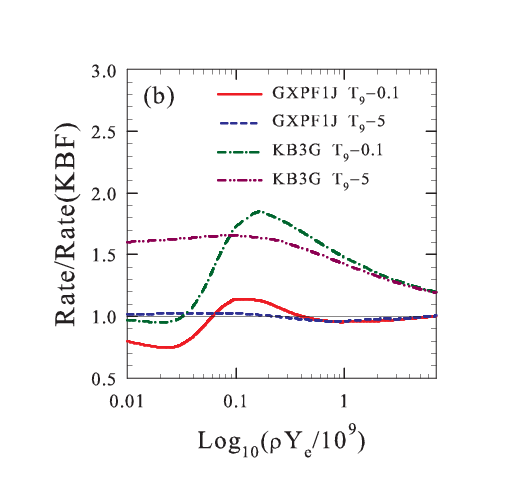,scale=1.0}
%energtnena23.eps, scale=1.1}
%\hspace*{-1.0cm}
%\epsfig{file=fig11_a20c.eps,scale=0.76}
\end{minipage}
\begin{minipage}[t]{16.5 cm}
%\vspace*{-0.5cm}
\caption{(a) Electron-capture rates on $^{56}$Ni obtained with GXPF1J, KBF and KB3G as a function of temperature at densities $\rho Y_e$ = 10$^{7}$, 10$^{8}$ and 10$^{9}$ g cm$^{-3}$ (from Ref.~\cite{Mori2016})  
(b) Ratios of e-capture rates for GXPF1J and KB3G relative to those for KBF as a function of $\rho Y_e$ at $T_{9}$ = 0.1 and 5, where $T$ =$T_{9}\times$10$^{9}$ K. \label{fig13}}
\end{minipage}
\end{center}
\end{figure}

The e-capture rates for $^{56}$Ni evaluated with the GXPF1J, KBF and KB3G are shown as a function of temperature and density ($\rho Y_e$) in Fig.~\ref{fig13}(a). 
The ratios of the rates for GXPF1J and KB3G relative to those of KBF are also shown in Fig.~\ref{fig13}(b) at $T_{9}$= 0.1 and 5.  
The KB3G gives the highest rates overall at high temperatures and densities, as expected from the difference in the fragmentation of the GT strengths. 
The difference between the KBF and the GXPF1J is as small as 10-20$\%$ in most cases and decreases at high temperatures. 

Here, we comment on possible effects of the contributions from excited states of parent nuclei.  
Excited states with $E_x <$ 2 MeV are taken into account in the present calculations. In case of $^{56}$Ni, the 2$^{+}$ state is located at $E_x$ = 2.70 MeV, and its contribution to the e-capture rates is insignificant. 
For $^{58}$Ni and $^{60}$Ni, 2$^{+}$ states are at $E_x$ = 1.45 and 1.33 MeV, respectively, and contribute to the rates to some extent at low densities \cite{Suzu2011}. 
In particular for $^{60}$Ni, the effects become non-negligible at $\rho Y_e$ $\leq$ 10$^{8}$ g cm$^{-3}$ because of a large negative $Q$ value for the e-capture process, $Q$ = -3.34 MeV. 
   
As for other $pf$-shell nuclei, the ratios of the e-capture rates of GXPF1J to those of KBF are compared for 11 iron-group nuclei considered in Ref.~\cite{Cole}. 
The ratios, $\lambda_{GXP}/\lambda_{KBF}$, 
for $T_{9}$ =3 and $\rho Y_e$ =10$^{7}$ g cm$^{-3}$, 
%are shown in Fig.~\ref{fig13}(b). 
%The ratios 
are found to be within the range of 0.4-2.4, which shows that the e-capture rates for GXPF1J and KBF are close to each other within a factor of 2.5.
The ratios, $\lambda_{GXP}/\lambda_{KB3G}$ are 0.6-4.0 for the 11 nuclei \cite{Cole}, which shows that $\lambda_{KB3G}$ deviates from $\lambda_{GXP}$ more strongly than $\lambda_{KBF}$.

\subsection{Nucleosynthesis of Iron-Group Elements in Type Ia SNe\label{sec:syn}}

The updated e-capture rates with the use of the GXPF1J are applied to study nucleosynthesis of iron-group elements in Type Ia SNe. 
In single-degenerate models, thermonuclear burning propagates outward as a subsonic flame front known as a deflagration wave. Rayleigh-Taylor instabilities at the flame front cause the enhancement of the burning in the surface area. 
In some cases, the deflagration is strong enough to undergo a deflagration to detonation transition.
The deflagration models and the delayed-detonation models \cite{Iwa} are used for the supernova explosions. 
Nuclear reaction network calculations are done for the central trajectories in the W7 deflagration and the WDD2 delayed-detonation explosion models. 
The final abundance ratios, relative to Fe relative to the solar abundance ratio, defined by 
\begin{equation}
R = \frac{Y_i/Y_{Fe}}{Y_{i,\odot}/Y_{Fe,\odot}} = \frac{Y_i/Y_{i,\odot}}{Y_{Fe}/Y_{Fe,\odot}}
\end{equation}
are obtained for GXPF1J and KBF.

\begin{figure}[tb]
\begin{center}
\begin{minipage}[t]{16.5 cm}
\hspace*{-1.2 cm}
\epsfig{file=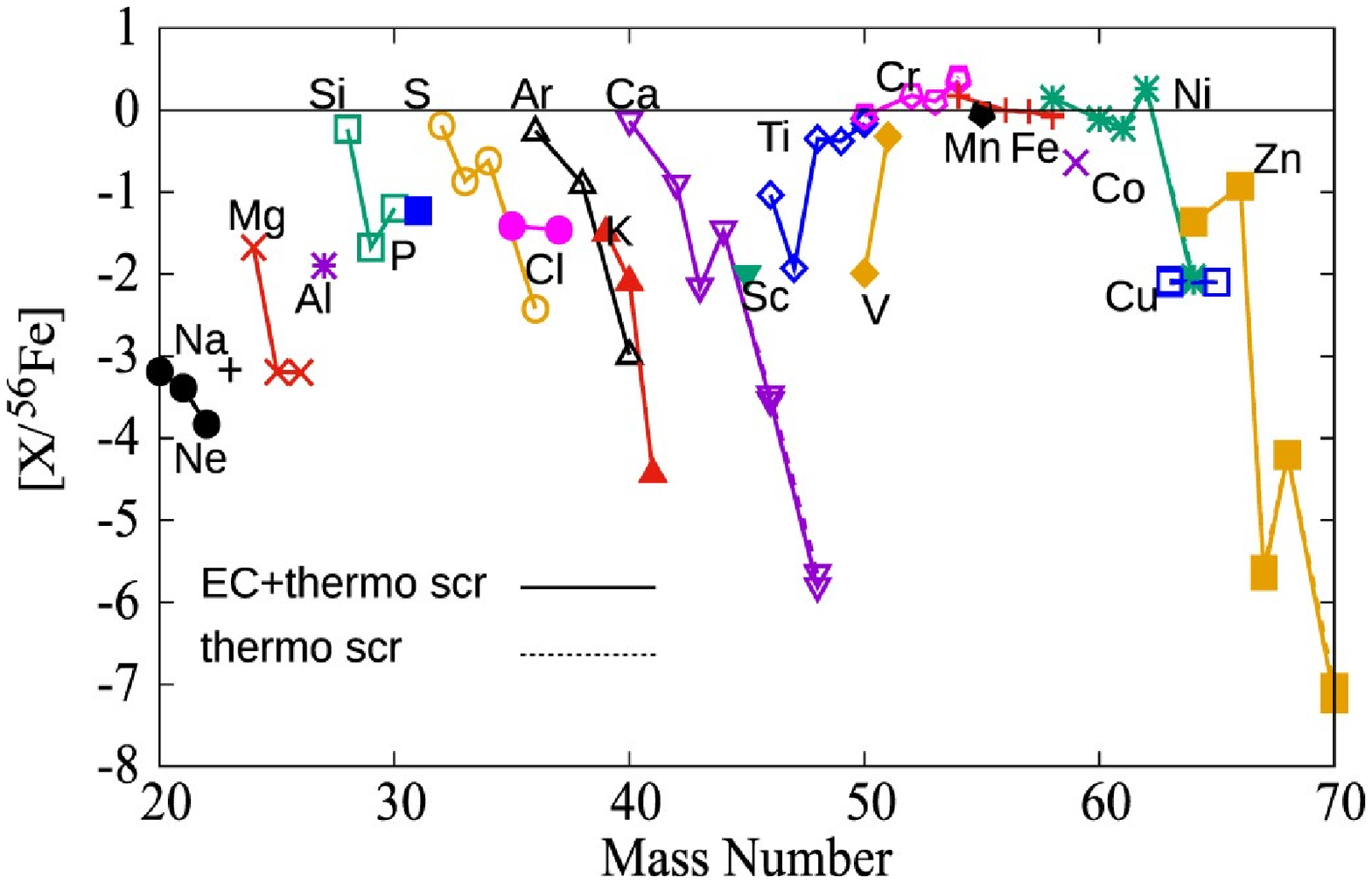,scale=0.32}
%ppnp_WDD2_vs_solar2.eps,scale=0.7}
%energtnane23.eps, scale=1.1}
\hspace*{-0.6 cm}
\epsfig{file=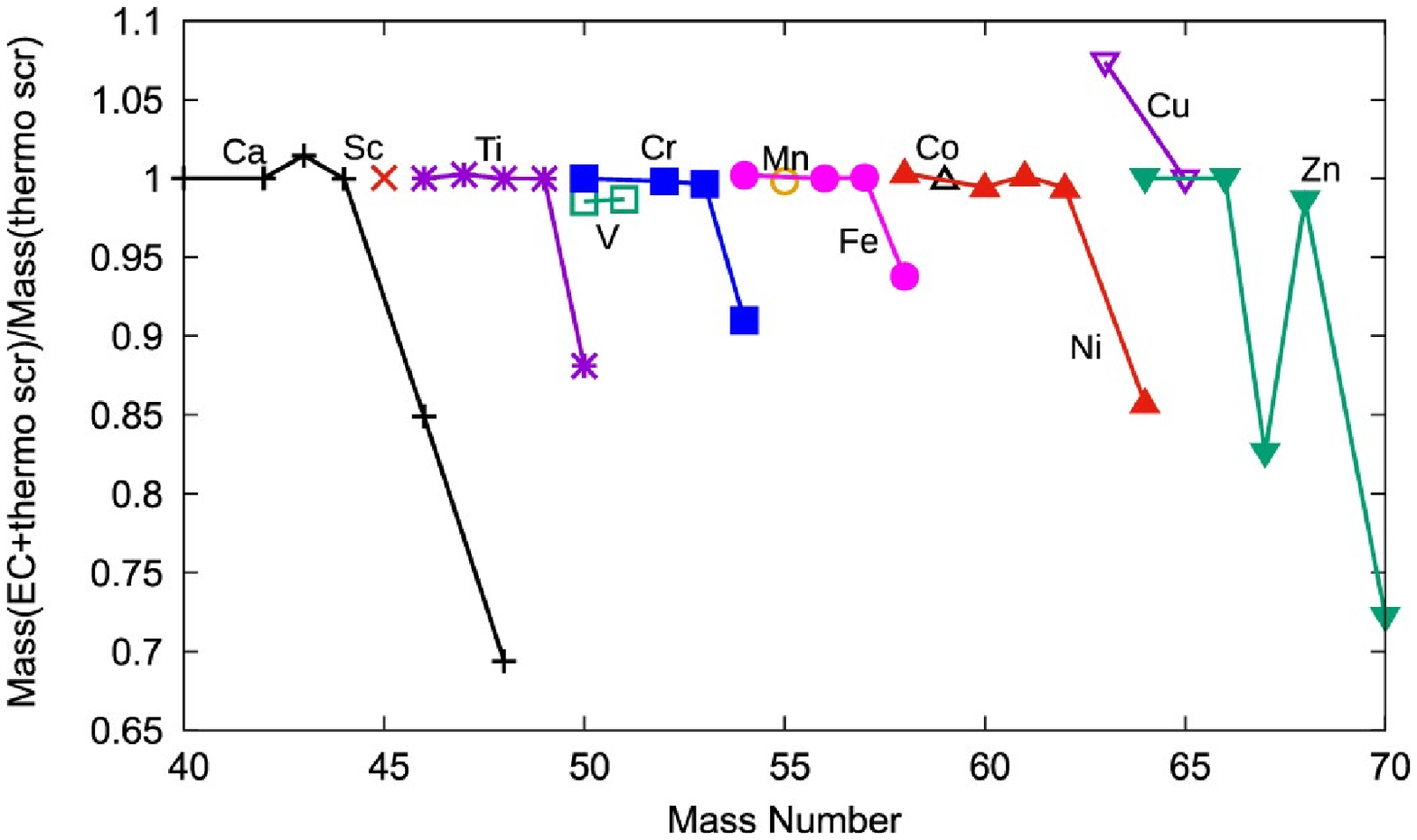,scale=0.325}
%ppnp_ratio_w7scr.eps,scale=0.37}
%energtnena23.eps, scale=1.1}
%\hspace*{-1.0cm}
%\epsfig{file=fig11_a20c.eps,scale=0.76}
\end{minipage}
\begin{minipage}[t]{16.5 cm}
%\vspace*{-0.5cm}
\caption{(Left panel) Abundances of nuclei produced in the WDD2 explosion model, normalized by the solar abundance and the $^{56}$Fe abundance. 
Electron-capture rates for $pf$-shell nuclei evaluated with the GXPF1J are used.
(Right panel) Abundance ratios between the cases with and without the Coulomb (screening) effects on the e-capture rates for the WDD2 explosion model. 
From Ref.~\cite{Mori2020}. \label{fig14}}
\end{minipage}
\end{center}
\end{figure}

The double ratios $R$ are obtained for the W7 and WDD2 models with the GXPF1J \cite{Mori2016}.
% are shown in Fig.~\ref{fig14}. 
Notable overabundance of $^{58}$Ni 
%and $^{54}$Fe 
compared to the production of lighter $Z$ nuclei is noticed for the W7 explosion model. 
For the WDD2 model, the double abundance ratios for Cr, Mn, Fe, Ni, Cu, and Zn isotopes are in good agreement with the solar abundance within a factor of two as shown in Fig.~\ref{fig14}(a).  
Overproduction factors for nucleosynthesis of neutron-excess isotopes such as $^{58}$Ni, $^{54}$Cr and $^{54}$Fe, noticed for the case with the FFN rates, are now found to be reduced to within a satisfactory range for the WDD2 model. 
Similar results are obtained with the KBF for the WDD2 model \cite{LanMart}.  
The difference of the ratios between GXPF1J and KBF is as small as 2$\%$ ans 4$\%$ for the WDD2 and W7 models, respectively. 

Various SNe models can be distinguished by using observational constraints on nucleosynthesis of iron-group elements, as the difference in the central densities of the WDs leads to different elemental and isotopic ratios in Type Ia SNe.
The two types of explosion models are found to be consistent with the observational constraints, in spite of their uncertainties.
One set of observations is consistent with the double-degenerate merger models for low central densities, while the other set favors the single-degenerate models for higher central densities (see Ref.~\cite{Mori2018} for more details).

We now discuss the Coulomb (screening) effects on the e-capture rates in $pf$-shell nuclei.
The effects are evaluated in the same way as for the $sd$-shell nuclei explained in Sect.\ref{sec:sdrate}. 
The weak rates including the screening effects are found to be reduced by about up to 20-40$\%$ compared to those without the screening effects, at the densities and temperatures shown in Fig.~\ref{fig13}(a). 
The abundance yields of the iron-group elements produced in Type Ia SNe with and without the screening effects are compared in Fig.~\ref{fig14}(b) for the WDD2 model \cite{Mori2020}. 
Here, thermonuclear electron screening effects based on Ref. \cite{WWW} are included for both cases. 
The abundances of neutron-rich isotopes, such as $^{48}$Ca, $^{50}$Ti, $^{54}$Cr, $^{58}$Fe, $^{64}$Ni and $^{67,70}$Zn, are found to be smaller by 10-30$\%$ for the case including the screening effects on the e-capture rates.
As the solar abundances of $^{54}$Cr and $^{58}$Fe are not small compared to $^{56}$Fe, the contributions of Type Ia SNe to the solar abundances of these nuclei can be as high as 50$\sim$150 $\%$. 
It is therefore important to take into account the screening effects in discussing the origin of these nuclei.

\section{Weak Rates for Two-Major-Shell Nuclei}

\subsection{Weak Rates in the Island of Inversion\label{sec:sdpf}}
Study of weak nuclear rates have been done mainly for nuclei in one-major shell by taking into account the contributions from Gamow-Teller transitions. Here we extend our study to nuclei that involve two-major shells and forbidden transitions. 
It has been pointed out that nuclear Urca processes are important in neutron star crusts \cite{Schatz}, where the density is as high as 3-6$\times 10^{10}$ g cm$^{-3}$.
Nuclear pairs that can contribute to the neutrino cooling are mostly in the regions between the closed neutron and proton shells, where nuclei are significantly deformed.
Therefore, two-major shells are involved in some of these nuclei, for example, which belong to the island of inversion. 
In this section, the weak rates for nuclei in the island of inversion, important for the Urca processes, are investigated by shell-model calculations including the $sd$-$pf$ shells. 

\begin{figure}[tb]
\begin{center}
\begin{minipage}[t]{16.5 cm}
\hspace*{-1.0 cm}
\epsfig{file=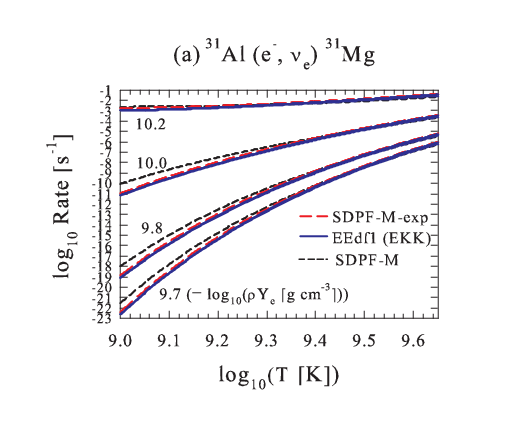,scale=1.10}
%energtnane23.eps, scale=1.1}
\hspace*{-1.0 cm}
\epsfig{file=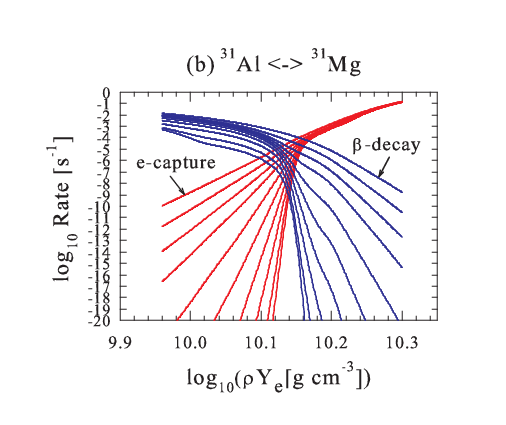,scale=1.10}
%energtnena23.eps, scale=1.1}
%\hspace*{-1.0cm}
%\epsfig{file=fig11_a20c.eps,scale=0.76}
\end{minipage}
\begin{minipage}[t]{16.5 cm}
%\vspace*{-0.5cm}
\caption{(a) Electron-capture rates for $^{31}$Al (e$^{-}$, $\nu_{e}$) $^{31}$Mg obtained with the SDPF-M and EEdf1 (denoted as EEdf1 (EKK)) Hamiltonians. The rates with the SDPF-M including available experimental data for the excitation energies are denoted by SDPF-M-exp. 
(b) Electron-capture and $\beta$-decay rates for the ($^{31}$Al, $^{31}$Mg) pair at various temperatures evaluated with the EEdf1 obtained by the EKK method.
From Ref.~\cite{OMEG15}. \label{fig15}} 
\end{minipage}
\end{center}
\end{figure}

Large $sd$-$pf$ shell admixtures are found in neutron-rich Ne, Na and Mg isotopes near $N$=20. Lowering of $2_{1}^{+}$ states and enhancement of E2 transition strengths show evidence for the breaking of the magicity at $N$=20 \cite{WBB}.
Energy levels of $2_{1}^{+}$ states and $B(\mbox{E2}: 2_{1}^{+} \rightarrow 0_{g.s.}^{+}$) values are successfully reproduced by the SDPF-M Hamiltonian \cite{Utsuno} with $sd$-$pf$-shell configurations. 
Important contributions from 2p-2h components are found in $^{30}$Ne and $^{32}$Mg. 

We discuss the weak rates for ($^{31}$Al, $^{31}$Mg) pair. The SDPF-M fails to reproduce the energy levels of $^{31}$Mg, that is, 7/2$^{-}$ state becomes the ground state while the experimental g.s. is 1/2$^{+}$.  
The Urca density can not be clearly assigned for the weak rates in the case of SDPF-M, as the transition from the g.s of $^{31}$Al (5/2$^{+}$) to the g.s. of $^{31}$Mg is forbidden.
This shortcoming can be improved for the effective interaction, EEdf1 \cite{Tsunoda}, constructed from a chiral effective field nucleon-nucleon interaction at N$^{3}$LO using the extended Kuo-Krenciglowa (EKK) method \cite{Tsunoda} and from the Fujita-Miyazawa three-nucleon forces \cite{FM}.
The EKK method can properly treat in-medium $\hat{Q}$-box calculations in two-major shells 
%free from 
without divergence problems \cite{Takaya,EKK}, which occurred in the degenerate treatment of the orbits in different major shells.
Energy levels of $^{31}$Mg are found to be well reproduced. The ground state proves to be a positive parity state, 1/2$^{+}$, and a low lying 3/2$^{+}$ state is found at $E_x$ =0.052 MeV close to the experimental value of 0.050 MeV.
More important roles of p-h excitations are noticed compared to the conventional SDPF-M case. 
The 4p-4h components are found to be as important as the 2p-2h components in $^{32}$Mg. 

The weak rates are evaluated with the EEdf1 in the $sd$-$pf$ shell,  
%obtained by the EKK method  the EKK method, 
and they prove to be close to those obtained by taking into account the experimental data as shown in Fig.~\ref{fig15}(a) for the e-capture rates on $^{31}$Al.
Here, the GT transitions $^{31}$Al (5/2$^{+}$, g.s.) $\rightarrow$ $^{31}$Mg (3/2$^{+}$) and $^{31}$Al (1/2$^{+}$, 3/2$^{+}$) $\rightarrow$ $^{31}$Mg (1/2$^{+}$, g.s.) are taken into account.   
For the EKK approach, the Urca density can be assigned to be log$_{10} (\rho Y_e$) =10.14 as shown in Fig.~\ref{fig15}(b), thus leading to the nuclear Urca process for the A=31 pair, and an appreciable cooling is expected to occur.   
An extension to other nuclei such as the A=33 nuclear pair, ($^{33}$Al, $^{33}$Mg), would be interesting.

\subsection{Weak Rates for $pf$-$sdg$-Shell Nuclei\label{sec:pfg}}

The e-capture rates in neutron-rich nuclei along and near $N$=50 are important for deleptonization in stellar core-collapse processes \cite{Sull}, where evaluations of forbidden transitions in the $pf$-$sdg$ shells become crucial.
When the electron-to-baryon ratio $Y_e$ rapidly changes from $\sim$0.41 to $\sim$0.28, nuclei with mass $A >$65 dominate the evolution. 
Neutron-rich nuclei with the $N$=50 closed neutron shell contribute most to the deleptonization.  
The weak rates for these nuclei are usually evaluated by RPA or QRPA methods \cite{Moller1997,Moller2003,Nabi,Paar,Dzh2010,Niu}, shell-model Monte-Carlo \cite{Dean}, or by using an effective rate formula with two parameters, the GT strength and the energy shift.
%In RPA calculations, .....

The effective e-capture rate is given by the following formula \cite{Sull},
\begin{eqnarray}
\lambda = \frac{ln2\cdot B}{K}(\frac{T}{m_ec^2})^6 [F_4(\eta)-2\xi F_3(\eta)+\xi^{2}F_2(\eta)]\nonumber\\
F_{k}(\eta) = -\Gamma(k+1) Li_{k+1}(-e^{\eta})
\end{eqnarray}
where $K$ =6146 s, $F_k$ are Fermi integrals of rank $k$ and $\eta=\xi + \mu_e/T$ with $\xi = (Q -\Delta E)/T$, and 
%$F_k(\eta)$ is equal to $-\Gamma(k+1) Li_{k+1}(-e^{\eta})$ where  
$Li_{s}$  is the polylogarithmic function of s-th order \cite{Gong}. 
The effective GT strength parameter is taken to be $B$ =4.6, and the effective energy difference parameter between final and initial excited states is taken to be $\Delta E$ =2.5 MeV \cite{Langanke}.  

Experimental studies of GT$_{+}$ strength in nuclei with $N$=50 have been done for $^{86}$Kr and $^{88}$Sr by ($t$, $^{3}$He) reactions. 
The GT$_{+}$ strength was found to be quite small due to Pauli-blocking of the $N$=50 core.
In $^{86}$Kr, a quite small GT strength up to $E_x$ ($^{86}$Br) = 5 Mev was obtained; 0.108+0.0631/-0.108 \cite{Titus}.
In $^{88}$Sr, no GT strength was found up to $E_x$ ($^{88}$Rb) = 8 MeV, and the strength amounted to 0.1$\pm$0.05 below $E_x$ = 10 MeV.
  
However,  the blocking of GT$_{+}$ strength by the $N$=50 shell gap is found to be overcome in high temperatures at $T >$ 10$^{10}$ K. 
Recently, thermal QRPA calculations based on thermofield dynamics (TFD) formalism have been applied to neutron-rich N=50 nuclei \cite{Dzh2019}.
Thermal quasiparticles are defined with respect to the thermal vacuum in the BCS approximation, and thermal phonons are constructed as linear superpositions of proton-neutron thermal two-quasiparticles.
The e-capture processes are treated as charge-exchange transitions from the thermal vacuum to the one-phonon states.
In the TFD formalism, both excitation and de-excitation processes at finite temperatures are naturally taken into account \cite{Dzh2010}.
The GT$_{+}$ strength is shifted toward lower excitation energies at high temperatures $\sim$1 MeV, and the contributions from the GT transitions are found to become larger than those from first forbidden transitions in e-capture cross sections on $^{78}$Ni at log$_{10}(\rho Y_e) \leq$ 10.
Similar unblocking of the GT$_{+}$ strength across the Z=40 proton and N=50 neutron shell gaps is found also for the e-captures on $^{82}$Ge, $^{86}$Kr and $^{88}$Sr \cite{Dzh2020}.
Correlations beyond thermal QRPA are shown to be important for further shift of the GT$_{+}$ strength to lower energy and enhancement of the e-capture rates in nuclei around $^{78}$Ni \cite{Litvinova2021}. 
Electron-capture rates for neutron-rich nuclei around $N$ =50 obtained by finite-temperature QRPA calculations are also found to be enhanced compared with those of large-scale shell-model calculations \cite{Giraud}.

\begin{figure}[tbh]
\begin{center}
\begin{minipage}[t]{16.5 cm}
%\hspace*{-1.0 cm}
\epsfig{file=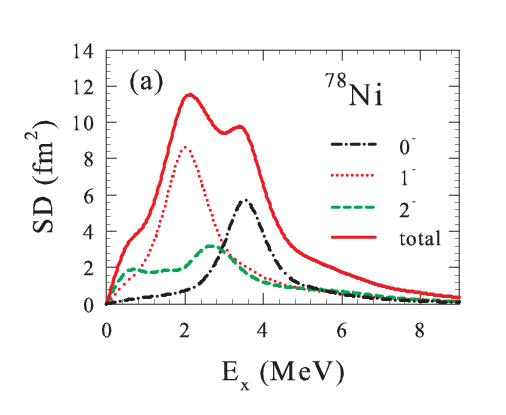,scale=0.9}
%energtnane23.eps, scale=1.1}
\hspace*{-1.0 cm}
\epsfig{file=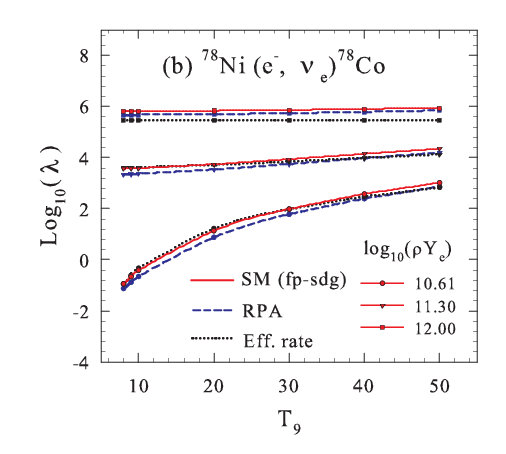,scale=0.8}
%ppnp_ecapnico78.eps,scale=1.0}
\epsfig{file=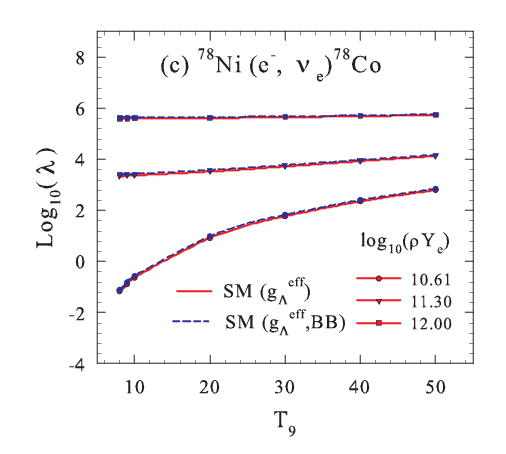,scale=0.8}
%ppnp_ecapnico78bbcvc.eps,scale=1.0}
%energtnena23.eps, scale=1.1}
%\hspace*{-1.0cm}
%\epsfig{file=fig11_a20c.eps,scale=0.76}
\end{minipage}
\begin{minipage}[t]{16.5 cm}
%\vspace*{-0.5cm}
\caption{(a) Spin-dipole strengths in $^{78}$Ni obtained by shell-model calculations with the modified A3DA interaction with $pf$-$sdg$ shells.
The strengths are folded with Lorenzians of width 0.5 MeV.
(b) Electron-capture rates on $^{78}$Ni obtained with the shell-model including $pf$-$sdg$ configurations and RPA calculations with the free $g_A$.
Rates obtained by the effective rate formula are also shown.  
(c) Electron-capture rates on $^{78}$Ni obtained with the shell-model including $pf$-$sdg$ configurations with the effective $g_A$.
The dashed curve denotes the rates based on the Behrens-B$\ddot{\mbox{u}}$hring method and includes terms that couple transition operators and distorted electron wave functions in the non-unique forbidden transitions.
\label{fig16}} 
\end{minipage}
\end{center}
\end{figure}

\begin{figure}[tbh]
\begin{center}
\begin{minipage}[t]{16.5 cm}
\hspace*{-1.0 cm}
\epsfig{file=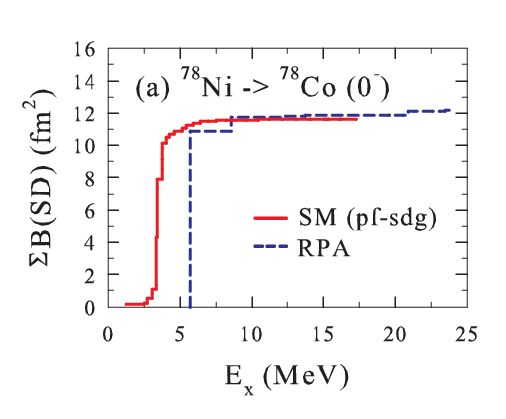,scale=1.0}
%0.701}
%energtnane23.eps, scale=1.1}
\hspace*{-0.50 cm}
\epsfig{file=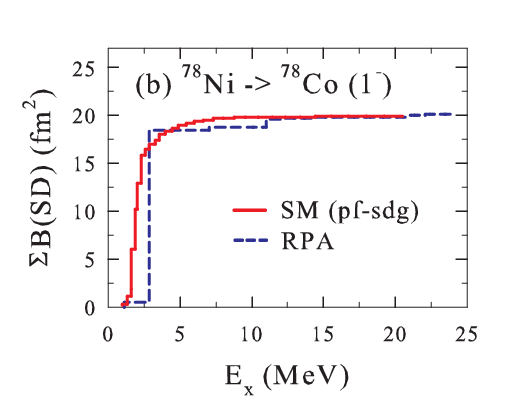,scale=1.0}
%0.701}
%energtnena23.eps, scale=1.1}
\hspace*{-1.0cm}
\epsfig{file=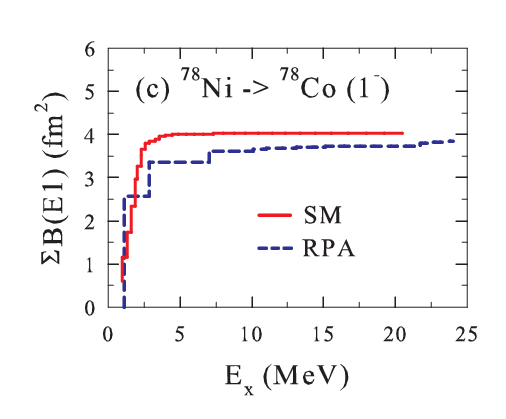,scale=1.0}
%0.701}
%\hspace*{-0.5cm}
\epsfig{file=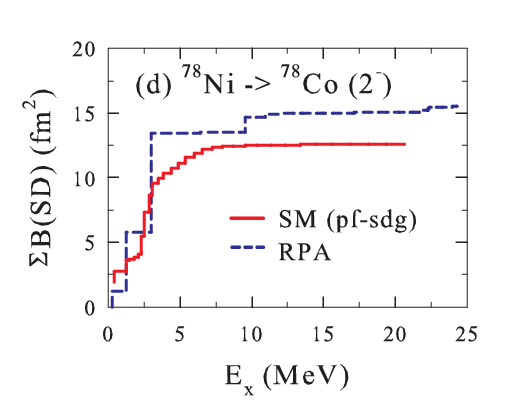,scale=1.0}
%0.701}
\end{minipage}
\begin{minipage}[t]{16.5 cm}
%\vspace*{-0.5cm}
\caption{Cumulative sum of the spin-dipole (SD) and electric dipole (E1) strengths in $^{78}$Ni obtained by shell-model calculations with the modified A3DA interaction including $pf$-$sdg$ shells, and RPA calculations.
\label{fig17}} 
\end{minipage}
\end{center}
\end{figure}

Here, we study transition strengths and e-capture rates in $^{78}$Ni as an example of $N$=50 nuclei.
%{\bf The main purpose here is to compare the rates.
The strengths and the rates are evaluated by shell-model calculations including $pf$-$sdg$ shells, and compared to the rates obtained by RPA.
%and the effective rate formula. 
The shell-model calculations are performed with the use of the modified A3DA interaction \cite{YTsunoda}, which was originally constructed for $pf$-$g_{9/2}d_{5/2}$ configurations. 
Here, up to 5p-5h excitations outside filling configurations of $^{78}$Ni are taken into account with full $pf$-$sdg$ shells.
Dominant contributions come from the spin-dipole transitions at low temperatures \cite{ni78}. 
The spin-dipole strengths in $^{78}$Ni are shown in Fig.~\ref{fig16}(a).
Sum of the strengths for $\lambda^{\pi}$ = 0$^{-}$, 1$^{-}$ and 2$^{-}$ are 11.60, 19.89 and 12.57 fm$^{2}$, respectively, which exhaust 95$\%$, 96$\%$ and 79$\%$ of the sum values, 
\begin{equation}
S^{\lambda} = \Sigma_{f,m} <g.s.|O_{m}^{\lambda \dagger} |f><f|O_{m}^{\lambda}|g.s.> = <g.s. | (O^{\lambda \dagger} \cdot O^{\lambda})^{0}|g.s.>
\end{equation}
where 
\begin{equation}
O_{m}^{\lambda} = [r Y^{1}(\hat{\vec{r}})\times\vec{\sigma}]^{\lambda}_{m}
\end{equation}
is the spin-dipole transition operator with the multipolarity $\lambda$,
respectively.
%For $pf$-$sdg$ shells, e-capture process in $^{78}$Ni is mainly induced by spin-dipole forbidden transitions due to full inclusion of $sdg$-shells. 
In case of restricted $pf$-$g_{9/2}d_{5/2}$ space, it was not possible to include the forbidden transitions in a satisfactory way.
The calculated sum of the strengths exhausts only 0.4$\%$, 8$\%$ and 19$\%$ of the sum values, respectively.  
As the contributions from the GT transitions from the ground state of $^{78}$Ni are quite small, they are not included here.
Inclusion of the shift of the GT$_{+}$ strength in the finite-temperature formalism is beyond the scope of the present study.
Though the rates  without the GT contributions are not realistic for core-collapse processes at log$_{10}(\rho Y_e) <$11, 
contributions from the spin-dipole transitions become dominant at log$_{10}(\rho Y_e) \geq$ 11 even at $T_9 >$ 10. 
Here, we compare calculated rates obtained by shell-model calculations, using the multipole expansion method of Walecka as well as the method of Behrens-B$\ddot{\mbox{u}}$hring, with those of RPA calculations and the effective rate formula in such a high-density region.

Electron-capture rates on $^{78}$Ni are evaluated by the multipole expansion formulae of Eqs. (10) and (11) using the shell-model in the $pf$-$sdg$ configuration space at densities $\rho Y_e$ $\sim$ 10$^{10}$-10$^{12}$ g cm$^{-3}$ and temperatures $T$ = (1-5)$\times 10^{10}$ K.
The transition matrix elements of the axial Coulomb and longitudinal operators are taken into account for 0$^{-}$ and 2$^{-}$ transitions, and also those of the axial electric and vector magnetic operators for 2$^{-}$.
For 1$^{-}$ transition, the Coulomb, longitudinal and transverse electric operators as well as the axial magnetic operator are taken into account.  
%The axial-vector coupling constant is taken to be a free value of $g_{A}$ = -1.26.
 
Calculated results are shown in Fig.~\ref{fig16}(b), and
they are compared with those of an RPA calculation with the SGII interaction \cite{SG}.
Here, only the contributions of the transitions from the ground state are considered. 
The same $Q$-value as for the shell-model is used for the RPA calculation.
Here, the free value $g_A$ = -1.26 is assumed for the axial-vector coupling constant..
Rates obtained by the shell-model are larger than RPA results by about 30-80$\%$, as we see from Fig.~\ref{fig16}(b).

Cumulative sums of the spin-dipole (SD) and electric-dipole (E1) strengths are shown in Fig.~\ref{fig17} for the shell-model amd RPA calculations.
The shell-model SD strengths are shifted toward lower energy region compared with the RPA in case of excitations of the 0$^{-}$ and 1$^{-}$ states, while the summed strength by the RPA exceeds the shell-model result at $E_x \geq$ 3 MeV for the 2$^{-}$ states.     
As for the E1 strength, 
the shell-model results are larger than the RPA results at $E_x >$ 2 MeV.  
In both cases, the dominant contributions come from 1$^{-}$ states.
These behaviors of the SD and E1 strengths lead to larger e-capture rates for the shell-model than for the RPA.

When we compare the rates calculated in RPA with those obtained by the effective rate formula, Eq. (29), the ratios $\lambda_{eff}$/$\lambda_{RPA}$ are found to be about 2 at lower temperatures and become close to 1 at $T_9$ =50, except for the case of log$_{10}(\rho Y_e)$ =12.0, where the ratio is 0.4-0.8.  
The shell-model rates are rather close to those of the effective rate formula.
The energy shift parameter $\Delta E$ =2.5 MeV taken in the effective rate formula is close to the energy of the peak position of the sum of the SD strengths in the shell-model calculations, as seen from Fig.~\ref{fig16}(a), while the peak for the RPA is around $E_x$ = 3 MeV as indicated by Fig.~\ref{fig17}.

Calculated rates for 1$^{-}$ and 2$^{-}$ transitions in the shell-model are reduced if we adopt the same universal quenching factor as for GT transitions, q =$g_{A}^{eff}$/$g_A$ =0.74.
As for the 0$^{-}$ transition, $g_A^{eff}$ is enhanced for the $\vec{\sigma}\cdot\vec{\nabla}/M$ term in the axial Coulomb multipole due to meson-exchange current effects \cite{KDR,Warburton,Brown}.
The enhancement factor of q =1.5 is taken for this term, while q =1 is adopted for the $\vec{\sigma}\cdot\vec{r}$ term in the axial longitudinal multipole.  
With these values of q, the shell-model rates are found to be reduced about by 40$\%$ as shown in Fig.~\ref{fig16}(c).
Experimental information on the quenching of $g_A$ for the SD transitions is not enough to determine a definite value for q.  
An effective spin $g$ factor $g_{s}^{eff}$ = 0.64$g_s$ is shown to reproduce the spin quadrupole (M2) strengths in $^{48}$Ca and $^{90}$Zr as well as the M1 strength in $^{48}$Ca obtained by backward electron scattering experiments \cite{Cosel}.
This may suggest that it is more reasonable to assume that strengths of SD transitions with $\lambda^{\pi}$ = 1$^{-}$ and 2$^{-}$ are also quenched similar to the GT transition. 

For the first-forbidden $\beta$-decays, it is common to use Behrens-B$\ddot{\mbox{u}}$hring (BB) formulae instead of Walecka's multipole expansion formulae.
In the BB formulae, in non-unique forbidden transitions there are additional terms which are absent in the Walecka's method.  
The shape form factors for the e-capture processes with $\lambda^{\pi}$ = 0$^{-}$ and 1$^{-}$ are given as follows in the low-momentum transfer limit in the Walecka method:
\begin{eqnarray}
C_{ecap}^{0^{-}} &=& (\xi'v -\frac{1}{3}w W_0)^2   \nonumber\\
C_{ecap}^{1^{-}} &=& [\xi'y +\frac{1}{3}(u+x) W_0]^2 + \frac{1}{18}(u-2x)^2 \nonumber\\
&+& W[-\frac{4}{3}\xi'yu-\frac{W_0}{9}(4x^2+5u^2)] + \frac{W^2}{9}(4x^2+5u^2)
\end{eqnarray}
where 
\begin{eqnarray}
\xi'v &=& -\frac{\sqrt{3}}{\sqrt{2J_i+1}} g_A <f||\frac{1}{M} [\vec{\sigma}\times\vec{\nabla}]^{(0)}||i> \nonumber\\
w &=& -\frac{\sqrt{3}}{\sqrt{2J_i+1}} g_A <f|| r[C^{1}(\Omega)\times\vec{\sigma}]^{(0)} ||i> \nonumber\\
%w' &=& -frac{\sqrt{3}}{\sqrt{2J_i+1}} g_A <f|| r[C^{1}(\Omega)\times\vec{\sigma}]^{(0)} I(1,1,1,1;r) ||i> \nonumber
\xi'y &=& \frac{1}{\sqrt{2J_i+1}} <f|| \frac{\vec{\nabla}}{M} ||i> \nonumber\\
x &=& \frac{1}{\sqrt{2J_i+1}} <f|| r C^{1}(\Omega) ||i> \nonumber\\
u &=& \frac{\sqrt{2}}{\sqrt{2J_i+1}} g_A <f|| r[C^{1}(\Omega)\times\vec{\sigma}]^{(1)} ||i>
\end{eqnarray}
with $W$ the electron energy, $W_0=|Q|$ where $Q$ is the $Q$-value for the reaction, and $J_i$ is the angular momentum of the initial state.    
%in the low momentum transfer limit in the Walecka method.
The relation, $\xi'y$ =-$\Delta E_{fi}x$ with $\Delta E_{fi} = E_f -E_i$, is satisfied because of the CVC.
One way to evaluate of the $\xi'y$ term is just to carry out derivative of wave functions.  
In this case, $\Delta E_{fi}$ is equal to 1 (-1) $\hbar\omega$ for $pf$ ($sdg$) $\rightarrow$ $sdg$ ($pf$) transitions in each matrix element when harmonic oscillator wave functions are used. 
Another way is to use the calculated excitation energies of 1$^{-}$ states in $^{78}$Co to obtain the value of -$\Delta E_{fi}x$.    
The former method is adopted here unless referred. 
    
The shape factors for the e-capture processes are related to those of the $\beta$-decay by
\begin{eqnarray}
C_{ecap}^{0^{-}} (\xi'v, w) = C_{\beta^{-}}^{0^{-}} (\xi'v, -w) \nonumber\\
C_{ecap}^{1^{-}} (\xi'y, x, u) = C_{\beta^{-}}^{1^{-}} (\xi'y, -x, u)
\end{eqnarray}

In case of the BB method, 
%\begin{eqnarray}
$\xi'v$ $\rightarrow$ $\xi'v -\xi w'$ for $\lambda^{\pi}$ = 0$^{-}$ and
$\xi'y$ $\rightarrow$ $\xi'y -\xi(u'-x')$ for $\lambda^{\pi}$ = 1$^{-}$,
where
\begin{eqnarray} 
w' &=& -\frac{\sqrt{3}}{\sqrt{2J_i+1}} g_A <f|| r[C^{1}(\Omega)\times\vec{\sigma}]^{(1)} I(1,1,1,1;r) ||i> \nonumber\\
x' &=& \frac{1}{\sqrt{2J_i+1}} <f|| r C^{1}(\Omega) I(1,1,1,1;r) ||i> \nonumber\\
u' &=& \frac{\sqrt{2}}{\sqrt{2J_i+1}} g_A <f|| r [C^{1}(\Omega)\times\vec{\sigma}]^{(1)} I(1,1,1,1;r) ||i> 
\end{eqnarray}
and $\xi$ = $\frac{\alpha Z}{2R}$.
The additional terms $\propto \xi$ come from the coupling of the transition operators with the distorted electron wave functions. 
%The capture rates are enhanced by about 10-15$\%$ when we include these additional terms.
% as shown in Fig.~\ref{fig18}. 
Their effects on the rates are insignificant in the present reaction,
%rather modest.
which is reasonable as $\xi \sim 0.1\frac{1}{M}$.    
Here, in case of the BB method, the $\xi'y$ term is evaluated in the form $-\Delta E_{fi} x$ by using calculated excitation energies for 1$^{-}$ states in $^{78}$Co.
Use of this CVC relation enhances the capture rates at low temperatures. 
When we compare the Walecka and BB methods, the rates for the latter are found to be enhanced compared with the former by about 20$\%$ at low temperatures, except for the high density of log$_{10}$($\rho Y_e$) =12.0 as shown in Fig.~\ref{fig16}(c).     

\section{$\beta$-Decay Rates for $N$=126 Isotones for
%and 
r-Process Nucleosynthesis}
%\subsection{$\beta$-Decay Half-Lives of $N$=126 Isotones \label{sec:sm126}}  

The origin of elements heavier than iron is still one of the important open questions in physics.
It has been known for more than half a century that about half of the elements heavier than iron are produced via rapid neutron capture (the r-process) \cite{Burbridge1957,Cowan,KBT}.
Recent studies confirm that r-process nucleosynthesis is the most promising answer to the question. However, the sites of the r-process are still under controversy though there are several candidates such as
magnetohydrodynamic-jet (MHDJ) CCSN \cite{Nishimura2006,Fujimoto2007,Fujimoto2008, Ono2012,Winteler2012,Nakamura2014,Nishimura2015} and binary neutron star mergers (NSM) \cite{Wanajo2014,Goriely2015}. 
In particular, much attention is paid to NSMs since the observation of a binary NSM through a gravitational wave event (GW170817) and its associated short gamma-ray burst (GRB17087A) \cite{Abbott2017a} and electromagnetic emission, kilonova (AT2017gfo) \cite{Abbott2017, Smartt2017}. 
Neutrino-driven wind ($\nu$DW) CCSN can be one of the candidates for weak r-process producing the elements with $A \approx$ 90-110 ($Z \approx$ 40) if an appropriate condition for $Y_{e}$, $Y_{e}<$ 0.5, is satisfied \cite{Woosley1994,Wanajo2013}.
However, recent core-collapse supernova simulations show that the difference between averaged energies of $\nu_e$ and $\bar{\nu}_{e}$ gives a too small neutron excess, yielding $Y_{e}>$ 0.5, which excludes $\nu$DW CCSN as an r-process site \cite{Hudepohl2010,Fischer2020}. 
Note that lower $\bar{\nu}_e$ energy in the $\bar{\nu}_e + p$ $\rightarrow$ $n + e^{+}$ reaction results in less production of neutrons. 
Even the appearance of a spherically symmetric $\nu$DW was not found in a recent long-term 3D supernova simulation \cite{Bollig2021}.
      
Abundances of the elements produced in the r-process depend sensitively on various nuclear inputs such as masses, neutron capture rates, fission yields, and $\beta$-decay rates especially at the waiting point nuclei \cite{Wanajo,Terasawa,Sasaqui,LanMart}.
Among them, the masses affect the abundances most \cite{Mumpower}.
When $\beta$-decay half-lives are changed by 1/10-10 times, the abundances can vary, that is, increase or decrease by an order or more of magnitude \cite{Mumpower}.
The abundances depend also on astrophysical conditions; electron-to-baryon number ratio $Y_e$, the entropy and temperature of the explosion environment, and neutrino processes \cite{Meyer,Terasawa2004,Grawe2007,LanMart}. 

Here, we focus on $\beta$-decay rates and half-lives of nuclei at the waiting points and study their effects on the r-process nucleosynthesis.
Studies of the $\beta$-decays of isotones with $N$=82 and $N$=126 have been done by various methods including the shell model \cite{Martinez-Pinedo1999,Suzuki2012,Zhi2013}, QRPA/finite-range-droplet model (FRDM) \cite{Moller1997,Moller2003}, QRPA/ETFSI \cite{Borozov2000}, HFB+QRPA \cite{Engel}, QRPA \cite{Fang} and CQRPA \cite{Borozov2003}. 
The half-lives obtained by these methods are found to be consistent to each other for $N$ = 82 isotones. 
For the case of $N$ =126 isotones, on the contrary, the calculated half-lives vary stronger than for $N$=82 \cite{LanMart}.
Moreover, experimental data for the masses, spectra and $\beta$-decays in this region of nuclei are quite rare.
The region near the waiting point nuclei at $N$=126 is therefore called the $^{\prime}$blank spot$^{\prime}$ region.  
First-forbidden (FF) transitions become important in addition to the GT transitions for the case of $N$=126 isotones.
Shell-model calculations have been done by including the contributions from both the GT and FF transitions \cite{Suzuki2012,Zhi2013}.

%Formulae
The $\beta$-decay rates $\lambda$ are obtained from the shape factors in Eqs. (32)-(35), and the half-life is given by $t_{1/2} =\frac{ln 2}{\lambda}$.

%Fig.19
\begin{figure}[tb]
\begin{center}
\begin{minipage}[t]{16.5 cm}
%\hspace*{3.0 cm}
%\epsfig{file=ppnp_sdnico78.eps,scale=1.10}
%energtnane23.eps, scale=1.1}
\hspace*{1.0 cm}
\epsfig{file=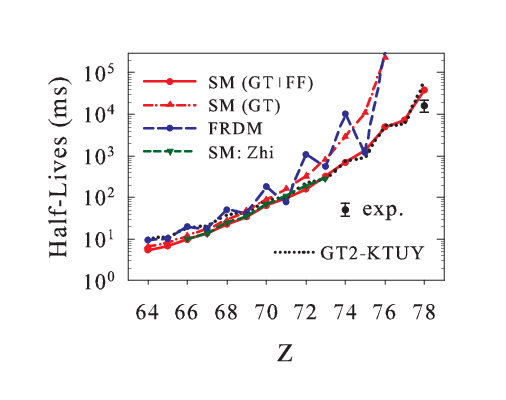,scale=1.5}
%energtnena23.eps, scale=1.1}
%\hspace*{-1.0cm}
%\epsfig{file=fig11_a20c.eps,scale=0.76}
\end{minipage}
\begin{minipage}[t]{16.5 cm}
\vspace*{-0.5cm}
\caption{
Calculated half-lives for the isotones with $N$=126 as well as the experimental half-life for $Z$=78 \cite{Morales}. 
Half-lives obtained by the shell-model (SM) calculations with GT+FF (GT only) contributions are shown by the solid (dash-dotted) curves \cite{Suzuki2018}, while those of FRDM \cite{Moller1997,Moller2003} are denoted by the dashed curve.
Results of another SM calculation \cite{Zhi2013} and GT2-KTUY \cite{Koura2005,Chiba2008} are also shown by short-dashed and dotted curves, respectively.
\label{fig18}}
\end{minipage}
\end{center}
\end{figure}

Half-lives of the isotones with $Z$ =64-78 have been evaluated by shell-model calculations with the use of a modified G-matrix \cite{Steer,Rydstrom}. 
A closed $N$=126 shell configurations is assumed for the parent nucleus.
For neutrons the 0$h_{9/2}$, 1$f_{5/2,7/2}$, 2$p_{1/2,3/2}$ and 0$i_{13/2}$ orbits outside $N$=82 core are taken as the model space, and a transition of a neutron to proton 1$g_{7/2}$, 1$d_{3/2,5/2}$, 2$s_{1/2}$ and 0$h_{11/2}$ hole orbits is considered.
The shell-model calculations are carried out with a certain truncation of the model space and quenching of axial- and vector-coupling constants (see Ref. \cite{Suzuki2018,Suzuki2012} for the details).
Calculated half-lives are shown in Fig.~\ref{fig18}, and compared with another shell-model result as well as those of FRDM and GT2-KTUY.
The half-lives obtained by the two shell-model calculations are consistent with each other and agree very well with the results of GT2-KTUY, and are found to be short compared with standard values of FRDM \cite{Moller1997,Moller2003}.
The contributions from the FF transitions, mainly the $\nu$i$_{13/2}$ $\rightarrow$ $\pi$h$_{11/2}$ transition, become more important for larger $Z$, and dominant at $Z >$72. 
They are crucial at $Z >$75 to get reasonable half-lives compatible with the observation. The calculated half-life for $Z$=78 ($^{204}$Pt), t$_{1/2}$ = 38.3 s, is found to be fairly consistent with the experimental vale; 16+6/-5 s \cite{Morales}. 
There are no even-odd staggerings in the shell-model half-lives, in contrast to the FRDM ones.
This comes from the monotonical dependence of the phase space factors on $Z$. 
The difference between the $Q$-value and the mean energy of the strength decreases monotonically as $Z$ increases.  
%In the study of $\beta$-decay properties of neutron-rich nuclei in the region along and near the neutron magic numbers $N$=82 and $N$=126 with the 
In another QRPA investigation, the half-lives of $N$=126 isotones are found to be free from the even-odd staggering and in good agreement with the shell-model results within a factor of $\sim$2 \cite{Fang}.  
%when the mass model of FRDM is used \cite{Fang}. 
The $\beta$-decay half-lives of almost all odd-mass and odd-odd nuclei on the neutron-rich side of stability are obtained by QRPA with the use of the finite-amplitude method \cite{Ney2020}.
Important roles of the first-forbidden transitions are found for $N$=126 isotones, while predicted half-lives are longer by several times to one order of magnitude compared with the present shell-model evaluations.

The $\beta$-decay half-lives of neutron-rich nuclides, $^{100}$Kr, $^{103-105}$Sr, $^{106-108}$Y, $^{108-110}$Zr, $^{111,112}$Nb, $^{112-115}$Mo, and$^{116,117}$Tc, which lie close to the astrophysical r-process path, have been measured at RIBF (Radioactive-Isotope Beam Factory), RIKEN \cite{Nishimura2011}. 
The results idicate an overestimation of the half-lives predicted by FRDM+QRPA model by a factor of 2 or more in the A=110 region. 
More satisfactory predictions of the half-lives are obtained from the GT2-KTUY model.
%A systematic enhancement of the measured $\beta$-decay rates of the Zr and Nb isotopes by a factor of 2 or more around A=110 compared with the predictions of the FRDM+QRPA model.
Evaluations of $\beta$-decay half-lives and $\beta$-delayed neutron emission probabilities have been done with a relativistic Hartree-Bogoliubov (RHB) and proton-neutron relativistic quasiparticle random phase approximation (pn-RPRPA) in a wide range of neutron-rich nuclei \cite{Marketin2016}. 
Shorter half-lives, as compared to the FRDM/QRPA model, and important roles of FF transitions are found for heavy nuclei in the region around $N$=126.
%Marketin  QRPA  shorter half-lives near Pb region than FRDM
The shorter half-lives are shown to slightly shift the position of the third peak of the r-process nucleosynthesis abundances for MHDJ CCSN and NSM 
%Possible effects of the shorter half-lives on r-process nucleosynthesis are discussed in 
\cite{Suzuki2018,Nishimura2016}. 

Multinucleon transfer (MNT) reactions in the $^{136}$Xe + $^{198}$Pt system at an energy above Coulomb barrier are shown to be promising to populate and characterize the $^{\prime}$blank spot$^{\prime}$, that is, neutron-rich isotopes around $N$=126 \cite{Watanabe2015}. 
Experimentally deduced cross sections for the production of the $N$=126 isotones wuth $Z\leq$77 are found to be much larger than those from the fragmentation experiments such as $^{208}$Pb +Be.    
The MNT reactions are used to produce neutron-rich isotopes $^{199}$Pt at the KEK Isotope Separation System (KISS), and the magnetic dipole moments and mean-square charge radius variations are measured by in-gas-cell laser ionization spectroscopy performed at KEK \cite{Hirayama2017}.
Spectroscopic studies of $^{197,198}$Os and $^{198}$Ir \cite{Hirayama2018} as well as $^{194,196}$Os \cite{Choi}, produced by MNT reaction at KISS, are performed to clarify the structure of these nuclei. 

$\beta$-delayed neutron emission is important as additional neutrons are added to the r-process, and affects the the final abundance pattern by modifyig the decay path back to stability and neutron flux after the freeze-out.
While there are theoretical studies with the shell-model \cite{Zhi2013}, FRDM/QRPA \cite{Moller1997,Moller2003}, QRPA \cite{Borozov2003} and relativistic HFB+QRPA \cite{Marketin2016}, the calculated values of the branching ratios for one- and two-neutron emissions ($P_{1n}$ and $P_{2n}$) are rather scattered.
Recently a systematic measurement of the one- and multi-neutron emission probabilties in neutron-rich nuclei started at RIKEN, using a high-efficiency array of $^{3}$He neutron countors (BRIKEN) \cite{Saldivia2017}, though there had been a number of experimental works \cite{Caballero2016,Pereira2009,Pereira2010,Gomez2014}.
For example $\beta$-delayed one- and two-neutron branching ratios ($P_{1n}$ and $P_{2n}$) have been measured in the r-process nuclei $^{86,87}$Ga at RIBF, RIKEN \cite{Yokoyama2019,Saldivia2017}.
An extension of the present shell-model study on half-lives of $N$ =126 isotones to nuclei in the $^{\prime}$blank spot$^{\prime}$ region and evaluation of neutron emission probabilities is a challenging future problem.   

\section{Neutrino-Nucleus Reactions relevant to Nucleosynthesis and Neutrino Detection}

\subsection{$\nu$-Nucleus Reactions on $^{12}$C and $^{56}$Fe\label{sec:nulab}}

In this section, we discuss $\nu$-nucleus reactions relevant to nucleosynthesis in SNe and $\nu$-detection.
$\nu$-induced reaction cross sections measured in laboratory experiments are rather scarce, only on $^{12}$C and $^{56}$Fe up to now \cite{LSND,KARMEN1,KARMEN2,LAMPF}.
We have updated $\nu$-nucleus reaction cross sections based on recent shell-model Hamiltonians for the $p$-shell and $pf$-shell.
Both charged- and neutral-current reaction cross sections in $^{12}$C induced by decay-at-rest (DAR) $\nu$ are successfully described by using GT and spin-dipole (SD) transition strengths obtained by the SFO Hamiltonian for $p$-shell \cite{Suzuki2006}.
Especially, the exclusive reaction cross sections $^{12}$C ($\nu_{e}$, e$^{-}$) $^{12}$N (1$^{+}$, g.s.) induced by the GT transition \cite{LSND} are quite well reproduced by the SFO \cite{Suzuki2003} at $E_{\nu}$ = 35-55 MeV, as shown in Fig. 4 in Ref.~\cite{Suzuki2013}. 
Cross sections induced by in-flight $\nu_{\mu}$ are also well described by RPA calculations \cite{Volpe}.

In case of $^{56}$Fe, the charged-current inclusive cross sections for $^{56}$Fe ($\nu_{e}$, e$^{-}$) $^{56}$Co induced by DAR $\nu$ are described well by hybrid models of shell-model and RPA \cite{Kolbe1999}. 
Shell-model calculations with GXPF1J and KB3G Hamiltonians are carried out for 0$^{+}$ and 1$^{+}$, while the RPA method is adopted for other multipoles.
Various evaluations of the charged-current cross section on $^{56}$Fe, such as RPA, QRPA and hybrid models, are found to give cross sections consistent with the experimental data \cite{PaarSuz}. 
The averaged theoretical cross section and the experimental one measured by KARMEN \cite{KARMEN1,KARMEN2} are $\sigma$ = (258$\pm$57) $\times$10$^{-42}$ cm$^{2}$ and $\sigma$ = (256$\pm$108$\pm$43) $\times$10$^{-42}$ cm$^{2}$, respectively. 
Thus, we can now describe $\nu$-nucleus reactions on light and medium-heavy nuclei, such as $^{12}$C and $^{56}$Fe, quite well within reasonable accuracies by shell-model and RPA calculations.
We have updated $\nu$-nucleus reaction cross sections on nuclei from $p$- shell to $pf$-shell based on recent advances in nuclear structure studies.
Global $\nu$-nucleus reaction cross sections obtained by QRPA are also available \cite{Paar-nu}.
In the folowing subsections we discuss nucleosynthesis by $\nu$-process, $\nu$ detection and $\nu$ properties based on these developments. 
%Some discussions are also made in a review article \cite{Balasi}.  

\subsection{Nucleosynthesis by $\nu$-Process\label{sec:nuproc}}

There are a few elements such as $^{7}$Li, $^{11}$B, $^{19}$F, $^{138}$La and $^{180}$Ta, which can not be produced by either s- nor r-processes.
In pioneering work by Ref.~\cite{Woos1990}, these elements are shown to be synthesized by $\nu$-process, where charged- and neutral-current reactions on nuclei play important roles \cite{Woos1990}. 
$\nu$-induced reaction cross sections for these nuclei were updated in Ref.~\cite{Heger2005}.
The use of accurate $\nu$-induced cross sections, including partial cross sections of particle emission channels, is important for reliable predictions of the production yields of the elements.
Here, branching ratios for each excited level are calculated for decay channels involving single and multi-particle emissions by the Hauser-Feshbach statistical model \cite{HF1952}.
The production yields of the elements are also sensitive to the neutrino spectra. This aspect should be carefully examined for a quantitative estimate.

Here, we first discuss light-element nucleosynthesis using the cross sections obtained from shell-model calculations with the SFO \cite{Suzuki2003}.      
Light elements, $^{7}$Li and $^{11}$B, are produced through neutral-current reactions ($\nu$, $\nu'p$) and ($\nu$, $\nu'n$) on $^{4}$He and $^{12}$C, respectively.
$^{11}$B is produced through $^{12}$C ($\nu$, $\nu$'p) $^{11}$B and $^{12}$C ($\nu$, $\nu$'n) $^{11}$C followed by $\beta^{-}$-decay.
$^{7}$Li is produced through $^{4}$He ($\nu$, $\nu$'p) $^{3}$H followed by $\alpha$-capture, and $^{4}$He ($\nu$, $\nu$'n) $^{3}$He followed by $\alpha$-capture and successive e-capture on $^{7}$Be. 
Most $^{7}$Li nuclei are produced in the He-layer in SN explosions (SNe), while $^{11}$B nuclei are produced mainly in the O/Ne, O/C and He layers.
The production yields of $^{7}$Li and $^{11}$B, evaluated with the recent updated $\nu$-induced cross sections, are found to be enhanced compared to those of Woosley \cite{Suzuki2006,YSC2008}. 
When the energy spectra of SN neutrinos  are taken to be the Fermi distributions with temperatures at 3.2, 5.0 and 6 MeV for $\nu_{e}$, $\bar{\nu}_e$ and $\nu_{\mu,\tau}$, respectively, the enhancement factor for $^{7}$Li and $^{11}$B is about 15$\%$ \cite{YSC2008}.
The temperature for 
%$\bar{\nu}_e$ and 
$\nu_{\mu,\tau}$ was assumed to be higher before \cite{Woos1990}; 
%T($\bar{\nu}_e$) = 6 MeV and 
T($\nu_{\mu,\tau}$) = 8 MeV.
Recent studies on galactic chemical evolution (GCE) with SNe support lower temperature of 6 MeV \cite{YK2005,YK2006}.

Evolutions of luminosities and average energies of neutrinos in SNe after core bounce are studied by numerical simulations for a variety of progenitor stellar masses, 13-50$M_{\odot}$, and metalicities, Z =0.02 and 0.004 \cite{Nakazato2013}.
Time integrated number spectra of neutrinos are found to be well fitted by the Fermi-Dirac distributions up to $E_{\nu}\sim$ 30 MeV, but have high energy tails originating in the accretion phase, where the low density outer region has high temperature due to shock heating. 
Mean energies of emitted neutrinos until 20 s after the core bounce are evaluated, and they are found to have a hierarchy, $<E_{\nu_{e}}>$ $<$ $<E_{\bar{\nu}_{e}}>$ $\leq$ $<E_{\nu_{x}}>$ with $x$ = $\mu, \tau, \bar{\mu}$ and $\bar{\tau}$, that is, the mean energies of $\bar{\nu}_{e}$ and $\nu_{x}$ are rather close.
For example, for $M$ = 20$M_{\odot}$ and Z = 0.02 ($M$ = 30$M_{\odot}$ and Z = 0.004), $<E_{\nu_{e}}>$, $<E_{\bar{\nu}_{e}}>$ and $<E_{\nu_{x}}>$ are 9.32, 11.1 and 11.9 MeV (17.3, 21.7 and 23.4 MeV), respectively \cite{Nakazato2013}. 
These averaged energies correspond to lower temperatures for $\bar{\nu}_e$ and $\nu_{\mu,\tau}$; $T_{\bar{\nu}_e} \approx T_{\mu,\tau} \approx$ 4 MeV for the solar metalicity.     
The effects of this low temperature, $T_{\bar{\nu}_e} = T_{\nu_{\mu,\tau}}$ = 4 MeV, on neutrino nucleosynthesis are investigated in Ref.~\cite{Siverd2018}.
The production yields of $^{7}$Li and $^{11}$B, as well as its ratio $^{7}$Li/$^{11}$B, are found to be considerably reduced compared with those for the standard case, $T_{\bar{\nu}_e}$ = 5 MeV and $T_{\nu_{\mu,\tau}}$ = 6 MeV.

The effects of the time-dependence of $\nu$ luminosities and $\nu$ energies on nucleosynthesis are investigated in Refs.~\cite{Siverd2019} by taking into account all the phases of SNe, that is, burst, accretion and cooling phases.
The burst and accretion phases, which give rise to high energy tails in the $\nu$ spectra, are important for the prediction of production yields of elements, in particular, for the case of $^{138}$La and $^{180}$Ta.
The use of time-dependent neutrino energies increases the production yields of the elements.
The assumption of constant neutrino energies averaged only in the cooling phase, as often used in previous literatures \cite{Woos1990,Heger2005}, leads to a rather noticeable difference in the yields \cite{Siverd2019}.
Inclusion of the time-dependence for low neutrino energies leads to a slight reduction of the yields of $^{138}$La and $^{180}$Ta by 10-20$\%$, compared to the case of neutrino spectra with higher energies as used here \cite{Siverd2019}.
Considerable reductions, obtained for the low neutrino energy spectra, remain for the yields of $^{7}$Li and $^{11}$B when the time-dependence is taken into account.

$^{11}$B can be produced also from $\nu$-induced reactions on $^{16}$O; $^{16}$O ($\nu$, $\nu$'$\alpha p$) $^{11}$B and $^{16}$O ($\nu_{e}$, e$^{-}$ $\alpha p$) $^{11}$C ( , e$^{-}$ $\bar{\nu}_{e}$) $^{11}$B.
% \cite{SCY2018}.
As the cross sections for $^{16}$O ($\nu$, $\nu$'$\alpha p$) $^{11}$B amount to be about 10$\%$ of those of $^{12}$C ($\nu$, $\nu'p$) $^{11}$B, the production of $^{11}$B from $^{16}$O through $\alpha p$ emission channel is not negligible.
The production yields of $^{11}$B and $^{11}$C in SNe from progenitors with mass of 20$M_{\odot}$ are estimated to be enhanced about by 16$\%$ and 8$\%$, respectively, by the inclusion of multi-particle emission channels in $^{16}$O \cite{SCY2018}.

The element $^{19}$F is produced through $^{20}$Ne ($\nu$, $\nu'p$) $^{19}$F and $^{20}$Ne ($\bar{\nu}_e$, e$^{+}n$) $^{19}$F. 
The magnitude of the $Q$-value of the reaction $^{20}$Ne ( $\bar{\nu}_e$, e$^{+}$) $^{20}$F is $|Q|$ = 7.024 MeV and the threshold energy for neutron production from $^{20}$F is $S_n$ = 6.601 MeV.
The summed energy $|Q|+S_n$ = 13.625 MeV is rather close to the threshold energy for proton production from $^{20}$Ne, $S_p$ = 12.84 MeV. 
Therefore, the ($\bar{\nu}_e$, e$^{+}n$) reaction on $^{20}$Ne can also contribute to the production of $^{19}$F, as the difference between the temperatures for $\bar{\nu}_e$ and $\nu_{\mu,\tau}$ is small now.          

%138La, 180Ta,  55Mn, 59Co,   92Nb, 98Tc  

$^{138}$La and $^{180}$Ta are produced mainly by charged-current reactions $^{138}$Ba ($\nu_{e}$, e$^{-}$) $^{138}$La and $^{180}$Hf ($\nu_{e}$, e$^{-}$) $^{180}$Ta, respectively.
The GT transition strengths to these nuclei are measured by ($^{3}$He, $t$) reaction up to neutron threshold \cite{Byelikov}.
The observed $B$(GT) as well as transition strengths for other multipolarities obtained by RPA are used to evaluate $\nu$-process yields of $^{138}$La and $^{180}$Ta in SNe of progenitors with M=15$M_{\odot}$ and 25$M_{\odot}$.
The production yield for $^{138}$La obtained with $T_{\nu_e}$ = 4 MeV is found to be consistent with the solar abundance, while $^{180}$Ta is over-produced.

The ground state (g.s.) of $^{180}$Ta (1$^{+}$) is unstable and undergoes $\beta$-decay with a half-life of 8.15 hr, and naturally abundant $^{180}$Ta is actually a metastable 9$^{-}$ isomer state at 77.1 keV with a half-life of $\ge$ 10$^{15}$ yr.
The g.s. (1$^{+}$) and the isomer stae (9$^{-}$) can couple via excitations of states at intermediate energies in astrophysical environments with finite temperature. 
In Refs. \cite{Hayakawa10a,Hayakawa10b}, the g.s. and isomer bands are treated as independent nuclear species, which are in thermal equilibrium amomg themselves separately. 
The two bands are assumed to be weakly connected by a few linking transitions. 
The time dependence of the population probabilities of the two bands is obtained by coupling them to each other with a time-dependent temperature of exponential form, $T$ = $T_0$ exp(-t/$\tau$).
The population of the isomer band decreases with decreasing temperature.
In the low-temperature freeze-out region ($T_9<$ 0.44), the two bands are decoupled and the isomer population ratio becomes $P_m$/($P_m$ +$P_{g.s.}$) = 0.39.
A similar branching ratio has been estimated in Ref.~\cite{Mohr}.  
The over-production problem of $^{180}$Ta is thus solved with this branching ratio.
     
Besides the nuclei discussed above, $^{55}$Mn, $^{59}$Co, $^{92}$Nb and $^{98}$Tc can be produced via $\nu$-process.
The production yields of Mn and Co produced in complete Si burning are affected by neutrino processes \cite{YUN2008}.
In addition to the reaction $^{54}$Fe ($p$, $\gamma$) $^{55}$Co, the neutral-current reaction $^{56}$Ni ($\nu$, $\nu$'p) $^{55}$Co produces $^{55}$Co.
$^{55}$Mn is produced through successive e-capture processes $^{55}$Co (e$^{-}$, $\nu_{e}$) $^{55}$Fe (e$^{-}$, $\nu_{e}$) $^{55}$Mn.  
As we discussed in Sect. 3, the GT strength in $^{56}$Ni has a two-peak structure with appreciable high-energy tail. 
The proton emission channel opens at $E_x$ = 10.1 MeV when the transition to the 1/2$^{+}$ (2.92 MeV) state of $^{55}$Co begins to contribute to the cross section by emitting s-wave protons. 
The calculated strength above $E_x$ = 10.1 MeV amounts to be 62$\%$ of the total GT strength for the GXPF1J Hamiltonian. 
Thus the cross section for $^{56}$Ni ($\nu$, $\nu'p$) $^{55}$Co is enhanced for the GXPF1J compared with conventional cross sections obtained by HW02 \cite{Woos1990} and KB3G, 
which leads to the enhancement of the production yield of $^{55}$Mn compared with those for other Hamiltonians \cite{SHY2009}.
The calculated values of the yields obtained in a SNe model of a population III star with $M$ =15$M_{\odot}$ are found to be consistent with the abundances observed in extremely metal-poor stars with [Fe/H] $\leq$ -3 \cite{Cayrel2004}.
The proton knocked out from $^{56}$Ni by the $\nu$-induced reaction enhances the production yield of $^{59}$Co through the reaction chain $^{58}$Ni ($p$, $\gamma$) $^{59}$Cu (e$^{-}$, $\nu_{e}$) $^{59}$Ni (e$^{-}$, $\nu_{e}$) $^{59}$Co,
though the neutrino processes are not sufficient to explain the observed data.

%92Nb, 98Tc  
Though the isotope $^{92}$Nb, which has a half-life of 3.47$\times$ 10$^{7}$ yr, does not exist in the current solar system, the initial abundance ratio for $^{92}$Nb/$^{93}$Nb has been measured in primitive meteorites, and is found to be $\sim$10$^{-5}$ \cite{Harpper,Schonbachler,Iizuka}.
A SN $\nu$-process origin of $^{92}$Nb, mainly by $^{92}$Zr ($\nu_{e}$, e$^{-}$) $^{92}$Nb, has been proposed, and the observed ratio is shown to be explained by the $\nu$-process \cite{Hayakawa2013}.

The isotope $^{98}$Tc with a half-life of 4.2$\times$ 10$^{6}$ yr could have been also produced at the time of solar system formation.
Production of $^{98}$Tc by $\nu$-process in SNe has been investigated with QRPA by taking into account both charged- and neutral-current reactions \cite{Hayakawa2018}.
The dominant contribution comes from the $^{98}$Mo ($\nu_{e}$, e$^{-}$) $^{98}$Tc reaction. 
The charged-current reactions induced by $\bar{\nu}_{e}$, $^{99}$Ru ($\bar{\nu}_{e}$, e$^{+} n$) $^{98}$Tc and $^{100}$Ru ($\bar{\nu}_{e}$, e$^{+} 2n$) $^{98}$Tc, can contribute about 20$\%$ of the total production of $^{98}$Tc.
The calculated production yield of $^{98}$Tc is lower than the one which corresponds to the measured upper limit of $^{98}$Tc/$^{98}$Ru $<$ 6$\times$10$^{-5}$ \cite{Becker}.
If the initial abundance were to be precisely measured, the $^{98}$Tc nuclear chronometer could be used to evaluate a much more precise value of the duration time from the last core-collpase SN until the formation of the solar system. 
Finally, we comment that the production yields of $^{92}$Nb and $^{98}$Tc are reduced by about 40$\%$ and 20$\%$, respectively, remaining in the same order of magnitude, if neutrino spectra with low temperatures, $T_{\bar{\nu}_e}$ = $T_{\nu_{\mu,\tau}}$ = 4 MeV, are used \cite{Siverd2018}.       

%\subsection{$\nu p$ Process\label{sec:nup}}
          
In proton-rich environment of SNe with $Y_e>$0.5, which occurs during a few seconds after core bounce \cite{Buras2003}, the $\alpha$ rich freeze-out of such proton-rich matter leads to the production of $\alpha$ nuclei such as $^{56}$Ni with some extra protons.
Subsequent proton captures induce rapid-proton capture process, proceeding along the proton-rich region and producing light p-process nuclides such as $^{64}$Zn \cite{Pruet2005,Frohlich2005}. 
However, the process to heavier nuclides is suppressed by the increasing Coulomb barrier of the produced elements, so-called waiting point nuclei such as $^{64}$Ge and $^{68}$Se.
In the presence of intense $\nu$ fluxes, $\bar{\nu}_{e}$-capture on protons produce free neutrons which can be captured by $N\sim Z$ neutron-deficient nuclei such as $^{64}$Ge. 
The $^{64}$Ge ($n, p$) $^{64}$Ga reaction permits the matter flow to continue to heavier nuclei with $A>$ 64 via subsequent proton capture 
%near proton dripline 
up to the mass region $A\sim$ 80-100 \cite{Frohlich2006,Pruet2006,Wanajo2006}.
Heavy p-nuclei such as $^{92,94}$Mo and $^{96,98}$Ru are produced by this rapid p-process.
This nucleosynthesis process in proton-rich environment is called the $\nu p$ process.

\subsection{Effects of Neutrino Oscillations on Nucleosynthesis\label{sec:osc}}

We now discuss the effects of $\nu$ flavor oscillations on nucleosynthesis. 
In SN explosions, the Mikheev-Smirnov-Wolfenstein (MSW) matter resonance $\nu$ oscillations \cite{MSW1,MSW2} become important in case of the normal (inverted) mass hierarchy for neutrinos (anti-neutrinos). 
The resonance condition for the MSW oscillation is determined by
\begin{eqnarray}
\rho Y_e &=& \frac{\Delta}{2\sqrt{2}E G_F} cos 2\theta \nonumber\\
&=& 6.55\times 10^6 \Bigl(\frac{\Delta m_{ij}^2}{1 eV^2}\Bigr)\Bigl(\frac{1 MeV}{E_\nu}\Bigr) cos 2\theta_{ij} \quad (\mbox{g cm}^{-3}) 
\end{eqnarray}

Values of the electron density $\rho Y_e$ are calculated to be $\rho Y_e $ = 300 -3000 g cm$^{-3}$ and 4-40 g cm$^{-3}$ for high-density and low-density resonances that correspond to $\theta_{13}$ and $\theta_{12}$, respectively. 
Thus the high-density resonance occurs in the O/C layer. 

In the case of a normal mass hierarchy with sin$^{2} \theta_{13} >$ 10$^{-3}$, where the adiabatic condition is satisfied in the resonance region, $\nu_{\mu}$ and $\nu_{\tau}$ convert to $\nu_{e}$ near O/C layer, and most $\nu_e$ in the He layer have high temperature inherited from $\nu_{\mu}$ and $\nu_{\tau}$.
A part of $\bar{\nu}_e$ in the He layer is converted to $\bar{\nu}_{\mu}$ and $\bar{\nu}_{\tau}$. 
This can be expressed as \cite{Dighe2000}
\begin{eqnarray}
N(\nu_e) &=& P N^0(\nu_e) + (1-P) N^0 (\nu_x) \nonumber\\
N(\bar{\nu}_e) &=& \bar{P} N^0(\bar{\nu}_e) + (1-\bar{P}) N^0(\bar{\nu}_x)
\end{eqnarray}
where $N^0$ and $N$ are initial and final neutrino numbers, $\nu_x$ = $\nu_{\mu}$ or $\nu_{\tau}$, $\bar{\nu}_x$ = $\bar{\nu}_{\mu}$ or $\bar{\nu}_{\tau}$, and ($P, \bar{P}$) = (0, cos$^2 \theta_{12}$) = (0, 0.68) with sin$^2 \theta_{12}$ =0.32.

In the case of an inverted mass hierarchy with sin$^2 \theta_{13} >$ 10$^{-3}$, $\bar{\nu}_{\mu}$ and $\bar{\nu}_{\tau}$ are converted to $\bar{\nu}_{e}$ near the O/C layer, and most $\bar{\nu}_e$ in the He layer are those converted from $\bar{\nu}_{\mu}$ and $\bar{\nu}_{\tau}$.
A part of $\nu_e$ in the He layer is converted to $\nu_{\mu}$ and $\nu_{\tau}$.
This is the case for ($P, \bar{P}$) = (0.32, 0) in Eq.(37). 
When the adiabatic condition is not satisfied, that is, sin$^2 \theta_{13} <$ 10$^{-3}$, considerable flavor changes do not occur in the O/C layer for both normal and inverted hierarchies. 

In the case of a normal hierarchy, increase of the rates of charged-current reactions, $^{4}$He ($\nu_e$, e$^{-} p$) $^{3}$He and $^{12}$C ($\nu_e$, e$^{-} p$) $^{11}$C, induced by more energetic $\nu_e$ in the He/C layer, leads to more production yields of $^{7}$Li and $^{11}$B through $^{4}$He ($\alpha, \gamma$) $^{7}$Li and $^{11}$C ( , e$^{+} \nu_e$) $^{11}$B. 

The dependence of the abundance ratio $^{7}$Li/$^{11}$B on the mixing angle $\theta_{13}$ and the mass hierarchies have been investigated for the $\nu$-induced reaction cross sections on $^{4}$He and $^{12}$C updated by the SFO-WBP Hamiltonian set, taking into account the ambiguities in neutrino spectra \cite{YSC2008}.
%Comparison of the calculated ratio with that deduced from previous $\nu$-induced cross sections \cite{Woos} is also done \cite{SK-JPG}. 
The ratio gets larger than 0.78 for sin$^2 \theta_{13} >$ 10$^{-3}$ for the normal hierarchy, while it remains a small value ($\sim$0.6) in the inverted case.
Recent long baseline accelerator experiments at T2K and MINOS, and reactor experiments at Daya Bay, Double CHOOZ and RENO, derived the value of $\theta_{13}$ as sin$^2 \theta_{13} \sim$ 0.1 \cite{T2K,MINOS,DAYA,DCHOOZ,RENO}.
Information on the ratio $^{7}$Li/$^{11}$B from pre-solar grains or supernova remnants can give reliable constraints on the mass hierarchy.

$^{11}$B and $^{7}$Li isotopes have been discovered in SiC X-grains of Marchison Meteorite \cite{Fujiya2011}. 
No significant enhancement of the ratio $^{7}$Li/$^{11}$B is found for the SN grains.
A statistical analysis of the meteorite data was done with the Bayesian method, and the inverted mass hierarchy was found to be more preferred with a probability of 74$\%$ \cite{Mathews2012}. 
On the contrary, recent accelerator experiments at T2K and NOvA suggest that a normal hierarchy is favored \cite{T2Kh,NOVA}. 
The mass hierarchy can be sensitive to the choice of neutrino spectra. The yield ratio $^{7}$Li/$^{11}$B is found to be significantly reduced when the spectra with lower energies \cite{Siverd2018} are used.
Further study on the sensitivity to the neutrino energies as well as their time dependence during SNe might be needed before drawing a definite conclusion.
%toward a resolution of this tension.} 

Besides the MSW oscillation, collective oscillations of neutrino flavors are induced by the $\nu -\nu$ scatterings in a sufficiently dense neutrino gas, such as the atmosphere above a proto-neutron star.
Some ideal cases including mean-field approximations have been studied \cite{Pastor,FQ,Fogli,DFCQ,Pretel}, and applied to the r-process nucleosynthesis in a SN \cite{BY,PBKY} and black hole accretion disc \cite{Malkus}. 

Collective $\nu$ oscillations affect the neutrino spectra in SNe. 
Bimodal instabilities lead to swapping and splitting of the neutrino spectra \cite{Duan2006,DFQ}.
%the following swapping and splitting of the $\nu$ spectra.
%Nothing occurs in case of normal hierarchy.
In case of inverted hierarchy for the two-flavor mixing, $\nu_e$ and $\nu_x$ swap each other at $E_{\nu}>$ $E_{split}$, resulting in more energetic $\nu_e$.
The ratio $^{7}$Li/$^{11}$B is then enhanced also for the inverted hierarchy. 
When both collective and MSW oscillations are taken into account, the dependence of the ratio $^{7}$Li/$^{11}$B on the mass hierarchy can be smaller and may lead to less probability for the inverted hierarchy.
 
Effects of collective $\nu$ flavor and MSW oscillation on supernova nucleosynthesis are
studied by taking into account the time-dependent neutrino spectra and electron density profiles \cite{Wu2015}.
The $\nu$ oscillations are shown to affect the production of $^{138}$La and $^{180}$Ta as well as light nuclei such as $^{7}$Li and $^{11}$B, but have little impact on the $\nu p$-process nucleosynthesis.

The abundances of $^{7}$Li, $^{11}$B, $^{92}$Nb, $^{98}$Tc, $^{138}$La and $^{180}$Ta produced by $\nu$-process in a core-collapse SN explosion are evaluated by taking into account both collective and MSW neutrino oscillations \cite{Ko2020}.
Time dependent neutrino spectra are obtained by considering the $\nu$ self-interaction near the neutrino sphere and the MSW effect in the outer layers.   
Abundances of $^{7}$Li and the heavy isotopes $^{92}$Nb, $^{98}$Tc and $^{138}$La are reduced by a factor of 1.5-2.0 by the $\nu$ self-interaction, while $^{11}$B is relatively insensitive to the $\nu$ self-interaction.
The abundance ratio, $^{138}$La/$^{11}$B, is found to be sensitive to the neutrino mas hierarchy, and the normal mass hierarchy is more likely to be consistent with the solar meteoric abundances. 
The ratio $^{7}$Li/$^{11}$B remains higher for normal hierarchy by a factor of 1.24 when both the $\nu$ self-interaction and MSW effects are included.  
Note that results are rather sensitive to initial neutrino parameters such as luminosities. Here, luminosities for $\nu_e$ and $\bar{\nu}_e$ are larger than those for heavy-flavor $\nu$.

The effects of collective neutrino oscillations on $\nu p$ process nucleosynthesis in proton-rich $\nu$-driven winds have been studied by combining three-flavor  multiangle simulations with nucleosynthesis network calculations \cite{Sasaki}.
Here, fluxes for $\nu_e$ and $\bar{\nu}_e$ are assumed to be more abundant than those of heavy flavor neutrinos at the neutrino sphere, in contrast to the case of the supernova model in Ref.~\cite{Wu2015}.  
In the early phase of $\nu$-driven wind, blowing at 0.6 s after core bounce, oscillation effects are prominent in inverted mass hierarchy and p-nuclei are synthesized up to $^{106,108}$Cd. 
In the later wind trajectory at 1.1 s after core bounce, abundances of p-nuclei are increased remarkably by $\sim$10-10$^{4}$ times in normal mass hierarchy, reaching heavier p-nuclei such as $^{124,126}$Xe and $^{130}$Ba. 
The averaged overproduction factor of p-nuclei is dominated by the later wind trajectories. 
The $\nu p$ process is shown to be strongly influenced by the collective oscillations, though the results depend on initial neutrino parameters and hydrodynamic quantities such as wind velocities. 
The effects of multi-azimuthal-angle instability \cite{Raffelt2011,Chakraborty2014,Zaizen}, caused by the violation of axial symmetry, on $\nu$ and $\nu p$ processes are left for future study. 

Possible effects of active-sterile neutrinos ($\nu_s$) of eV mass scale on supernova explosion and nucleosynthesis are investigated for an 8.8$M_{\odot}$ star \cite{Wu2014}.
Conversions of $\nu_e \rightarrow \nu_s$ and $\bar{\nu}_e \rightarrow \bar{\nu}_s$ by the MSW oscillations occur in the resonance region $Y_e \approx$ 1/3, whose feedback is found to lead to further enhancement of the $\nu_e \rightarrow \nu_s$ conversion and the reduction of $Y_e$ at outer regions.
This results in the production of heavier nuclei with $Z$ =38-48 in ECSN, which was not possible without the active-sterile flavor conversion.
An extension of the work has been done by including the $\alpha$-effect and $\nu$-$\nu$ interaction \cite{Pllumbi2015}.  
The $\alpha$-effect is shown to affect the $Y_e$ evolution in a subtle way.
Effects of active-sterile neutrino oscillations on the dynamics and nucleosynthesis of $\nu$DW CCSNe are studied by taking into account the feedback of the oscillations on the wind profiles \cite{Xiong2019}.
For heavier $\nu_s$ mass case of $m_{\nu_s} >$ 1 eV, the oscillations are found to reduce the mass-loss rate and the wind velocity by a factor of 1.6$\sim$2.7, and change $Y_e$ significantly in favor of the production of heavier elements such as $^{86}$Kr and $^{90}$Zr.

Assuming that there occurs the so-called fast flavor oscillations caused by multiangle instability just near the proto-neutron star surface \cite{Sawyer2009,Dasgupta2017}, possible effects of such oscillations on the $\nu$DW CCSNe are studied for progenitors of 8.8M$_{\odot}$ and 27M$_{\odot}$ \cite{Xiong2020}.
The oscillations are found to enhance the total mass loss by a factor 1.5$\sim$1.7, and leads to more proton-rich conditions, enhancing the production of $^{64}$Zn and light $p$-nuclei such as $^{74}$Se, $^{78}$Kr and $^{84}$Sr.

\subsection{Neutrino-Nucleus Reactions for $\nu$ Detection\label{sec;detect}}

 Recently, $\nu$-induced reaction cross sections that are important for neutrino detection have been updated, for example, for $^{13}$C, $^{16}$O and $^{40}$Ar.
$^{13}$C is an attractive target for detection of very low energy neutrinos, $E_{\nu}\le$ 10 MeV.
As the threshold energy for $\nu$-induced reactions on $^{12}$C is $\sim$13 MeV, neutrinos with energy less than 13 MeV can interact only with the $^{13}$C isotope in natural carbon or carbon-based scintillators. 
Natural isotope abundance of $^{13}$C is 1.07$\%$, and $\nu$-$^{13}$C interactions are non-negligible in precision experiments. 
$^{13}$C-enriched targets can be useful detectors for solar $\nu$ ($E_{\nu}<$ 15 MeV) and reactor $\bar{\nu}$ ($E_{\bar{\nu}}<$ 8 MeV).
Cross sections for charged- and neutral-current reactions to low-lying negative and positive parity states of $^{13}$N and $^{13}$C have been updated \cite{SBK} with the use of the SFO Hamiltonian for $p$-$sd$ shell, which proved to be quite successful in $\nu$-$^{12}$C reactions.
Cross sections for GT transitions are compared with those obtained by the Cohen-Kurath (CK) Hamiltonian for $p$-shell. 
A moderate quenching for the axial-vector coupling constant $g_A^{eff}/g_A$ =0.95 is enough, in contrary to the CK case, which needs a substantial quenching factor of 0.69. 

An extension of the study to SN $\nu$ energy region is carried out for $^{13}$C \cite{SBKC}.
The partial cross sections for various $\gamma$ and particle emission channels are obtained by using the Hauser-Feshbach model.
In the charged-current reaction $^{13}$C ($\nu_e$, e$^{-}$) $^{13}$N, the proton emission becomes dominant above the particle-emission threshold energy.
This partial cross section can be observed if the scintillator is capable of pulse-shape discrimination. 
The reaction to the ground state of $^{13}$N is the second dominant channel, while the neutron emission cross section is quite small. 

In the neutral-current reaction, $\gamma$ and neutron emission channels give large contributions to the total cross section except for the elastic coherent scattering.
Neutrons are  detectable via the 2.2 MeV photons emitted from thermal proton capture reaction, $n$ + $p$ $\rightarrow$ $d$ + $\gamma$, and can provide a useful signal.
Note that neutron emission cross sections are much larger for the neutral-current reaction on $^{13}$C with the threshold energy of $E_{th}$ = 4.95 MeV compared to the reaction on $^{12}$C ($E_{th}$ =18.72 MeV) by about 10$^{3}$-10$^{1}$ at $E_{\nu}$ = 22-40 MeV.    

One-neutron emission leading to $^{12}$C (0$^{+}$, g.s.) is the dominant channel among the one-neutron knock-out, while the channel leading to $^{12}$C (2$^{+}$, 4.44 MeV) followed by $\gamma$ (4.44 MeV) emission ($E_{th}$ = 9.4 MeV) gives a contribution smaller by an order of magnitude or more at $E_{\nu}<$ 30 MeV.
Short-baseline reactor neutrino experiments identified a shape distortion in 5-7 MeV range in the measured $\nu$ spectrum \cite{Balant}, which appears as an excess over the predicted spectra.
A beyond the Standard Model solution to resolve this issue was proposed \cite{Berryman}: non-standard neutrino interactions induce the reaction $^{13}$C ($\bar{\nu}$, $\bar{\nu}'n$) $^{12}$C (2$^{+}$) followed by a prompt 4.44 MeV photon emission.
The produced neutrons would then be captured by protons, yielding scintillation light.
It was proposed that this scintillation light along with the prompt photon would mimic the spectral distortion around 5 MeV. 
The cross section obtained by the Standard Model here would help to assess further investigation of such processes.
The Standard Model cross section is too small to be compatible with the solution proposed above. 

Now, we discuss elastic coherent cross sections.
The coherent scattering induced by the weak neutral current is an important tool to test the standard model such as the value of the Weinberg angle, as well as the possibility for non-standard weak neutrino-nucleon interactions.
It is also a background for detection of dark matter by neutrino scattering.
Here, we focus more on nuclear physics aspects. 
The coherent scattering can be a good probe of neutron density distributions in nuclei \cite{Patton}. 

\begin{figure}[tbh]
\begin{center}
\begin{minipage}[t]{16.5 cm}
%\vspace*{-3.0 cm}
%\epsfig{file=ppnp_sdnico78.eps,scale=1.10}
%energtnane23.eps, scale=1.1}
\hspace*{-0.50 cm}
\epsfig{file=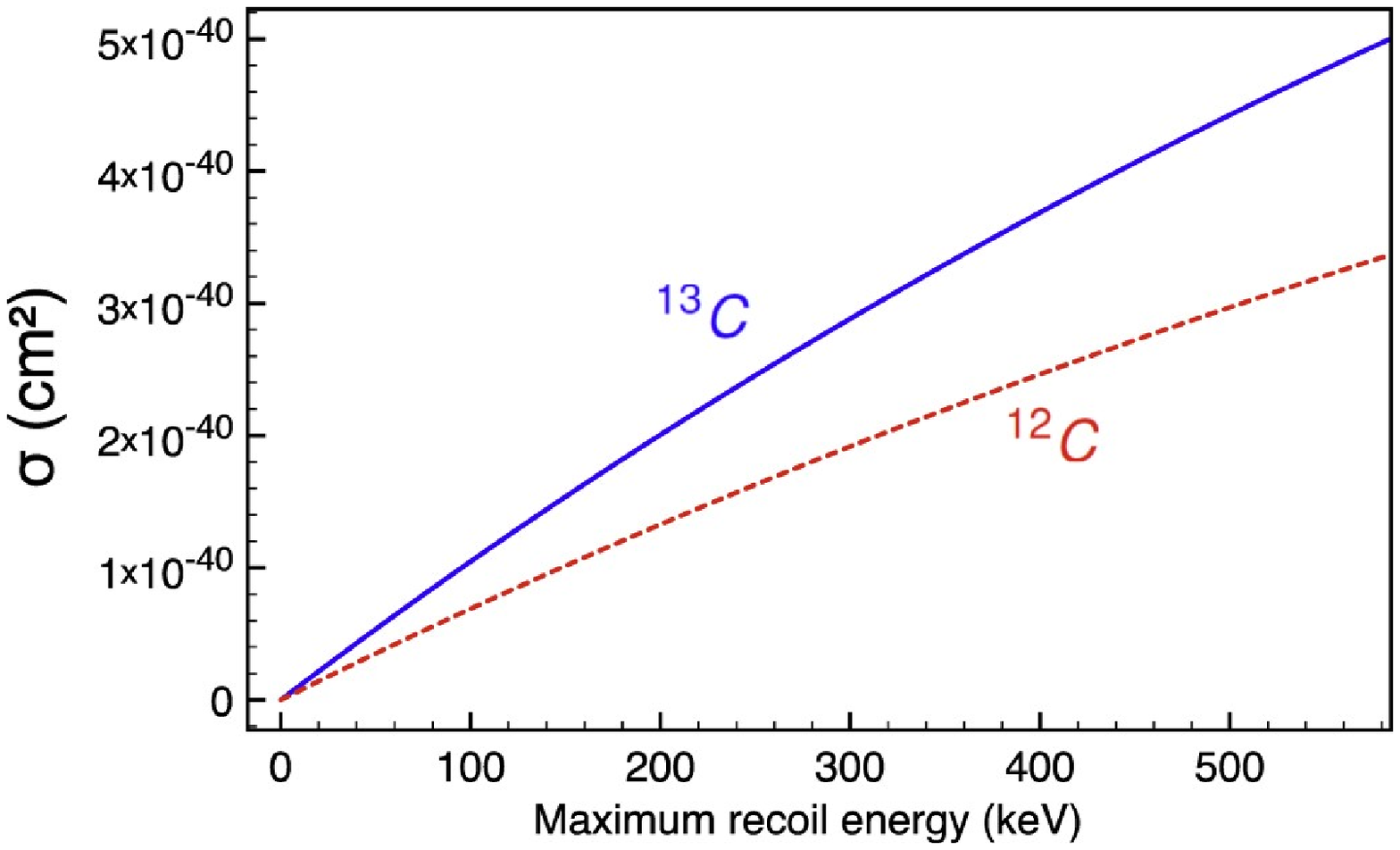,scale=0.30}
%energtnena23.eps, scale=1.1}
\hspace*{-0.5cm}
%\vspace*{1.0cm}
\epsfig{file=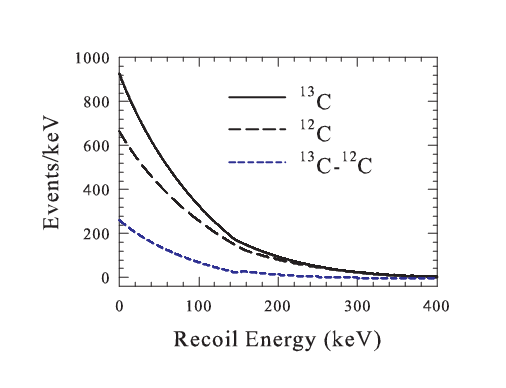,scale=1.08}
%\vspace*{2.0cm}
\end{minipage}
\begin{minipage}[t]{16.5 cm}
%\vspace*{-0.5cm}
\caption{
(Left) Neutrino elastic scattering cross sections on $^{13}$C (solid line) and $^{12}$C (dashed line) as functions of the maximum nuclear recoil energy, $T_{max}$. Taken from Ref. \cite{SBKC}. 
(Right) Event rates for coherent elastic scatterings on $^{13}$C and $^{12}$C induced by DAR neutrinos, as well as their difference, are shown as functions of the nuclear recoil energy (see the text for the details).  
\label{fig19}}
\end{minipage}
\end{center}
\end{figure}

The neutrino-nucleus coherent elastic scattering cross section is given by \cite{Freedman,FST,Drukier}
\begin{equation}
\frac{d\sigma}{dT} = \frac{G_F^2}{8\pi}M [2-\frac{2T}{T_{max}} +(\frac{T}{E})^2] Q_W^2 [F(Q^2)]^2
\end{equation}
where $T$ is the recoil energy of the nucleus, $E$ is the energy of the incoming neutrino, $M$ is the mass of the target nucleus, $Q^2=2MT+T^2$ is the square of the momentum transfer, $T_{max}$ = $\frac{2E^2}{2E+M}$ is the maximum nuclear recoil energy, and 
\begin{equation} 
Q_W = N - (1 -4sin^2\theta_W) Z
\end{equation}
is the weak charge of the nucleon.
The effect of the finite size of the nucleus is given by the form factor
\begin{equation}
F(Q^2) = \frac{1}{Q_W} \int [\rho_n -(1-4sin^2\theta_W)\rho_p] r^2\frac{sin(Qr)}{Qr} dr
\end{equation}
where $\rho_n$ and $\rho_p$ are neutron and proton density distributions in the nucleus, respectively.
The deviation of $F(Q^2)$ from unity reflects the finite size of the density distributions of the nucleus.
The coherent scattering was experimentally observed for the first time only very recently using a CsI scintillator \cite{Akimov2017}.
The total elastic scattering cross section is given by
\begin{equation}
\sigma(E) = \int_{0}^{T_{max}} \frac{d\sigma}{dT}(E,T) dT.
\end{equation}
The total elastic cross sections on $^{13}$C and $^{12}$C as functions of $T_{max}$ are compared in Fig.~\ref{fig19}.
We see that even a single extra neutron appreciably increases the coherent elastic cross section.

\begin{figure}[tbh]
\begin{center}
\begin{minipage}[t]{16.5 cm}
%\vspace*{-3.0 cm}
%\epsfig{file=ppnp_sdnico78.eps,scale=1.10}
%energtnane23.eps, scale=1.1}
\hspace*{-0.50 cm}
\epsfig{file=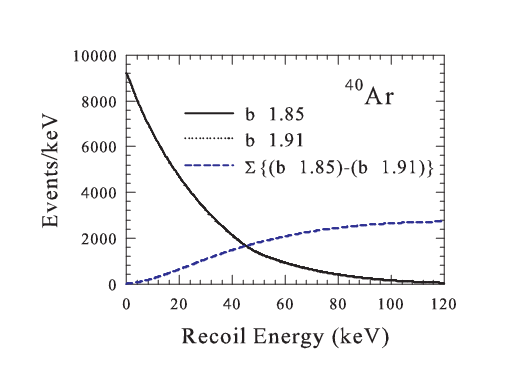,scale=1.05}
%energtnena23.eps, scale=1.1}
\hspace*{-1.0cm}
%\vspace*{1.0cm}
\epsfig{file=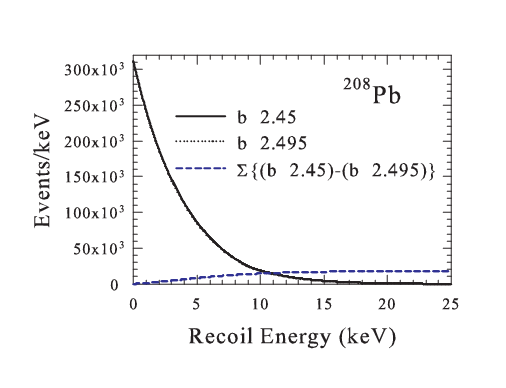,scale=1.05}
%\vspace*{2.0cm}
\end{minipage}
\begin{minipage}[t]{16.5 cm}
%\vspace*{-0.5cm}
\caption{
Event rates for coherent elastic scatterings on $^{40}$Ar (left panel) and $^{208}$Pb (right panel) induced by the DAR neutrinos obtained for two cases of neutron distributions in nuclei.  
The r.m.s. radii of the neutron distributions are set to differ by 0.1 fm. 
Cumulative sums of the difference of the event rates for the two cases are also shown (short-dashed line).
\label{fig20}}
\end{minipage}
\end{center}
\end{figure}

Event rates for the coherent elastic scatterings on $^{13}$C and $^{12}$C induced by DAR neutrinos are estimated for one-year $\nu$-fluxes of 3$\times10^{7}$/cm$^{2}$/s and 1 ton of the targets.
Three flavors of the DAR neutrinos, $\nu_{\mu}$, $\nu_e$ and $\bar{\nu}_{\mu}$, are taken into account. 
Experiments on natural carbon with 1.07$\%$ mixture of $^{13}$C and $^{13}$C-enriched target would give the event rates for each carbon isotopes separately.
The event rates for $^{13}$C and $^{12}$C as well as their difference are shown in Fig.~\ref{fig19}. 
The difference of the event rates between $^{13}$C and $^{12}$C is large enough to distinguish one-extra neutron in $^{13}$C.        
Cumulative sum of the event numbers amounts to be about 11000, 16000, 18000 and 19000 up to the nuclear recoil energy ($T_{rec}$) of 50, 100, 150 and 200 keV.
 
Event rates for the coherent elastic scatterings on $^{40}$Ar and $^{208}$Pb induced by the DAR neutrinos are also estimated for the same conditions of the $\nu$-fluxes and target.
The difference between the event numbers for two neutron density distributions, whose r.m.s. radii differ by 0.1 fm, as well as the event rates are shown in Fig.~\ref{fig20}.
The cumulative sum of the difference in the event numbers amounts to be about 1800, 2600 and 2700 at $T_{rec}$ = 50, 100 and 150 keV, respectively, for $^{40}$Ar, and about 8000, 14000 and 17000 at $T_{rec}$ = 5, 10 and 15 keV, respectively, for $^{208}$Pb.
Coherent elastic scattering would be able to distinguish the radius of neutron distribution in the nucleus to the accuracy of 0.1 fm. 
When the radius of proton distribution in the nucleus is measured by electron scattering, the neutron skin of the nucleus can be obtained and gives a crucial information on the equation of state of nuclear matter.      

While hydrogen (proton) contained in water detects $\bar{\nu}_e$ by the inverse $\beta$-decay process, $\bar{\nu}_e +p$ $\rightarrow$ $n +e^{+}$, $^{16}$O in water can detect both $\nu_e$ and $\bar{\nu}_e$ as well as heavy-flavor neutrinos via charged- and neutral-current reactions.
Neutrino-induced reaction cross sections on $^{16}$O have been updated by shell-model calculations with the use of a new Hamiltonian, SFO-tls \cite{SO2008}, which can describe well the spin-dipole transition strength in $^{16}$O \cite{SCY2018}.
The dominant contributions to the cross sections come from the spin-dipole transitions with $\lambda^{\pi}$ = 0$^{-}$, 1$^{-}$ and 2$^{-}$.   
The total strength of the spin-dipole operator
\begin{equation}
S^{\lambda}_{\mp,\mu} = r[Y^{1}\times\vec{\sigma}]^{\lambda}_{\mu} t_{\mp}
\end{equation}
given by
\begin{equation}
B(SD\lambda)_{\mp} = \frac{1}{2J_i +1} \sum_{f}|<f||S^{\lambda}_{\mp}||i>|^2
\end{equation}
is roughly proportional to $2\lambda$+1, and exactly so if $^{16}$O is assumed to be an LS-closed core.

The strength distribution of each multipole is determined by the momentum and spin dependent part of the nuclear Hamiltonian.
The 2$^{-}$ strength is at the lowest energy side, while the 1$^{-}$ strength is shifted toward the higher energy region.  
This can be understood from the energy-weighted sum (EWS) of the strength defined by
\begin{eqnarray}
EWS^{\lambda}_{\pm} &=& \sum_{\mu} |<\lambda, \mu| S^{\lambda}_{\pm,\mu}|0>|^2 (E_{\lambda}-E_0), \nonumber\\
EWS^{\lambda} &=& EWS^{\lambda}_{-} + EWS^{\lambda}_{+} \nonumber\\
&=&\frac{1}{2}<0| [S^{\lambda^{\dagger}}_{-}, [H,S^{\lambda}_{-}]] + [[S^{\lambda^{\dagger}}_{+},H],S^{\lambda}_{+}] |0>. 
\end{eqnarray}
The EWS rule values for the kinetic energy term $H=K=p^2/2m$ and one-body spin-orbit potential $V_{LS} = -\xi\sum_{i} \vec{\ell}_i\cdot\vec{\sigma}_i$ are given by \cite{Suzuki2018,Suzuki-NP}
\begin{eqnarray}
EWS^{\lambda}_{K} &=& \frac{3}{4\pi}(2\lambda +1) \frac{\hbar^2}{2m}A [1+\frac{f_{\lambda}}{3A}<0|\sum_{i}\vec{\sigma}_i\cdot\vec{\ell}_i|0>] \nonumber\\
EWS^{\lambda}_{LS} &=& \frac{3}{4\pi}(2\lambda +1)\frac{f_{\lambda}}{3}\xi <0|\sum_{i} (r_i^2 + g_{\lambda} r_i^2 \vec{\ell}_i\cdot\vec{\sigma}_i)0>.
\end{eqnarray}
where $f_{\lambda}$ = 2, 1 and -1, and $g_{\lambda}$ = 1, 1 and -7/5 for $\lambda^{\pi}$ = 0$^{-}$, 1$^{-}$ and 2$^{-}$, respectively. 
For an LS-closed core, $<0|\sum_{i} \vec{\sigma}_i\cdot\vec{\ell}_i|0>$ and 
$<0|\sum_{i} r_i^2 \vec{\ell}_i\cdot\vec{\sigma}_i|0>$ vanish, and $EWS^{\lambda}_{-}$ = $EWS^{\lambda}_{+}$. 
Both $EWS^{\lambda}_{K}$ and $EWS^{\lambda}_{LS}$ are proportional to $2\lambda +1$, and $EWS^{2}_{LS}$ is negative while $EWS^{0,1}_{LS}$ are positive ($\xi>$0).
The centroid energy for the strength distribution can be defined as $\bar{E}^{\lambda}$ = $EWS^{\lambda}_{-}$/$NEWS^{\lambda}_{-}$ where $NEWS^{\lambda}_{-}$ = $B(SD\lambda)_{-}$ with $|i>=|0>$.    
It is noticed from the sum rules discussed above that $\bar{E}^{2} < \bar{E}^{1} < \bar{E}^{0}$. 
Spin-dependent interactions, especially the tensor interaction, further affect the strength distributions.
The tensor interaction is attractive (repulsive) in $\lambda^{\pi}$ =0$^{-}$ and 2$^{-}$ (1$^{-}$), and shifts the centroid energy to lower (higher) energy region for $2^{-}$ and 0$^{-}$ (1$^{-}$) leading to the energy order, $\bar{E}^{2} < \bar{E}^{0} <\bar{E}^{1}$.
     
Charged- and neutral-current reaction cross sections are obtained with $g_A^{eff}$/$g_A$ =0.95 for the SFO-tls in both GT and SD transitions. 
The calculated total cross sections are rather close to those by CRPA \cite{Langanke2002} except at low neutrino energy below 30 MeV. 
Partial cross sections for various particle and $\gamma$ emission channels are evaluated with the branching ratios to each excited states obtained by the Hauser-Feshbach statistical model.
For ($\nu_e$, $e^{-}X$) and ($\nu$, $\nu'X$) reactions, the proton emission channel gives the dominant contributions, while for ($\bar{\nu}_e$, $e^{+}X$) reaction the neutron emission channel and the transition to the ground state of $^{16}$N give the dominant contributions.
The cross sections for $\alpha p$ emission channels are found to be rather large in the shell-model calculations, in contrast to the case of the CRPA.
This leads to sizeable production yields of $^{11}$B and $^{11}$C from $^{16}$O in supernova explosions.  This point was discussed in Sect.\ref{sec:nuproc}.     

A possible signal for supernova $\nu_{\mu}$ and $\nu_{\tau}$ neutrinos in water Cerenkov detectors induced by neutral-current reactions on $^{16}$O has been suggested \cite{Langanke1996}.
Photons with energies between 5 and 10 MeV, generated by ($\nu$, $\nu'p\gamma$) and ($\nu$, $\nu'n\gamma$) reactions on $^{16}$O, are shown to constitute a signal for a unique identification of the SN neutrinos. 
The yields of such $\gamma$ events are evaluated by using the CRPA cross sections, and a few hundred events are estimated to be detected in Super-Kamiokande for a SN at 10 kpc distance.    

Event spectra of $\nu$-$^{16}$O charged-current reactions at Super-Kamiokande (SK) are evaluated for a future SN neutrino burst \cite{Nakazato2018}.
The evaluations of the spectra as functions of electron or positron energy are performed with and without the MSW $\nu$-oscillations for an ordinary SN model with ($M$, Z) = (20$M_{\odot}$, 0.02) and a black-hole forming collapse model with ($M$, Z) = (30$M_{\odot}$, 0.004).
Here, $M$ is the progenitor mass and Z is the metallicity, and time-integrated $\nu$ spectra obtained in SN explosion simulations are used \cite{Nakazato2013}.
The expected event numbers of $^{16}$O ($\nu_e$, $e^{-}$) X and $^{16}$O ($\bar{\nu}_e$, $e^{+}$) X reactions are estimated for a SN at a distance of 10 kpc with the use of the shell-model cross sections.
The event numbers increase with the $\nu$-oscillation in the ($\nu_e$, $e^{-}$) and ($\bar{\nu}_e$, $e^{+}$) channels, respectively, for the normal and inverted mass hierarchy case for the ordinary SN model.
In case of the black-hole forming model, the impact of the $\nu$-oscillation is found to be smaller as the total energy of heavy-flavor neutrinos is much less than the one carried by $\nu_e$ and $\bar{\nu}_e$. 
In the case of ($\bar{\nu}_e$, $e^{+}$), the inverse $\beta$-decay channel has a large cross section and becomes the main background for the ($\bar{\nu}_e$, $e^{+}n$) channel. 
The spectral investigation of $\nu$-$^{16}$O charged-current events is challenging even in SK-Gd.  
  
Liquid argon detectors have excellent potentialities to detect core-collapse supernova neutrinos, especially by the $^{40}$Ar ($\nu_e$, $e^{-}$) $^{40}$K$^{\ast}$ reaction with tagging photons from excited states of $^{40}$K.
A liquid argon time projection chamber (TPC), proposed by ICARUS Collaborations \cite{Icarus}, can provide three-dimensional full particle track reconstruction with good energy resolution. 
Direct measurements of the charged-current reaction cross sections on $^{40}$Ar are accessible by using a liquid argon TPC detector and a spallation neutron source for neutrinos \cite{Cavanna}.

Gamow-Teller strength in $^{40}$Ar is studied by shell-model calculations \cite{SHAr40} with monopole-based universal interaction (VMU) \cite{VMU,OSFGA}, which includes tensor components from $\pi$+$\rho$-meson exchanges. 
The SDPF-M \cite{Utsuno} and GXPF1J interactions are used for the $sd$ shell and $pf$ shell, respectively, and the VMU is adopted for the $sd$-$pf$ cross shell part. 
Configurations within 2$\hbar\omega$ excitations, ($sd$)$^{-2}$($pf$)$^{2}$, are taken into account with a quenching of $g_A^{eff}$/$g_A$ = 0.775 \cite{Ormand}.
The calculated GT strength is found to be consistent with the experimental data obtained by ($p, n$) reactions \cite{Bhattacha}. 
Neutrino capture reaction on $^{40}$Ar for solar $^{8}$B are found to be enhanced compared with previous calculations \cite{Ormand}, where the GT strength obtained from $\beta^{+}$-decay of $^{40}$Ti was used.
The $\beta$-decay, $^{40}$Ti $\rightarrow$ $^{40}$Sc is an analog transition to the ($p, n$) reaction, $^{40}$Ar $\rightarrow$ $^{40}$K. 
Mirror symmetry is violated between the observed GT strengths.
The observed asymmetry can be explained by shell-model calculations with the use of an interaction that includes the Coulomb part and violates isospin symmetry \cite{Karakoc}.
Thus, the GT strength extracted from the $^{40}$Ar ($p, n$) data is recommended for the calculation of neutrino capture reactions on $^{40}$Ar.                

The reaction cross sections for multipoles other than 0$^{+}$ and 1$^{+}$ are obtained by RPA. Their contributions become important for neutrino energies larger than 50 MeV.
The calculated total cross section obtained \cite{SHAr40} is found to be rather close to that in Ref.~\cite{Kolbe} obtained by RPA for all the multipoles.
The cross section by the present hybrid model is enhanced by about 20-40$\%$ compared to Ref. ~\cite{Kolbe} at $E_{\nu}$ = 20-40 MeV, where the GT contributions dominate.
 
Exclusive cross sections for $\gamma$ emission and various particle emission channels in $^{40}$Ar ($\nu_e$, e$^{-}$) $^{40}$K$^{*}$ are evaluated by using the Hauser-Feshbach statistical model \cite{Gardiner2021}. 
Two de-excitation modes, $\gamma$ emission and single neutron emission, are found to be dominant at $E_{\nu} <$ 100 MeV. 
Tagging events involving neutron emission could substantially improve energy reconstruction of supernova $\nu_e$.      

\section{Summary}

In this work, nuclear weak rates in stellar environments are obtained based on recent advances in shell-model studies of both stable and unstable nuclei. 
Dominant contributions to the weak rates come from spin-dependent transitions such as GT transition strengths, whose evaluations become more precise owing to new shell-model Hamiltonians that can describe spin degree's of freedom in nuclei quite well with proper account of spin-dependent interactions. 
The weak rates, e-capture and $\beta$-decay rates, as well as $\nu$-nucleus reaction cross sections are thus improved and used to study evolution of stars and nucleosynthesis in SN explosion (SNe). 

The e-capture and $\beta$-decay rates in the $sd$-shell are evaluated with the use of the USDB Hamiltonian \cite{Toki,Suzu2016}, and applied to study cooling and heating of the O-Ne-Mg cores in stars with 8-10 $M_{\odot}$. 
The e-capture rates increase while the $\beta$-decay rates decrease as the density increases.  There is thus a density where both rates coincide. 
When this density is almost independent of the temperature, it can be assigned as the 'Urca density' where the cooling becomes most efficient by emitting both neutrinos and anti-neutrinos. 
The nuclear Urca processes are found to be important for nuclear pairs with $A$=23 and 25, ($^{23}$Na, $^{23}$Ne) and ($^{25}$Mg, $^{25}$Na) pairs \cite{Toki,Suzu2016}. 
Once the Urca processes are ignited in the core, first by the $A$=25 pair then by the $A$=23 pair, cooling of the O-Ne-Mg core proceeds by emitting $\nu_e$ and $\bar{\nu}_e$. 

In later stage of the evolution, double e-captures on $^{24}$Mg and $^{20}$Ne occur and the core is heated by the emitted $\gamma$'s \cite{Pinedo}. 
There is a competition between the contraction of the core caused by e-captures as $Y_e$ increases, and the heating of the core toward thermonuclear explosion.
The standard scenario is that the core contraction continues, and there occurs ECSNe and a NS is formed.
Recently, it was pointed out that the e-capture reaction $^{20}$Ne ($e^{-}$, $\nu_e$) $^{20}$F (2$_1^{+}$) induced by the second-forbidden transition may have important effects on the final fate of the core, i.e., whether thermonuclear explosion or core collapse occurs \cite{Kirsebom}. 
We find that the e-capture rates in $^{20}$Ne are considerably enhanced by the second-forbidden transition at densities log$_{10}$($\rho Y_e$) = 9.4-9.7 and temperatures log$_{10} T\leq$8.8 \cite{Suzuki2019,Kirsebom2}.
The multipole expansion method of Walecka \cite{Walecka,Ocon} and the Behrens-B$\ddot{\mbox{u}}$hring method \cite{bb1971} are used to evaluate the weak rates for the second-forbidden transition.
% \cite{Suzuki2019}, while the Behrens-Bu$\ddot{\mbox{u}}$ring method \cite{bb1971} is adopted in Refs.~\cite{Kirsebom2}. 
The relation between the two methods is explained and differences in the calculated rates are shown.
When the CVC relation is used for the evaluation of the transverse E2 matrix element, the difference in the calculated rates is insignificant.
The difference becomes large at log$_{10}(\rho Y_e)$ $\approx$ 9.6  for the case without the use of the CVC relation.  

%add BB vs Walecka, and CVC
The inclusion of the second-forbidden transition in $^{20}$Ne leads to an oxygen ignition, where the energy production and neutrino energy loss become equal, earlier than in the usual case without the second-forbidden contributions. 
The oxygen deflagration occurs at a certain density and temperature somewhat higher than for the case of the oxygen ignition due to the convection effects. 
Starting from the deflagration, explosion-collapse bifurcations are examined by multi-dimensional simulations, and the critical density $\rho_{crt}$ for the bifurcation is obtained \cite{Zha2019}.    
If the central density at the deflagration $\rho_{c,def}$ is lower (higher) than the critical density $\rho_{crt}$, thermonuclear (collapsing) ECSNe takes place.
By taking into account semiconvection and convection effects, the value for $\rho_{c,def}$ is estimated to be higher than $\rho_{crt}$.
Thus, the core is likely to collapse with a remnant NS \cite{Zha2019,Leung2019a}.
Note that in Ref.~\cite{Zha2019} use was made of the rates calculated in Ref.~\cite{Suzuki2019}.
%, in which the Walecka method is used without the CVC relation for the transverse E2 matrix element.
%It is an interesting problem to see if the present final fate, collapsing ECSNe, remains when the rates with the CVC relation are used.} 
When the effects of the convection are assumed to be small, and $\rho_{c,def}$ is taken to be close to the density at the oxygen ignition, the opposite conclusion is drawn: thermonuclear ECSNe is more likely to occur with a remnant O-Ne-Fe WD \cite{Jones2019,Kirsebom}.
Further investigation to clarify the transition from the oxygen ignition till the deflagration is necessary in future.

Nuclear weak rates in $pf$-shell are updated by shell-model calculations with the use of the GXPF1J Hamiltonian, which can describe the GT strengths in Ni and Fe isotopes quite well \cite{Suzu2011}.
In particular, the experimental GT strength in $^{56}$Ni is well reproduced by GXPF1J \cite{Sasano}.
Much amount of Ni and Fe nuclides are produced in the last stage of the stars, that is, in supernova explosions. Precise evaluations of the e-capture rates in $pf$-shell nuclei are important for the study of nucleosynthesis of iron-group nuclei in SNe. 
Here, the single-degenerate progenitor model is used for Type Ia SNe. 
Synthesis of elements is evaluated by taking into account production yields of nuclides by SNe and galactic chemical evolution (GCE).  
%Type Ia   ....
There was an overproduction problem of neutron-rich iron-group nuclei in Type Ia SNe, when previous FFN e-capture rates \cite{FFN} obtained by simple shell-model calculations were used. 
In this case, the production yields of neutron-rich nuclei, such as $^{58}$Ni, $^{54}$Cr and $^{54}$Fe, exceed the solar abundances by more than several times \cite{Iwa}. 
Once more accurate e-capture rates, obtained by improved shell-model calculations, are employed, this overproduction is found to be considerably suppressed \cite{Mori2016}.
The production yields of these neutron-rich isotopes are now close to the solar abundances within a factor of two.
The screening effects on the weak rates are also investigated. Their effects on the production yields of iron-group elements in Type Ia SNe are rather minor, as small as 10-30$\%$. 
The e-capture and $\beta$-decay rates, as well as neutrino energy-loss  and gamma-energy production rates, evaluated by the GXPF1J with and without the screening effects are tabulated \cite{Mori2020}, so that they can be used for astrophysical applications.   
       
We also discuss the weak rates for nuclei that concern two-major shells, such as $sd$-$pf$ and $pf$-$sdg$ shells.
Some nuclear pairs in two-major shells are found to be important for the nuclear Urca processes in neutron star crusts using QRPA evaluations \cite{Schatz}. 
The nuclear pair ($^{31}$Mg, $^{31}$Al) is such an example in the island of inversion.
While the ground state of $^{31}$Mg is 1/2$^{+}$, it was predicted to be 7/2$^{-}$ by the $sd$-$pf$ shell model Hamiltonian, SDPF-M, which was successful in describing spectroscopic properties of even nuclei in the island of inversion \cite{Utsuno}.
Then, the g.s. to g.s. transitions become forbidden 
%transitions 
and the Urca process does not occur.
This situation 
%in the shell-model calculation 
is improved by using an effective interaction in $sd$-$pf$ shell obtained by the extended Kuo-Krenciglowa (EKK) method \cite{Tsunoda,Takaya}, which is free from the divergence problem in G-matrix calculations in two-major shells \cite{EKK}. 
Thus the transitions between the low-lying states of $^{31}$Mg and $^{31}$Al are of GT-type, which leads to cooling by the nuclear Urca process \cite{OMEG15}.

The weak rates for isotones with $N$=50 were pointed out to be important for the core-collapse processes in SNe \cite{Sull}. 
The e-capture rates for nuclei at and around $N$=50 region are usually evaluated with the RPA and QRPA methods. The spin-dipole transitions were found to give important contributions.
The full $pf$-$sdg$ model space is necessary to evaluate spin-dipole transitions in nuclei with $N$=50 in shell-model calculations.
The spin-dipole strengths and e-capture rates for $^{78}$Ni are evaluated by the shell-model with the use of a $pf$-$sdg$ shell Hamiltonian, and compared with RPA calculations and the effective rate formula \cite{ni78}. 
%{The main purpose here is to compare the calculated rates of various methods as the effects of unblocking of the GT contributions in high temperature mentioned below are not taken into account.}
The rates obtained in the shell-model are enhanced compared with those of the RPA by 30-80$\%$, while they are found to be rather close to those obtained by the effective rate formula.
The dependence of the results on the quenching of the axial-vector coupling constant $g_A$ is also investigated.  
Comparison of the multipole expansion method of Walecka with the Behrens-B$\ddot{\mbox{u}}$hring method is made, and the difference in the rates is found to be about 20$\%$.
Recently, the thermal QRPA method, as well as the method including correlations beyond QRPA, were applied to obtain the e-capture rates in neutron-rich $N$=50 nuclei, and unblocking of the GT transition strength was found at high temperatures \cite{Dzh2019,Dzh2020,Litvinova2021}.     
%{\bf A considerable enhancement of the rates from GT$_{+}$ transitions is found with correlations beyond thermal QRPA \cite{Litvinova2021}.
These effects are not taken into account in this work.           

More than half of the elements heavier than iron are produced by rapid neutron capture process (r-process), but the sites of the r-process are still under controversy.
Though the r-process is sensitive to various nuclear physics inputs, 
%such as masses, neutron capture rates and $\beta$-decay rates.
here $\beta$-decay half-lives at the waiting point nuclei are studied by shell-model calculations \cite{Zhi2013,Suzuki2012}. 
In particular, focus is put on the half-lives of the isotones with $N$ =126.
%, which are important for the third peak of the r-process element abundances.
In case of $N$=126, the contributions from the first-forbidden transitions are important in addition to the GT transitions, in contrast to the cases of $N$ =50 and 82. 
Calculated half-lives are short compared with those obtained by the FRDM \cite{Moller1997,Moller2003}, which are conventionally used as the standard values.
%The half-life of $^{204}$Pt obtained by the shell-model is also consistent with the observation.
According to the measurements at RIBF, RIKEN, an overestimation of the half-lives predicted by the FRDM is also found in neutron-rich nuclides in the $A$=110 region \cite{Nishimura2011}.
Effects of the shorter half-lives obtained in the shell-model 
on r-process nucleosynthesis are investigated in Refs.~\cite{Suzuki2018,Nishimura2016}.
%slightly shift the position of the third peak of the r-process abundances toward higher mass region in all models of the $\nu$DW-CCSNe, MHDJ SNe and binary NSMs \cite{Suzuki2018}.
Study of neutron-rich nuclei around $N$=126, called $^{\prime}$blank spot$^{\prime}$ nuclei, is in progress by multinucleon transfer reactions \cite{Watanabe2015,Hirayama2017,Hirayama2018,Choi}. 
Another important nuclear physics input, the $\beta$-delayed neutron emission probability, is also under investigation in several facilities of the world \cite{Saldivia2017,Caballero2016,Pereira2009,Pereira2010,Gomez2014,Yokoyama2019}.

Neutrino-nucleus reaction cross sections relevant to nucleosynthesis and neutrino detection are updated with the use of new shell-model Hamiltonians.
Neutrino-induced reactions on $^{12}$C and $^{56}$Fe at DAR energies are now well described by the SFO in the $p$-$sd$ shell \cite{Suzuki2006} and by the GXPF1J in the $pf$-shell \cite{PaarSuz}, respectively.
The hybrid model combined with RPA for higher multipolarities is used for $^{56}$Fe and other $pf$-shell nuclei.
Important roles of neutral-current $\nu$-processes in producing $^{7}$Li and $^{11}$B in SNe are pointed out \cite{Suzuki2006}. 
Production yields of $^{7}$Li and $^{11}$B evaluated with the updated cross sections are enhanced by $\sim$15 $\%$ compared with previous calculations \cite{Woos1990}.
$^{138}$La and $^{190}$Ta are produced mainly by charged-current reactions on $^{138}$Ba and $^{180}$Hf, respectively \cite{Byelikov}. 
The coupling between the unstable ground state (1$^{+}$) and the metastable isomer state (9$^{-}$) of $^{180}$Ta in astrophysical environments at finite temperatures solves the overproduction problem of $^{180}$Ta, constrained by the measured GT transition strength \cite{Hayakawa10a,Hayakawa10b}.
Production of $^{55}$Mn, $^{59}$Co, $^{92}$Nb and $^{98}$Tc via $\nu$-process is also examined.
Flavor dependence and hierarchy of average $\nu$ energies and $\nu$ luminosities produced in SNe as well as their time dependence following the burst, accretion and cooling phases are important issues to obtain precise production yields of the elements \cite{Nakazato2013,Siverd2018,Siverd2019}.

Effects of $\nu$ oscillations on nucleosynthesis in SNe are also investigated. 
The MSW matter oscillations occur for normal $\nu$-mass hierarchy.
In the case of normal hierarchy, increase of the rates of charged-current reactions on $^{4}$He and $^{12}$C induced by more energetic $\nu_e$ leads to higher production yields of $^{7}$Li and $^{11}$B. 
The abundance ratio $^{7}$Li/$^{11}$B is found to be enhanced for the normal mass hierarchy \cite{YSC2008}. 
The inverted hierarchy is found to be more favorable according to a statistical analysis of the meteoric data in SiC X-grains \cite{Mathews2012}.
On the contrary, recent accelerator experiments at T2K and NOvA suggest a normal hierarchy \cite{T2Kh,NOVA}.
As the production yields of $^{7}$Li and $^{11}$B are sensitive to the choice of neutrino energy spectra and their time-dependence \cite{Siverd2018,Siverd2019}, one should be careful for drawing a definite conclusion.
Element synthesis by $\nu$-process in CCSNe has been studied by taking into account both the MSW oscillations and the collective $\nu$ oscillations induced by $\nu$-$\nu$ self interactions \cite{Wu2015,Ko2020}.  
The abundance ratio $^{138}$La/$^{11}$B is found to be sensitive to the mass hierarchy, and the normal hierarchy is more likely to be consistent with the solar meteorite abundances \cite{Ko2020}.
The ratio $^{7}$Li/$^{11}$B remains higher for normal hierarchy \cite{Ko2020}.
%{\bf One should note that results are rather sensitive to the initial $\nu$ parameters such as luminosities. 
Effects of the collective oscillations on $\nu p$-process nucleosynthesis are also studied. The production of $p$-nuclei is found to be enhanced by the collective oscillations, which depend sensitively on initial $\nu$ parameters and hydrodynamical quantities in SNe \cite{Sasaki}.

Nuclei such as $^{13}$C, $^{16}$O and $^{40}$Ar are attractive targets for the detection of SN and reactor neutrinos.
Neutrino-induced reaction cross sections on $^{13}$C in various $\gamma$ and particle emission channels are evaluated by the Hauser-Feshbach statistical model with the use of the SFO, which proved to be successful in $^{12}$C.
The $^{13}$C target is useful for detection of neutrinos with energies lower than 10 MeV, that is, reactor neutrinos, since the threshold energy for $\nu$-$^{12}$C reaction is as high as 15 MeV \cite{SBK}. 
As for $^{16}$O, the partial cross sections for the various channels are evaluated with a modified version of the SFO, 
%SFO-tls Hamiltonian, whose $p$-$sd$ cross shell part is improved from the SFO by taking into account proper tensor components.
which can well describe the spin-dipole strengths in $^{16}$O \cite{SCY2018}.
% are well described by the SFO-tls.
Properties of SN neutrinos and $\nu$ oscillations can be studied by the measurement of $\gamma$'s as well as emitted neutrons. 
Event spectra of $\nu$-$^{16}$O charged-current reactions at Super-Kamiokande are evaluated for a future SN neutrino burst with and without the MSW $\nu$-oscillation effects \cite{Nakazato2018}.
Liquid argon detectors have a powerful potential to detect CCSN neutrinos.
Charged-current reaction cross sections for $^{40}$Ar ($\nu_e$, e$^{-}$) $^{40}$K$^{*}$ are updated by a new shell-model Hamiltonian, which can describe well the experimental GT strength in $^{40}$Ar obtained by ($p$, $n$) reaction \cite{SHAr40}.

Coherent elastic $\nu$-nucleus scattering became accessible recently using a CsI scintillator \cite{Akimov2017}.
While the coherent scattering can be used to test the standard model, it can be a good probe of neutron distributions in nuclei.
Comparing cross sections and event numbers of the coherent scattering on $^{12}$C and $^{13}$C, one can distinguish the effect of one extra neutron in $^{13}$C \cite{SBKC}.
Coherent elastic scattering would be able to distinguish the radius of the neutron distributions in nuclei to the accuracy of 0.1 fm.

We have shown that the refinements of the weak rates and cross sections have important impacts on the study of various astrophysical processes.
For nuclei around $N$=50, inclusion of transitions from excited states is needed for high temperature astrophysical conditions of CCSNe, and its shell-model study is a challenge for future.
In the region of nuclei around $N$ =126, an extension of the shell-model calculations to various nuclei, such as in the $^{\prime}$blank spot$^{\prime}$ region, and evaluation of neutron emission probabilities is also an interesting challenge.
% for future.   
Further progress in the evaluation of the rates in various nuclear regions as well as in more extreme astrophysical environments would open up new advances in nuclear astrophysics.

\vspace*{1cm}      
{\bf Acknowledgements}
\\
The author would like to express his sincere thanks to K. Nomoto, S. Zha, S.-C. Leung, H. Toki, M. Honma, K. Mori, T. Kajino, B. A. Balantekin, S. Chiba, M. Famiano, N. Tsunoda, K. Shimizu, Y. Tsunoda, T. Otsuka, T. Yoshida, S. Shibagaki, K. Nakazato, and M. Sakuda for collaborations of the works reported here.  
He would also thank O. S. Kirsebom for providing him the numerical data of the  e-capture rates for the second-forbidden transition in $^{20}$Ne, which were shown in the figures in Refs.~\cite{Kirsebom,Kirsebom2}.  
He is grateful to W. Bentz for his careful reading of the manuscript.
This work was supported by JSPS KAKENHI grant No. JP19K03855.
He also thanks the NAOJ Japan for accepting him a visiting researcher during the period of the works. 


\vspace*{1cm}

\begin{thebibliography}{99}
\itemsep -2pt 
\bibitem{RMP2020} T. Otsuka, A. Gade, O. Sorlin, T. Suzuki, and Y. Utsuno, \Journal{\RMP} {92}{015002} {2020}. doi:10.1103/RevModPhys.92.015002
\bibitem{LMRMP} E. Caurier, G. Mart\'inez-Pinedo, F. Nowacki, A. Poves, and A. P. Zuker, \Journal{\RMP} {77}{427} {2005}. doi:10.1103/RevModPhys.77.427
\bibitem{FFN} G. M. Fuller, W. A. Fowler, and M. J. Newman, \Journal{\ApJS} {42}{447} {1980}. doi:10.1086/190657; \Journal{\ApJ} {252}{715} {1982}. doi:10.1086/159597; \Journal{\ApJS} {48}{279} {1982}. doi:10.1086/190779
%\bibitem{Fuller1980} G. M. Fuller, W. A. Fowler, and M. J. Newman, \Journal{\ApJS} {42}{447} {1980}
%\bibitem{Fuller1982a} G. M. Fuller, W. A. Fowler, and M. J. Newman, \Journal{\ApJ} {252}{715} {1982}
%\bibitem{Fuller1982b} G. M. Fuller, W. A. Fowler, and M. J. Newman, \Journal{ApJS} {48}{279} {1982}
\bibitem{Oda} T. Oda, M. Hino, K. Muto, M. Takahara, and K. Sato \Journal{\ADNDT} {56}{231} {1994}. doi:10.1006/adnd.1994.1007
\bibitem{LM} K. Langanke and G. Mart\'inez-Pinedo, \Journal{\ADNDT}{79}{1}{2001}.  doi:10.1006/adnd.2001.0865
\bibitem{HL} W. R. Hix, O. E. B. Messer, A. Mezzacappa, M. Liebendorfer, J. Sampaio, K. Langanke, D. J. Dean, and G. Mart\'inez-Pinedo, \Journal{\PRL} {91}{201102} {2003}. doi:10.1103/PhysRevLett.91.201102;\\
K. Langanke, G. Mart\'inez-Pinedo, J. M. Sampaio, D. J. Dean, W. R. Hix, O. E. B. Messer, A. Mezzacappa, M. Liebendorfer, H.-Th. Janka, and M. Rampp , \Journal{\PRL} {90}{241102} {2003}. doi:10.1103/PhysRevLett.90.241102
\bibitem{Miyaji} S. Miyaji, K. Nomoto, K. Yokoi, and D. Sugimoto, \Journal{\PASJ} {32}{303} {1980}. https://ui.adsabs.harvard.edu/abs/1980PASJ...32..303M
\bibitem{Nomoto1984} K. Nomoto, \Journal{\ApJ} {277}{791} {1984}. doi.10.1086/161749; \Journal{\ApJ}{322}{206} {1987}. doi:10.1086/165716
%\bibitem{Nomoto1987}
\bibitem{Nomoto1988} K. Nomoto and M. Hashimoto, \Journal{\PREP} {163}{13} {1988}. doi:10.1016/0370-1573(88)90032-4
%\bibitem{Nomoto2013} K. Nomoto, C. Kobayashi, and N. Tominaga, \Journal{\ARAA} {51}{457} {2013}. https://doi.org/10.1146/annurev-astro-082812-140956
\bibitem{Gutierrez1996} J. Gutierrez, E. Garcia-Berro, I. Iben, Jr., J. Isern, J. Labay, and R. Canal, \Journal{\ApJ} {459}{701} {1996}. doi:10.1086/176934 
\bibitem{Jones2016} S. Jones, F. K. Ropke, R. Pakmor, I. R. Seitenzahl, S. T. Ohlmann, and P. V. F. Edelmann, \Journal{\AA} {593}{A72} {2016}. doi:10.1051/0004-6361/201628321
\bibitem{Isern1991} J. Isern, R. Canal, and J. Labay, \Journal{\ApJ} {372}{L83} {1991}.  doi:10.1086/186029 
\bibitem{Canal1992} R. Canal, J. Isern, and J. Labay, \Journal{\ApJ} {398}{L49}{1992}. doi:10.1086/186574
\bibitem{Gutierrez2005}  J. Gutierrez, R. Canal, and E. Garcia-Berro, \Journal{\AA} {435}{231} {2005}. doi:10.1051/0004-6361:20042254
\bibitem{SMN2021} D. F. Str$\ddot{\mbox{o}}$mberg, G. Mart$\acute{\mbox{i}}$nez-Pinedo and F. Nowacki,
arXiv:2104.02614 (2021)
\bibitem{Pinedo} G. Mart\'inez-Pinedo, Y. H. Lam, K. Langanke, R. G. Zegres, and C. Sullivan, \Journal{\PRC} {89}{045806} {2014}. doi:10.1103/PhysRevC.89.045806
\bibitem{USDB} B. A. Brown and W. A. Richter, \Journal{\PRC} {74}{034315} {2006}. doi:10.1103/PhysRevC.74.034315
\bibitem{USD} B. A. Brown and B. H. Wildenthal, \Journal{\PRC} {28}{2397} {1983}. doi:10.1103/PhysRevC.28.2397
\bibitem{BW} B. A. Brown and B. H. Wildenthal, \Journal{\ARNPS} {38}{29}{1988}. doi:10.1146/annurev.ns.38.120188.000333
\bibitem{Suzu2011} T. Suzuki, M. Honma, H. Mao, T. Otsuka, and T. Kajino, \Journal{\PRC} {83}{044619} {2011}. doi:10.1103/PhysRevC.83.044619
\bibitem{Juod} A. Juodagalvis, K. Langanke, W. R. Hix, G. Mart\'inez-Piedo, and J M. Sampaio, \Journal{\NPA} {848}{454} {2010}. doi:10.1016/j.nuclphysa.2010.09.012
\bibitem{Toki} H. Toki, T. Suzuki, K. Nomoto, S. Jones, and R. Hirschi, \Journal{\PRC} {88}{015806}{2013}. doi:10.1103/PhysRevC.88.015806
\bibitem{Suzu2016} T. Suzuki, H. Toki, and K. Nomoto, \Journal{\ApJ} {817}{163} {2016}. doi:10.3847/0004-637X/817/2/163
\bibitem{Bravo} E. Bravo and D. Garcia-Senz, \Journal{\MNRAS} {307}{984} {1999}. doi:10.1046/j.1365-8711.1999.02694.x
%\bibitem{Itoh} N. Itoh, N. Tomizawa, M. Tamamura, and S. Wanajo, \Journal{\ApJ} {579}{380} {2002}. doi:10.1086/342726 
\bibitem{Slattery} W. L. Slattery, G. D. Doolen, and H. E. DeWitt, \Journal{\PRA} {26}{2255} {1982}. doi:10.1103/PhysRevA.26.2255
\bibitem{Ichimaru} S. Ichimaru, \Journal{RMP} {65}{255} {1993}. doi:10.1103/RevModPhys.65.255
%\bibitem{Itoh} N. Itoh, N. Tomizawa, M. Tamamura, and S. Wanajo, \Journal{\ApJ} {579}{380} {2002}. doi:10.1086/342726
\bibitem{Yakov} D. G. Yakovlev and D. A. Shalybkov, \Journal{\ASPRv} {7}{311} {1989}. 
https://ui.adsabs.harvard.edu/abs/1989ASPRv...7..311Y
\bibitem{Itoh} N. Itoh, N. Tomizawa, M. Tamamura, and S. Wanajo, \Journal{\ApJ} {579}{380} {2002}. doi:10.1086/342726
\bibitem{Richter} W. A. Richter, S. Mkhize, and B. A. Brown, \Journal{\PRC} {78}{064302} {2008}. doi:10.1103/PhysRevC.78.064302 
%\bibitem{Takahara} M. Takahara, M. Hino, T. Oda, K. Muto, A. A. Wolters, P. W. M. Glaudemans, and K. Sato, \Journal{\NPA} {504}{167} {1989}. doi:10.1016/0375-9474(89)90288-1
%\bibitem{Pinedo} G. Martinez-Pinedo, Y. H. Lam, K. Langanke, R. G. Zegres, and C. Sullivan, \Journal{\PRC} {89}{045806} {2014}. doi:10.1103/PhysRevC.89.045806
\bibitem{NNDC} National Nuclear Data Center, https://www.nndc.bnl.gov/
%\bibitem{Kirsebom} O. S. Kirsebom et al., \Journal{\PRL}{123}{262701}{2019}. doi:10.1103/PhysRevLett.123.262701
\bibitem{Endt} P. M. Endt, \Journal{\NPA} {633}{1} {1998}. doi:10.1016/S0375-9474(97)00613-1
\bibitem{Tilley} D. R. Tilley, C. M. Cheves, J. H. Kelley, S. Raman, and H. R. Weller, \Journal{\NPA} {636}{249} {1998}. doi:10.1016/S0375-9474(98)00129-8
\bibitem{Fuller} G. M. Fuller, W. A. Fowler and M. J. Newman, \Journal{\ApJ} {293}{1} {1985}. doi.10.1086/163208
\bibitem{Paxton2015} B. Paxton, P. Marchant, J. Schwab et al., \Journal{\ApJS} {220}{15} {2015}. doi:10.1088/0067-0049/220/1/15
\bibitem{Jones} S. Jones, R. Hirschi, K. Nomoto, T. Fischer, F. X. Timmes, F. Herwig, B. Paxton, H. Toki, T. Suzuki, G. Mart\'inez-Pinedo, Y. H. Lam, and M. G. Bertolli, \Journal{\ApJ} {772}{150} {2013}. doi:10.1088/0004-637X/772/2/150
%\bibitem{Walecka} J. D. Walecka, in {\it Muon Physics}, Vol. II, ed. V. W. Hughes and C. S. Wu (New York, Academic, 1975) 
%\bibitem{bb1971} H. Behrens and W. B$\ddot{\mbox{u}}$hring, \Journal{\NPA} {162}{111} {1971}. doi:10.1016/0375-9474(71)90489-1
%\bibitem{Ocon} J. S. O'Connell, T. W. Donnelly, J. D. Walecka, \Journal{\PRC} {6}{719} {1972}. doi:10.1103/PhysRevC.6.719  
\bibitem{Takahara} M. Takahara, M. Hino, T. Oda, K. Muto, A. A. Wolters, P. W. M. Glaudemans, and K. Sato, \Journal{\NPA} {504}{167} {1989}. doi:10.1016/0375-9474(89)90288-1
\bibitem{Kirsebom} O. S. Kirsebom et al., \Journal{\PRL}{123}{262701}{2019}. doi:10.1103/PhysRevLett.123.262701
\bibitem{Walecka} J. D. Walecka, in {\it Muon Physics}, Vol. II, ed. V. W. Hughes and C. S. Wu (New York, Academic, 1975)
\bibitem{bb1971} H. Behrens and W. B$\ddot{\mbox{u}}$hring, \Journal{\NPA} {162}{111} {1971}. doi:10.1016/0375-9474(71)90489-1
\bibitem{Ocon} J. S. O'Connell, T. W. Donnelly, J. D. Walecka, \Journal{\PRC} {6}{719} {1972}. doi:10.1103/PhysRevC.6.719
\bibitem{Paar} N. Paar, G. Colo, E. Khan, and D. Vretenar, \Journal{\PRC}{80}{055801}{2009}. doi:10.1103/PhysRevC.80.055801
\bibitem{Vretenar} A. F. Fantina, E. Khan, G. Colo, N. Paar, and D. Vretenar, \Journal{\PRC} {86}{035805} {2012}. doi:10.1103/PhysRevC.86.035805
\bibitem{Kura} T. Kuramoto, M. Fukugita, Y. Kohyama, and K. Kubodera, \Journal{NPA} {512}{711} {1990}. doi:10.1016/0375-9474(90)90232-B
\bibitem{Schopper} H. Schopper, {\it Weak Interactions and Nuclear Beta Decays} (North-Holland, Amsterdam, 1966)
\bibitem{Forest} T. de Forest Jr. and J. D. Walecka, \Journal{\ADV}{15}{1}{1966}. doi:10.1080/00018736600101254
%\bibitem{Adler} S. L. Adler, \Journal{\PREV} {139}{B1638} {1965}. 
\bibitem{Eichler} J. Eichler, \Journal{\ZP}{171}{463}{1963}. doi:10.1007/BF01377868
\bibitem{BlinStoyle} R. J. Blin-Stoyle, {\it Fundamental Interactions and the Nucleus} (North-Holland, Amsterdam, 1973) 
\bibitem{Fujita1} J. Fujita, \Journal{\PREV}{126}{202}{1962}. doi:10.1103/PhysRev.126.202
\bibitem{Fujita2} J. Fujita, \Journal{\PRO}{28}{338}{1962}. doi:10.1143/PTP.28.338
%\bibitem{BB} H. Behrens and W. Buring, {\it Electron Radial Wave Functions and Nuclear Beta-Decay} (Clarenton, Oxford, 1982) 
%\bibitem{bb1971} H. Behrens and W. B$\ddot{\mbox{u}}$hring, \Journal{\NPA} {162}{111} {1971}. doi:10.1016/0375-9474(71)90489-1 
\bibitem{buhring} W. B$\ddot{\mbox{u}}$hring, \Journal{\NP} {40}{472} {1963}. doi:10.1016/0029-5582(63)90290-6
\bibitem{behrens} H. Behrens and J. Janecke, {\it Numerical Tables for Beta-Decay and Electron Capture}, Landolt-Bornstein, New Series, Group I, Vol. 4 (Springer-Verlag, Berlin, 1969)
%\bibitem{Suzuki2019} T. Suzuki, S. Zha, S.-C. Leung, and K. Nomoto, \Journal{\ApJ} {881}{64} {2019}. doi:10.3847/1538-4357/ab2b93
\bibitem{Kirsebom2} O. S. Kirsebom et al., \Journal{\PRC} {100}{065805} {2019}. doi:10.1103/PhysRevC.100.065805
\bibitem{Suzuki2019} T. Suzuki, S. Zha, S.-C. Leung, and K. Nomoto, \Journal{\ApJ} {881}{64} {2019}. doi:10.3847/1538-4357/ab2b93
\bibitem{Zha2019} S. Zha, S.-C. Leung, T. Suzuki, and K. Nomoto, \Journal{\ApJ} {886}{22} {2019}. doi:10.3847/1538-4357/ab4b4b
\bibitem{NomotoKondo1991} K. Nomoto and Y. Kondo, \Journal{\ApJL} {367}{L19} {1991}. doi:10.1086/185922 
\bibitem{Timmes1992} F. X. Timmes and S. E. Woosley, \Journal{\ApJ} {396}{649} {1992}. doi:10.1086/171746
%\bibitem{Jones2016} S. Jones, F. K. Ropke, R. Pakmor, I. R. Seitenzahl, S. T. Ohlmann, and P. V. F. Edelmann, \Journal{\AA} {593}{A72} {2016}. doi:10.1051/0004-6361/201628321
\bibitem{LeungNomoto2019a} S.-C. Leung and K. Nomoto, \Journal{\PASA}{36}{e006} {2019}. doi:10.1017/pasa.2018.49
\bibitem{Fryer1999} C. Fryer, W. Benz, M. Herant, and S. A. Colgate, \Journal{\ApJ} {516}{892} {1999}. doi:10.1086/307119
\bibitem{Kitaura2006} F. S. Kitaura, H.-T. Janka, and W. Hillebrandt, \Journal{\AA} {450}{345} {2006}. doi:10.1051/0004-6361:20054703
\bibitem{Radice2017} D. Radice, A. Burrows, D. Vartanyan,M. A. Skinner, and J. C. Dolence, \Journal{\ApJ} {850}{43} {2017}. doi:10.3847/1538-4357/aa92c5
\bibitem{NomotoLeung2017a} K. Nomoto and S.-C. Leung, in Handbook of Supernovae, Vol. I ed. A. W. Alsabti and P. Murdin (Berlin, Springer, 2017), 483
\bibitem{Leung2019a} S.-C. Leung, K. Nomoto, and T. Suzuki, \Journal{\ApJ} {889}{34} {2020}. doi:10.3847/1538-4357/ab5d2f
\bibitem{Seitenzahl2009} I. R. Seitenzahal, D. M. Townsley, F. Peng, and J. W. Truran, \Journal{\ADNDT} {95}{96} {2009}. doi:10.1016/j.adt.2008.08.001 
\bibitem{Schwab2015} J. Schwab, E. Quataert, and L. Bildsten, \Journal{\MNRAS} {453}{1910} {2015}. doi:10.1093/mnras/stv1804
\bibitem{Schwab2017a} J. Schwab, L. Bildsten, and E. Quataert, \Journal{\MNRAS} {472}{3390} {2017}. doi:10.1093/mnras/stx2169
\bibitem{Takahashi2019} K. Takahashi, K. Sumiyoshi, S. Yamada, H. Umeda, and T. Yoshida, \Journal{\ApJ} {871}{153} {2019}. doi:10.3847/1538-4357/aaf8a8
\bibitem{SchwabRocha2019} J. Schwab and K. A. Rocha, \Journal{\ApJ} {872}{131} {2019}. doi:10.3847/1538-4357/aaffdc
\bibitem{MiyajiNomoto1987} S. Miyaji and K. Nomoto, \Journal{\ApJ} {318}{307} {1987}. doi:10.1086/165368 
\bibitem{Kippenhahn2012} R. Kippenhahn, A. Weigert, and A. Weiss, Stellar Structure and Evolution (2nd ed.; Berlin Springer, 2012)
%\bibitem{Zha2020} S. Zha, private communications
\bibitem{Jones2019} S. Jones, F. K. Ropke, C. Fryer et al., \Journal{\AA} {622}{A74} {2019}.  doi:10.1051/0004-6361/201834381
\bibitem{Nittler2018} L. R. Nittler, O'D. Alexander, N. Lau, and J. Wang, \Journal{\ApJL}{856}{L24}{2018}. doi:10.3847/2041-8213/aab61f
\bibitem{Poela2007} A. J. T. Poelarends, PhD Thesis, Astronomical Institute Utrecht (2007)
\bibitem{Doherty2015} C. L. Doherty, P. Gil-Pons, L. Siess, J. C. Lattanzio, and H. H. B. Lau, \Journal{\MNRAS}{446}{2599}{2015}. doi:10.1093/mnras/stu2180
\bibitem{Poela2008} A. J. T. Poelarends, F. Herwig, N. Langer, and A. Heger, \Journal{\ApJ} {675}{614}{2008}. doi:10.1086/520872 
\bibitem{Wanajo2011} S. Wanajo, H.-T. Janka, and B. M$\ddot{\mbox{u}}$ller, \Journal{\ApJL} {726}{L15} {2011}. doi:10.1088/2041-8205/726/2/L15 
\bibitem{Wanajo2013a} S. Wanajo, H.-T. Janka, and B. M$\ddot{\mbox{u}}$ller, \Journal{\ApJL} {767}{L26} {2013}; \Journal{\ApJL}{774}{L6} {2013}. doi:10.1088/2041-8205/767/2/L26
%\bibitem{Wanajo2013b}
%\bibitem{Nittler2018} L. R. Nittler, O'D. Alexander, N. Lau, and J. Wang, \Journal{\ApJL}{856}{L24}{2018}
\bibitem{Zhang2020} J. Zhang, X. Wang, V. Jozsef et al., \Journal{\MNRAS}{498}{84}{2020}.  doi:10.1093/mnras/staa2273
\bibitem{Hira2021} D. Hiramatsu, D. A. Howell, S. D. Van Dyk et al., \Journal{\NATAS}{5}{}{2021}. doi:10.1038/s41550-021-01384-2
\bibitem{Iwa} K. Iwamoto, F. Brachwitz,K. Nomoto, N. Kishimoto, H. Umeda, W. R. Hix, and F.-K. Thielemann, \Journal{\ApJS} {125}{439} {1999}. doi:10.1086/313278
\bibitem{LanM}  K. Langanke and G. Mart\'inez-Pinedo, \Journal{\NPA} {673}{481} {2000}. doi:10.1016/S0375-9474(00)00131-7
%\bibitem{LMRMP} E. Caurier, G. Martinez-Pinedo, F. Nowacki, A. Poves, and A. P. Zuker, \Journal{\RMP} {77}{427} {2005}. doi:10.1103/RevModPhys.77.427
\bibitem{KBF} E. Caurier, K. Langanke, G. Mart\'inez-Pinedo, and F. Nowacki, \Journal{\NPA} {653}{439}{1999}. doi:10.1016/S0375-9474(99)00240-7  
\bibitem{GX1J} M. Honma, T. Otsuka, T. Mizusaki, M. Hjorth-Jensen, and B. A. Brown, \Journal{\JPC}{20}{7} {2005}. doi:10.1088/1742-6596/20/1/002 
\bibitem{GX1} M. Honma, T. Otsuka, B. A. Brown, and T. Mizusaki, \Journal{\PRC} {65}{061301(R)}{2002}. doi:10.1103/PhysRevC.65.061301; \Journal{\PRC}{69}{034335}{2004}. doi:10.1103/PhysRevC.69.034335
\bibitem{KB3G} A. Poves, J. Sanche-Solano, E. Caurier, and F. Nowacki, \Journal{\NPA} {694}{157} {2001}. doi:10.1016/S0375-9474(01)00967-8
\bibitem{Caurier} G. Mart\'inez-Pinedo, A. Poves, E. Caurier, and A. P. Zuker, \Journal{PRC} {53}{R2602}{1996}. doi:10.1103/PhysRevC.53.R2602
\bibitem{PTPS} T. Suzuki, M. Honma, T. Otsuka, and T. Kajino, \Journal{\PTPS} {196}{382} {2012}. doi:10.1143/PTPS.196.382  
\bibitem{Suzuki2013} T. Suzuki and T. Kajino, \Journal{\JPG}{40}{083101}{2013}. doi:10.1088/0954-3899/40/8/083101
\bibitem{Sasano} M. Sasano et al., \Journal{\PRL} {107}{202501} {2012}. doi:10.1103/PhysRevLett.107.202501;
\Journal{\PRC} {86}{034324} {2012}. doi:10.1103/PhysRevC.86.034324
%\bibitem{Cole} A. L. Cole et al., \Journal{\PRC} {86}{015809} {2012}. doi:10.1103/PhysRevC.86.015809 
\bibitem{Mori2016} K. Mori, M. A. Famiano, T. Kajino, T. Suzuki, J. Hidaka, M. Honma, K. Iwamoto, K. Nomoto, and T. Otsuka. \Journal{\ApJ} {833}{179} {2016}. doi:10.3847/1538-4357/833/2/179 
\bibitem{Cole} A. L. Cole et al., \Journal{\PRC} {86}{015809} {2012}. doi:10.1103/PhysRevC.86.015809
\bibitem{LanMart} K. Langanke and G. Mart\'inez-Pinedo, \Journal{\RMP} {75}{819} {2003}. doi:10.1103/RevModPhys.75.819
\bibitem{Mori2018} K. Mori, M. A. Famiano, T. Kajino et al., \Journal{\ApJ} {863}{176} {2018}. doi:10.3847/1538-4357/aad233 
\bibitem{Mori2020} K. Mori, T. Suzuki, M. Honma, M. A. Famiano, T. Kajino, M. Kusakabe, and A. B. Balantekin, \Journal{\ApJ} {904}{29} {2020}. doi:10.3847/1538-4357/abbb32
\bibitem{WWW} R.K. Wallace, S. E. Woosley, and T. A. Weaver, \Journal{\ApJ} {258}{696} {1982}. doi:10.1086/160119
\bibitem{Schatz} H. Schatz, S. Gupta, P. M$\ddot{\mbox{o}}$ller et al., \Journal{\NAT} {505}{62} {2014}. doi:10.1038/nature12757 
\bibitem{WBB} E. K. Warburton, J. A. Becker, and B. A. Brown, \Journal{\PRC} {41}{1147} {1990}. doi:10.1103/PhysRevC.41.1147
\bibitem{Utsuno} Y. Utsuno, T. Otsuka, T. Mizusaki, and M. Honma, \Journal{PRC} {60}{054315} {1999}. doi:10.1103/PhysRevC.60.054315
\bibitem{Tsunoda} N. Tsunoda, T. Otsuka, N. Shimizu, M. Hjorth-Jensen, K. Takayanagi, and T. Suzuki, \Journal{\PRC} {95}{021304(R)} {2017}. doi:10.1103/PhysRevC.95.021304
\bibitem{FM} J. Fujita and H. Miyazawa, \Journal{\PRO} {17}{360} {1957}. doi:10.1143/PTP.17.360  
\bibitem{Takaya} K. Takayanagi, \Journal{\NPA} {852}{61}{2011}. doi:10.1016/j.nuclphysa.2011.01.003; \Journal{\NPA}{864}{91}{2011}. doi:10.1016/j.nuclphysa.2011.06.025  
\bibitem{EKK} N. Tsunoda, K. Takayanagi, M. Hjorth-Jensen, and T. Otsuka, \Journal{\PRC}{89}{024314}{2014}. doi:10.1103/PhysRevC.89.024313 
\bibitem{OMEG15} T. Suzuki, N. Tsunoda, Y. Tsunoda, N. Shimizu, and T. Otsuka, \Journal{\EPJWC} {165}{01048}{2017}. doi:10.1051/epjconf/201716501048  
\bibitem{Sull} C. Sullivan, E. O'Connor, R. G. T. Zegres, T. Grubb, and S. M. Austin, \Journal{\ApJ} {816}{44} {2016}. doi:10.3847/0004-637X/816/1/44 
\bibitem{Moller1997} P. M$\ddot{\mbox{o}}$ller, J. Nix, and K.-L. Kratz, \Journal{ADNDT} {66}{131} {1997}. doi:10.1006/adnd.1997.0746
\bibitem{Moller2003} P. M$\ddot{\mbox{o}}$ller, B. Pfeiffer, and K.-L. Kratz, \Journal{\PRC} {67}{055802} {2003}. doi:10.1103/PhysRevC.67.055802 
\bibitem{Nabi} J.-U. Nabi and H. V. Klapdor-Kleingrothaus, \Journal{ADNDT} {88}{237} {2004}. doi:10.1016/j.adt.2004.09.002 
%\bibitem{Paarb}
\bibitem{Dzh2010} A. A. Dzhioev, A. I. Vdovin, V. Y. Ponomarev, J. Wambach, K. Laganke, and G. Mart$\acute{\mbox{i}}$nez-Pinedo, \Journal{\PRC} {81}{015804} {2010}. doi:10.1103/PhysRevC.81.015804
\bibitem{Niu} Y. F. Niu, N. Paar, D. Vretenar, and J. Meng, \Journal{PRC} {83}{045807} {2011}. doi:10.1103/PhysRevC.83.045807
\bibitem{Dean} K. Langanke, E. Kolbe, and D. J. Dean, \Journal{\PRC} {63}{032801(R)} {2001}. doi:10.1103/PhysRevC.63.032801; 
D. J. Dean, K. Langanke, L. Chatterjee, P. N. Radha, and M. R. Strayer, \Journal{\PRC} {58}{536} {1998}. doi:10.1103/PhysRevC.58.536
\bibitem{Gong} Z. Gong, L. Zejda, W. Dappen, and J. M. Aparicio, \Journal{\CPC} {136}{294}{2001}. doi:10.1016/S0010-4655(01)00145-X  
\bibitem{Langanke} K. Langanke, G. Mart\'inez-Pinedo, J. M. Sampaio, D. J. Dean, W. R. Hix, O. E. B. Messer, A. Mezzacappa, M. Liebendorfer, H.-T, Janka, and M. Rampp, \Journal{\PRL} {90}{241102} {2003}. doi:10.1103/PhysRevLett.90.241102
\bibitem{Titus} R. Titus, C. Sullivan, R. G. T. Zegers, B. A. Brown, and B. Gao, \Journal{\JPG}{45}{014004}{2018}. doi:10.1088/1361-6471/aa98c1 
\bibitem{Dzh2019} A. A. Dzhioev, A. I. Vdovin, and Ch. Stoyanov, \Journal{\PRC} {100}{025801} {2019}. doi:10.1103/PhysRevC.100.025801
\bibitem{Dzh2020} A. A. Dzhioev, K. Langanke, G. Mart\'inez-Pinedo, A. I. Vdovin, and Ch. Stoyanov, \Journal{\PRC} {101}{025805} {2020}. doi:10.1103/PhysRevC.101.025805
\bibitem{Litvinova2021} E. Litvinova and C. Robin, \Journal{\PRC}{103}{024326}{2021}. doi:10.1103/PhysRevC.103.024326
\bibitem{Giraud} S. Giraud, R. G. T. Zegeres, B. A. Brown et al., arXiv:2112.01626 (2021).
\bibitem{YTsunoda} Y. Tsunoda, T. Otsuka, N. Shimizu, M. Honma, and Y. Utsuno, \Journal{\PRC} {89}{031301(R)} {2014}. doi:10.1103/PhysRevC.89.031301    
\bibitem{ni78} T. Suzuki, S. Chiba, T. Yoshida, A. B. Balantekin, T. Kajino, M. Honma, Y. Tsunoda, N. Tsunoda, and N. Shimizu, \Journal{\JPSC} {31}{011039}{2020}. doi:10.7566/JPSCP.31.011039
\bibitem{SG}  N. Van Giai and H. Sagawa, \Journal{\PLB} {106}{379} {1981}. doi:10.1016/0370-2693(81)90646-8
\bibitem{KDR} K. Kubodera, J. Delorme, and M. Rho, \Journal{\PRL} {40}{755} {1978}. doi:10.1103/PhysRevLett.40.755
%\bibitem{Brown} E. K. Warburton, I. S. Towner, and B. A. Brown, \Journal{\PRC} {49}{824} {1994}. doi:10.1103/PhysRevC.49.824
\bibitem{Warburton} E. K. Warburton, \Journal{\PRC} {44}{233}{1991}. doi:10.1103/PhysRevC.44.233
\bibitem{Brown} E. K. Warburton, I. S. Towner, and B. A. Brown, \Journal{\PRC} {49}{824} {1994}. doi:10.1103/PhysRevC.49.824
%\bibitem{Baumann} P. Baumann et al., \Journal{\PRC}{58}{1970} {1998}. https://doi.org/10.1103/PhysRevC.58.1970
\bibitem{Cosel} P. von Neumann-Cosel, A. Poves, J. Retamosa, and A. Richter, \Journal{\PLB}{443}{1}{1998}. doi:10.1016/S0370-2693(98)01298-2
%\bibitem{Dzh2019} A. A. Dzhioev, A. I. Vdovin, and Ch. Stoyanov, \Journal{\PRC} {100}{025801} {2019}. doi:10.1103/PhysRevC.100.025801            
%\bibitem{Dzh2020} A. A. Dzhioev, K. Langanke, G. Martinez-Pinedo, A. I. Vdovin, and Ch. Stoyanov, \Journal{\PRC} {101}{025805} {2020}. doi:10.1103/PhysRevC.101.025805
%\bibitem{Litvinova2021} E. Litvinova and C. Robin, \Journal{\PRC}{103}{024326}{2021}. doi:10.1103/PhysRevC.103.024326
%\bibitem{Giraud} S. Giraud, R. G. T. Zegeres, B. A. Brown et al., arXiv:2112.01626 (2021).  
\bibitem{Burbridge1957} E. M. Burbridge, G. R. Burbridge, W. A. Fowler, and F. Hoyle, \Journal{\RMP} {29}{547} {1957}. doi:10.1103/RevModPhys.29.547
\bibitem{Cowan} J. J. Cowan, F.-K. Thielemann, and J. W. Truran \Journal{\PREP} {208}{267} {1991}. doi:10.1016/0370-1573(91)90070-3
\bibitem{KBT} K.-L. Kratz, J. Bitouzet, F.-K. Thielemann, P. M$\ddot{\mbox{o}}$ller, and B. Pfeiffer, \Journal{\ApJ} {403}{216} {1993}. doi:10.1086/172196
\bibitem{Nishimura2006} S. Nishimura, K. Kotake, M. Hashimoto et al., \Journal{\ApJ} {642}{410} {2006}. doi:10.1086/500786
\bibitem{Fujimoto2007}  S. Fujimoto, M. Hashimoto, K. Kotake, and S. Yamada, \Journal{\ApJ}{656}{382} {2007}. doi:10.1086/509908
\bibitem{Fujimoto2008} S. Fujimoto, N. Nishimura, and M. Hashimoto, \Journal{\ApJ} {680}{1350} {2008}. doi:10.1086/529416
\bibitem{Ono2012} M. Ono, M. Hashimoto, S. Fujimoto, K. Kotake, and S. Yamada, \Journal{\PRO} {128}{741} {2012}. doi:10.1143/PTP.128.741
\bibitem{Winteler2012} C. Winteler, R. Kappeli, A. Perego et al., \Journal{\ApJL} {750}{L22} {2012}. doi:10.1088/2041-8205/750/1/L22
\bibitem{Nakamura2014} K. Nakamura, T. Kajino, G. J. Mathews, S. Susumu, and S. Harikae, \Journal{\INT} {22}{1330022} {2013}. doi:10.1142/S0218301313300221
\bibitem{Nishimura2015} N. Nishimura, T. Takiwaki, and F.-K. Thielemann, \Journal{\ApJ} {810}{109} {2015}. doi:10.1088/0004-637X/810/2/109
\bibitem{Wanajo2014} S. Wanajo, Y. Sekiguchi, N. Nishimura, K. Kiuchi, K. Kyuotoku, and M. Shibata, \Journal{\ApJL} {789}{L39} {2014}. doi:10.1088/2041-8205/789/2/L39
\bibitem{Goriely2015}  S. Goriely, A. Bauswein, O. Just, E. Pllumbi, and H.-T. Janka, \Journal{\MNRAS} {452}{3894} {2015}. doi:10.1093/mnras/stv1526
%\bibitem{Nishimura2016} N. Nishimura, Z. Podolyak, D.-L. Fang, and T. Suzuki, \Journal{\PLB}{756}{273}{2016}. doi:10.1016/j.physletb.2016.03.025
%\bibitem{Wanajo2014} S. Wanajo, Y. Sekiguchi, N. Nishimura, K. Kiuchi, K. Kyuotoku, and M. Shibata, \Journal{\ApJL} {789}{L39} {2014}. doi:10.1088/2041-8205/789/2/L39
%\bibitem{Goriely2015}  S. Goriely, A. Bauswein, O. Just, E. Pllumbi, and H.-T. Janka, \Journal{\MNRAS} {452}{3894} {2015}. doi:10.1093/mnras/stv1526
%\bibitem{Woosley1994} S. E. Woosley, J. R. Wilson, G. J. Mathews, R. D. Hoffman, and B. S. Meyer \Journal{\ApJ} {433}{229} {1994}. doi:10.1086/174638
%\bibitem{Wanajo2013} S. Wanajo, \Journal{\ApJL} {770}{L22} {2013}. doi:10.1088/2041-8205/770/2/L22
\bibitem{Abbott2017a} B. P. Abbott et al., \Journal{\PRL} {119}{161101} {2017}. doi:10.1103/PhysRevLett.119.161101
\bibitem{Abbott2017} B. P. Abbott, R. Abbott, T. D. Abbott et al., \Journal{\ApJL} {848}{L12} {2017}. doi:10.3847/2041-8213/aa91c9
\bibitem{Smartt2017} S. Smartt, T. W. Chen, A. Jerkstrand et al., \Journal{\NAT} {551}{75} {2017}. doi:10.1038/nature24303
\bibitem{Woosley1994} S. E. Woosley, J. R. Wilson, G. J. Mathews, R. D. Hoffman, and B. S. Meyer \Journal{\ApJ} {433}{229} {1994}. doi:10.1086/174638
\bibitem{Wanajo2013} S. Wanajo, \Journal{\ApJL} {770}{L22} {2013}. doi:10.1088/2041-8205/770/2/L22
\bibitem{Hudepohl2010} L. H$\ddot{\mbox{u}}$depohl, B. M$\ddot{\mbox{u}}$ller, H.-T. Janka, A. Marek, and G. G. Raffelt, \Journal{\PRL}{104}{251101}{2010}. doi:10.1103/PhysRevLett.104.251101
\bibitem{Fischer2020} T. Fischer, G. Guo, A. A. Dzhioev, G. Mart\'inez-Pinedo, M.-R. Wu, A. Lohs, and Y.-Z. Qian, \Journal{\PRC}{101}{025804}{2020}. doi:10.1103/PhysRevC.101.025804
\bibitem{Bollig2021} R. Bollig, N. Yadav, D. Kresse, H.-T. Janka, B. M$\ddot{\mbox{u}}$ller, and A. Heger, \Journal{\ApJ}{915}{28}{2021}. doi:10.3847/1538-4357/abf82e
%\bibitem{KBT} K.-L. Kratz, J. Bitouzet, F.-K. Thielemann, P. Moller, and B. Pfeiffer, \Journal{\ApJ} {403}{216} {1993}. doi:10.1086/172196
\bibitem{Wanajo} S. Wanajo, S. Goriely, M. Samyn, and N. Itoh, \Journal{\ApJ} {606}{1057} {2004}. doi:10.1086/383140
\bibitem{Terasawa} M. Terasawa, K. Sumiyoshi, T. Kajino, G. J. Mathews, and I Tanihata, \Journal{ApJ} {562}{470} {2001}. doi:10.1086/323526
\bibitem{Sasaqui} T. Sasaqui, T. Kajino, G. Mathews, K. Otsuki, and T. Nakamura, \Journal{\ApJ} {634}{1173} {2005}. doi:10.1086/497061
%\bibitem{LanMart} K. Langanke and G. Martinez-Pinedo, \Journal{\RMP} {75}{819} {2003}.  
\bibitem{Mumpower} M. R. Mumpower, R. Surman, G. C. McLaughlin, and A. Aprahamian, \Journal{\PPNP} {86}{86} {2016}. doi:10.1016/j.ppnp.2015.09.001; 
M. R. Mumpower, R. Surman, and A. Aprahamian, \Journal{\JPC} {599}{012031} {2015}. doi:10.1088/1742-6596/599/1/012031
\bibitem{Meyer} B. S. Meyer, G. C. McLaughlin, and G. M. Fuller, \Journal{\PRC} {58}{3696} {1998}. doi:10.1103/PhysRevC.58.3696
\bibitem{Terasawa2004} M. Terasawa, K. Langanke, T. Kajino, and G.Mathews, \Journal{\ApJ} {608}{470} {2004}. doi:10.1086/386359
\bibitem{Grawe2007} H. Grawe, K. Langanke, and G. Mart\'inez-Pinedo, \Journal{\RPPh} {70}{1525} {2007}. doi:10.1088/0034-4885/70/9/R02
\bibitem{Martinez-Pinedo1999} G. Mart\'inez-Pinedo and K. Langanke, \Journal{\PRL} {83}{4502} {1999}. doi:10.1103/PhysRevLett.83.4502 
\bibitem{Suzuki2012} T. Suzuki, T. Yoshida, T. Kajino, and T. Otsuka, \Journal{\PRC} {85}{015802} {2012}. doi:10.1103/PhysRevC.85.015802 
\bibitem{Zhi2013} Q. Zhi, E. Caurier, J. J. Cuenca-Garcia, K. Langanke, G. Mart\'inez-Pinedo, and K. Sieja, \Journal{\PRC} {87}{025803} {2013}. doi:10.1103/PhysRevC.87.025803
\bibitem{Borozov2000} I. N. Borozov and S. Goriely, \Journal{\PRC} {62}{035501} {2000}. doi:10.1103/PhysRevC.62.035501
\bibitem{Engel} J. Engel, M. Bender, J. Dobaczewski, W. Nazarewicz, and R. Surman, \Journal{\PRC} {60}{014302} {1999}. doi:10.1103/PhysRevC.60.014302
\bibitem{Fang} D.-L. Fang, B. A. Brown, and T. Suzuki, \Journal{\PRC} {88}{034304} {2013}. doi:10.1103/PhysRevC.88.034304 
\bibitem{Borozov2003} I. N. Borozov, \Journal{\PRC} {67}{025802} {2003}. doi:10.1103/PhysRevC.67.025802      
%\bibitem{LangMart} K. Langanke and G. Martinez-Pinedo, \Journal{\RMP}{75}{819}{2003}   
\bibitem{Steer} S. J. Steer, Zs. Podolyak, S. Pietri et al., \Journal{\PRC} {78}{061302} {2008}. doi:10.1103/PhysRevC.78.061302
\bibitem{Rydstrom} L. Rydstrom, J. Blomqvist, R. J. Liotta, and C. Pomar, \Journal{\NPA} {512}{217} {1980}. doi:10.1016/0375-9474(90)93152-V
%\bibitem{Morales} A. I. Morales, J. Benlliure, T. Kurtukian-Nieto et al., \Journal{\PRL} {113}{022702} {2014}. doi:10.1103/PhysRevLett.113.022702
\bibitem{Suzuki2018} T. Suzuki, S. Shibagaki, T. Yoshida, T. Kajino, and T. Otsuka, \Journal{\ApJ} {859}{133} {2018}. doi:10.3847/1538-4357/aabfde 
\bibitem{Morales} A. I. Morales, J. Benlliure, T. Kurtukian-Nieto et al., \Journal{\PRL} {113}{022702} {2014}. doi:10.1103/PhysRevLett.113.022702
\bibitem{Koura2005} H. Koura, T. Tachibana, M. Uno, and M. Yamada, \Journal{\PRO} {113}{305} {2005}. doi:10.1143/PTP.113.305 
\bibitem{Chiba2008} S. Chiba, H. Koura, T. Maruyama et al., in {\it AIP Conf. Proc} {\it 1016}, {\it Origin of Matter and Evolution of Galaxies}, ed. T. Suda et al. (Melville, NY: AIP), 162. doi:10.1063/1.2943568   
%\bibitem{Steer} S. J. Steer, Zs. Podolyak, S. Pietri et al., \Journal{\PRC} {78}{061302} {2008}. doi:10.1103/PhysRevC.78.061302 
%\bibitem{Rydstrom} L. Rydstrom, J. Blomqvist, R. J. Liotta, and C. Pomar, \Journal{\NPA} {512}{217} {1980}. doi:10.1016/0375-9474(90)93152-V
\bibitem{Ney2020} E. M. Ney, J. Engel, T. Li, and N. Schunck, \Journal{\PRC}{102}{034326}{2020}. doi:10.1103/PhysRevC.102.034326
\bibitem{Nishimura2011} S. Nishimura, Z. Li, H. Watanabe et al. \Journal{\PRL} {106}{052502} {2011}. doi:10.1103/PhysRevLett.106.052502
\bibitem{Marketin2016} T. Marketin, G. Huther, and G. Mart\'inez-Pinedo, \Journal{\PRC} {93}{025805} {2016}. doi:10.1103/PhysRevC.93.025805
\bibitem{Nishimura2016} N. Nishimura, Z. Podolyak, D.-L. Fang, and T. Suzuki, \Journal{\PLB}{756}{273}{2016}. doi:10.1016/j.physletb.2016.03.025
\bibitem{Watanabe2015} Y. X. Watanabe, Y. H. Kim, S. C. Jeong et al., \Journal{\PRL} {115}{172503} {2015}. doi:10.1103/PhysRevLett.115.172503
\bibitem{Hirayama2017} Y. Hirayama, M. Mukai, Y. X. Watanabe et al., \Journal{\PRC} {96}{014307} {2017}. doi:10.1103/PhysRevC.96.014307 
\bibitem{Hirayama2018} Y. Hirayama, Y. X. Watanabe, M. Mukai et al., \Journal{\PRC} {98}{014321} {2018}. doi:10.1103/PhysRevC.98.014321
\bibitem{Choi} H. Choi, Y. Hirayama, S. Choi et al., \Journal{\PRC} {102}{034309} {2020}. doi:10.1103/PhysRevC.102.034309
\bibitem{Saldivia2017} A. Tarifeno-Saldivia et al., (The BRIKEN collaboration) \Journal{\JINST} {12} {P04006} {2017}. doi:10.1088/1748-0221/12/04/P04006    
\bibitem{Caballero2016} R. Caballero-Folch, C. Domingo-Pardo, J. Agramunt et al., \Journal{\PRL} {117}{012501} {2016}. doi:10.1103/PhysRevLett.117.012501
\bibitem{Pereira2009} J. Pereira et al., \Journal{\PRC} {79}{035806} {2009}. doi:10.1103/PhysRevC.79.035806
\bibitem{Pereira2010} J. Pereira et al., \Journal{\NIMA} {618}{275}{2010}. doi:10.1016/j.nima.2010.02.262
\bibitem{Gomez2014} M. B. Gomez-Hornillos et al., \Journal{\HYP} {223}{185} {2014}. doi:10.1007/s10751-012-0617-4         
\bibitem{Yokoyama2019} R. Yokoyama, R. Grzywacz, B. C. Rasco et al., \Journal{\PRC} {100}{031302(R)} {2019}. doi:10.1103/PhysRevC.100.031302  
%\bibitem{MeyerBrown1997} B. S. Meyer and J. S. Brown, \Journal{\ApJS} {112}{199} {1997}. doi:10.1086/313032
%\bibitem{Otsuki2003} K. Otsuki, G. J. Mathews, and T. Kajino, \Journal{\NEWA} {8}{767} {2003}. doi:10.1016/S1384-1076(03)00065-4 
%\bibitem{Shibagaki2016} S. Shibagaki, T. Kajino, G. J. Mathews, S. Chiba, S. Nishimura, and G. Lorusso, \Journal{\ApJ} {816}{79} {2016}. doi:10.3847/0004-637X/816/2/79
%\bibitem{Wasserburg1996} G. J. Wasserburg, M. Busso, and R. Galliano, \Journal{\ApJL} {466}{L109} {1996}. doi:10.1086/310177
%%\bibitem{Woosley1994} S. E. Woosley, J. R. Wilson, G. J. Mathews, R. D. Hoffman, and B. S. Meyer \Journal{\ApJ} {433}{229} {1994}. doi:10.1086/174638  
%%\bibitem{Wanajo2013} S. Wanajo, \Journal{\ApJL} {770}{L22} {2013}. doi:10.1088/2041-8205/770/2/L22
%\bibitem{Nishimura2006} S. Nishimura, K. Kotake, M. Hashimoto et al., \Journal{\ApJ} {642}{410} {2006}. doi:10.1086/500786  
%\bibitem{Fujimoto2007}  S. Fujimoto, M. Hashimoto, K. Kotake, and S. Yamada, \Journal{\ApJ}{656}{382} {2007}. doi:10.1086/509908
%\bibitem{Fujimoto2008} S. Fujimoto, N. Nishimura, and M. Hashimoto, \Journal{\ApJ} {680}{1350} {2008}. doi:10.1086/529416
%\bibitem{Ono2012} M. Ono, M. Hashimoto, S. Fujimoto, K. Kotake, and S. Yamada, \Journal{\PRO} {128}{741} {2012}. doi:10.1143/PTP.128.741
%\bibitem{Winteler2012} C. Winteler, R. Kappeli, A. Perego et al., \Journal{\ApJL} {750}{L22} {2012}. doi:10.1088/2041-8205/750/1/L22
%\bibitem{Nakamura2014} K. Nakamura, T. Kajino, G. J. Mathews, S. Susumu, and S. Harikae, \Journal{\INT} {22}{1330022} {2013}. doi:10.1142/S0218301313300221 
%\bibitem{Nishimura2015} N. Nishimura, T. Takiwaki, and F.-K. Thielemann, \Journal{\ApJ} {810}{109} {2015}. doi:10.1088/0004-637X/810/2/109
%\bibitem{Wanajo2014} S. Wanajo, Y. Sekiguchi, N. Nishimura, K. Kiuchi, K. Kyuotoku, and M. Shibata, \Journal{\ApJL} {789}{L39} {2014}. doi:10.1088/2041-8205/789/2/L39
%\bibitem{Goriely2015}  S. Goriely, A. Bauswein, O. Just, E. Pllumbi, and H.-T. Janka, \Journal{\MNRAS} {452}{3894} {2015}. doi:10.1093/mnras/stv1526
%\bibitem{Nishimura2016} N. Nishimura, Z. Podolyak, D.-L. Fang, and T. Suzuki, \Journal{\PLB}{756}{273}{2016}. doi:10.1016/j.physletb.2016.03.025 
%\bibitem{Wanajo2011} S. Wanajo, H.-T. Janka, and B. Muller, \Journal{\ApJL} {726}{L15} {2011}. https://doi.org/10.1088/2041-8205/726/2/L15 
%\bibitem{Abbott2017a} B. P. Abbott et al., \Journal{\PRL} {119}{161101} {2017}. doi:10.1103/PhysRevLett.119.161101 
%\bibitem{Abbott2017} B. P. Abbott, R. Abbott, T. D. Abbott et al., \Journal{\ApJL} {848}{L12} {2017}. doi:10.3847/2041-8213/aa91c9  
%\bibitem{Smartt2017} S. Smartt, T. W. Chen, A. Jerkstrand et al., \Journal{\NAT} {551}{75} {2017}. doi:10.1038/nature24303
%\bibitem{Kasen2017} D. Kasen, B. Metzger, J. Barnes, E Quataert, and E. Ramirez-Ruiz, \Journal{\NAT} {551}{80} {2017}. doi:10.1038/nature24453
%\bibitem{Tanaka2018} M. Tanaka, D. Kato, G. Gaigalas et al., \Journal{\ApJ} {852}{109} {2018}. doi:10.3847/1538-4357/aaa0cb  
%\bibitem{Wu2018} M.-R. Wu, J. Barnes, G. Martinez-Pinedo, and B. D. Metzger, \Journal{\PRL} {122}{062701} {2019}. doi:10.1103/PhysRevLett.122.062701  
%\bibitem{Watson2019} D. Watson, C. J. Hansen, J. Selsing et al., \Journal{\NAT} {574}{497} {2019}. doi:10.1038/s41586-019-1676-3 
%\bibitem{Goriely1999} S. Goriely, \Journal{\AA} {342}{881} {1999}. https://ui.adsabs.harvard.edu/abs/1999A\&A...342..881G/abstract
%\bibitem{Sneden} C. Sneden, J. J. Cowan, and R. Gallino, \Journal{\ARAA}{46}{241}{2008}. doi:10.1146/annurev.astro.46.060407.145207 
%\bibitem{Mathews1992} G. J. Mathews, G. Bazan, and J. J. Cowan, \Journal{\ApJ}{391}{719}{1992}. doi:10.1086/171383 
%\bibitem{Argast2000} D. Argast, M. Samland, O. E. Gerhard, and F.-K. Thielemann, \Journal{\AA} {356}{873}{2000}. https://ui.adsabs.harvard.edu/abs/2000A\&A...356..873A/abstract
%\bibitem{Reichert2019} M. Reichert, M. Obergaulinger, M. Eichler, A. Aloy, and A. Arcones, \Journal{\MNRAS}{501}{5733} {2021}. doi:10.1093/mnras/stab029  
%\bibitem{Cote2019} B. Cote, M. Eichler, A. Arcones et al., \Journal{\ApJ} {875}{106}{2019}. doi:10.3847/1538-4357/ab10db 
%\bibitem{Kobayashi2020} C. Kobayashi, A. Karakas, and M. Lugaro, \Journal{\ApJ} {900}{179} {2020}. doi:10.3847/1538-4357/abae65
%%arXiv:2008.046601 
%\bibitem{Siegel2019} D. M. Siegel, J. Barnes, and B. D. Metzger, \Journal{\NAT}{569}{241}{2019}. doi:10.1038/s41586-019-1136-0
%\bibitem{Collap} Y. Yamazaki, T. Kajino, G. J. Mathews, X. Tamg, J. Shi, and M. A. Famiano, arXiv:2102.05891
%\bibitem{Wanajo2021} S. Wanajo, Y. Hirai, and N. Prantzos, 
%\Journal{\MNRAS}{}{}{2021}, 
%arXiv:2106.03707 
\bibitem{LSND} C. Athanassopoulos et al., \Journal{\PRC}{55}{2078}{1997}. doi:10.1103/PhysRevC.55.2078 
\bibitem{KARMEN1}  B. E. Bodmann et al., \Journal{\PLB}{332}{251}{1994}. doi:10.1016/0370-2693(94)91250-5
\bibitem{KARMEN2}  R. Maschuw. \Journal{\PPNP}{40}{183}{1998}. doi:10.1016/S0146-6410(98)00024-6
\bibitem{LAMPF} D. A. Krakauer et al., \Journal{\PRC}{45}{2450}{1992}. doi:10.1103/PhysRevC.45.2450  
\bibitem{Suzuki2006} T. Suzuki, S. Chiba, T. Yoshida, T. Kajino, and T. Otsuka, \Journal{\PRC}{74}{034307}{2006}. doi:10.1103/PhysRevC.74.034307
\bibitem{Suzuki2003} T. Suzuki, R. Fujimoto, and T. Otsuka, \Journal{\PRC}{67}{044302}{2003}. doi:10.1103/PhysRevC.67.044302 
%\bibitem{Suzuki2013} T. Suzuki and T. Kajino, \Journal{\JPG}{40}{083101}{2013}. doi:10.1088/0954-3899/40/8/083101
\bibitem{Volpe} C. Volpe, A. Auerbach, G. Colo, T. Suzuki, and N. Van Giai, \Journal{\PRC}{62}{015501}{2000}. doi:10.1103/PhysRevC.62.015501   
\bibitem{Kolbe1999} E. Kolbe, K. Langanke, and G. Mart\'inez-Pinedo, \Journal{\PRC}{60}{052801}{1999}. doi:10.1103/PhysRevC.60.052801 
\bibitem{PaarSuz} N. Paar, T. Suzuki, M. Honma, T. Marketin, and D. Vretenar, \Journal{\PRC}{84}{047305}{2011}. doi:10.1103/PhysRevC.84.047305
\bibitem{Paar-nu} N. Paar, H. Tutman, T. Marketin, and T. Fischer, \Journal{\PRC}{87}{025801}{2013}. doi:10.1103/PhysRevC.87.025801
\bibitem{Woos1990} S. E. Woosley, D. H. Hartmann, R. D. Hoffman, and W. C. Haxton, \Journal{\ApJ}{356}{272}{1990}. doi:10.1086/168839 
\bibitem{Heger2005} A. Heger, E. Kolbe, W. C. Haxton, K. Langanke, G. Mart\'inez-Pinedo, and S. E. Woosley, \Journal{\PLB}{606}{258}{2005}. doi:10.1016/j.physletb.2004.12.017
\bibitem{HF1952} W. Hauser and H. Feshbach, \Journal{\PREV}{87}{366}{1952}. doi:10.1103/PhysRev.87.366
\bibitem{YSC2008} T. Yoshida, T. Suzuki, S. Chiba, T. Kajino, H. Yokomakura, K. Kimura, A. Takamura, and D. H. Hartmann, \Journal{\ApJ}{686}{448}{2008}. doi:10.1086/591266
\bibitem{YK2005} T. Yoshida, T. Kajino, and D. H. Hartmann, \Journal{\PRL}{94}{231101}{2005}. doi:10.1103/PhysRevLett.94.231101 
\bibitem{YK2006} T. Yoshida, T. Kajino, H. Yokomakura, K. Kimura, A. Takamura, and D. H. Hartmann, \Journal{\PRL}{96}{091101}{2006}. doi:10.1103/PhysRevLett.96.091101; 
ibid. \Journal{\ApJ}{649}{319}{2006}. doi:10.1086/506374
\bibitem{Nakazato2013} K. Nakazato, K. Sumiyoshi, and H. Suzuki, \Journal{\ApJS}{205}{2}{2013}. doi:10.1088/0067-0049/205/1/2  
\bibitem{Siverd2018} A. Siverding, G. Mart\'inez-Pinedo, L. Huther, K. Langanke, and A. Heger, \Journal{\ApJ}{865}{143}{2018}. doi:10.3847/1538-4357/aadd48
\bibitem{Siverd2019}  A. Siverding, K. Langanke, G. Mart\'inez-Pinedo, R. Bollig, H.-T. Janka, and A. Heger, \Journal{\ApJ}{876}{151}{2019}. doi:10.3847/1538-4357/ab17e2
\bibitem{SCY2018} T. Suzuki, S. Chiba, T. Yoshida, K. Takahashi, and H. Umeda, \Journal{\PRC}{98}{034613}{2018}. doi:10.1103/PhysRevC.98.034613
\bibitem{Byelikov} A. Byelikov, T. Adachi, H. Fujita et al., \Journal{\PRL}{98}{082501}{2007}. doi:10.1103/PhysRevLett.98.082501
\bibitem{Hayakawa10a} T. Hayakawa, T. Kajino, S. Chiba, and J. Mathews, \Journal{\PRC}{81}{052801(R)}{2010}. doi:10.1103/PhysRevC.81.052801
\bibitem{Hayakawa10b} T. Hayakawa, P. Mohr, T. Kajino, S. Chiba, and J. Mathews, \Journal{\PRC}{82}{058801}{2010}. doi:10.1103/PhysRevC.82.058801 
\bibitem{Mohr} P. Mohr, F. Kappeler, and R. Gallino, \Journal{\PRC}{75}{012802(R)}{2007}. doi:10.1103/PhysRevC.75.012802 
\bibitem{YUN2008} T. Yoshida, H. Umeda, and K. Nomoto, \Journal{\ApJ}{672}{1043}{2008}. doi:10.1086/523833 
\bibitem{SHY2009} T. Suzuki, M. Honma, K. Higashiyama, T. Yoshida, T. Kajino, T. Otsuka, H. Umeda, and K. Nomoto, \Journal{\PRC}{79}{061603(R)}{2009}. doi:10.1103/PhysRevC.79.061603
\bibitem{Cayrel2004} R. Cayrel et al., \Journal{\AA} {416}{1117}{2004}. doi:10.1051/0004-6361:20034074     
\bibitem{Harpper} C. L. Harper, Jr., \Journal{\ApJ}{466}{437}{1996}. doi:10.1086/177523
\bibitem{Schonbachler} M. Schonbachler, M. Rehkamper, A. N. Halliday et al., \Journal{\SCI}{295}{1705}{2002}. doi:10.1126/science.1067400
\bibitem{Iizuka} T. Iizuka, Y. J. Laib, W. Akram, Y. Amelin, and M. Schonbachler, \Journal{\EPSL}{439}{172}{2016}. doi:10.1016/j.epsl.2016.02.005  
\bibitem{Hayakawa2013} T. Hayakawa, K. Nakamura, T. Kajino, S. Chiba, N. Iwamoto, M. K. Cheoun, and G. J. Mathews, \Journal{\ApJL}{779}{L9}{2013}. doi:10.1088/2041-8205/779/1/L9 
\bibitem{Hayakawa2018}  T. Hayakawa, H. Ko, M. K. Cheoun et al., \Journal{\PRL}{121}{102701}{2018}. doi:10.1103/PhysRevLett.121.102701
\bibitem{Becker} H. Becker and R. J. Walker, \Journal{\CG}{196}{43}{2003}. doi:10.1016/S0009-2541(02)00406-0
\bibitem{Buras2003} R. Buras, M. Rampp, H.-Th. Janka, K. Kifonidis, \Journal{\PRL}{90}{241101}{2003}. doi:10.1103/PhysRevLett.90.241101   
\bibitem{Pruet2005} J. Pruet, S. E. Woosley, R. Buras, H.-T. Janka, and R. D. Hoffman, \Journal{\ApJ}{623}{325}{2005}. doi:10.1086/428281  
\bibitem{Frohlich2005} C. Frohlich et al., \Journal{\ApJ}{637}{415}{2005}. doi:10.1086/498224
\bibitem{Frohlich2006} C. Frohlich, G. Mat\'inez-Pinedo et al., \Journal{\PRL}{96}{142502}{2006}. doi:10.1103/PhysRevLett.96.142502
\bibitem{Pruet2006} J. Pruet, R. D. Hoffmann, S. E. Woosley, H.-T. Janka, and R. Buras, \Journal{\ApJ}{644}{1028}{2006}. doi:10.1086/503891 
\bibitem{Wanajo2006} S. Wanajo, \Journal{\ApJ}{647}{1323}{2006}. doi:10.1086/505483    
\bibitem{MSW1} L. Wolfenstein, \Journal{\PRD}{17}{2369}{1978}. doi:10.1103/PhysRevD.17.2369; \Journal{\PRD}{20}{2634}{1979}. doi:10.1103/PhysRevD.20.2634 
\bibitem{MSW2} S. P. Mikheyev and A. Y. Smirnov, \Journal{\SJNP}{42}{913}{1985}; \Journal{JETP}{64}{4}{1986}  
\bibitem{Dighe2000}  A. S. Dighe and A. Y. Smirnov, \Journal{\PRD}{62}{033007}{2000}. doi:10.1103/PhysRevD.62.033007
%\bibitem{YSC2008} T. Yoshida, T. Suzuki, S. Chiba, T. Kajino, H. Yokomakura, K. Kimura, A. Takamura, and D. H. Hartmann, \Journal{\ApJ}{686}{448}{2008}. doi:10.1086/591266
\bibitem{T2K} K. Abe et al. (T2K Collaboration), \Journal{\PRL}{107}{041801}{2011}. doi:10.1103/PhysRevLett.107.041801
\bibitem{MINOS} P. Adamson et al. (MINOS Collaboration), \Journal{\PRL}{107}{181802}{2011}. doi:10.1103/PhysRevLett.107.181802 
\bibitem{DAYA} F. An et al. (Daya Bay Collaboration), \Journal{\PRL}{108}{171803}{2012}. doi:10.1103/PhysRevLett.108.171803
\bibitem{DCHOOZ} Y. Abe et al. (Double-Chooz Collaboration), \Journal{\PRL}{108}{131801}{2012}. doi:10.1103/PhysRevLett.108.131801 
\bibitem{RENO} J. K. Ahn et al. (RENO Collaboration), \Journal{\PRL}{108}{191802}{2012}. doi:10.1103/PhysRevLett.108.191802
\bibitem{Fujiya2011} W. Fujiya, P. Hoppe, and U. Ott, \Journal{\ApJL}{730}{L3}{2011}. doi:10.1088/2041-8205/730/1/L7  
\bibitem{Mathews2012} G. J. Mathews, T. Kajino, W. Aoki, W. Fujiya, and J. B. Pitts, \Journal{\PRD}{85}{105023}{2012}. doi:10.1103/PhysRevD.85.105023 
\bibitem{T2Kh} K. Abe et al. (T2K Collaboration), \Journal{\PRL}{121}{171802}{2018}. doi:10.1103/PhysRevLett.121.171802;\\ 
\Journal{\PRD}{97}{072001}{2018}. doi:10.1103/PhysRevD.97.072001 
\bibitem{NOVA} M. A. Acero et al. (NOvA Collaboration), \Journal{\PRL}{123}{151803}{2019}. doi:10.1103/PhysRevLett.123.151803    
\bibitem{Pastor} S. Pastor and G. Raffelt, \Journal{\PRL}{89}{191101}{2002}. doi:10.1103/PhysRevLett.89.191101
\bibitem{FQ}
G. M. Fuller and Y. Qian, \Journal{\PRD}{73}{023004}{2006}. doi:10.1103/PhysRevD.73.023004
\bibitem{Fogli}
G. L. Fogli, E. Lisi, A. Marrone, and A. Mirizzi, \Journal{\JCAP}{12}{010}{2007}. doi:10.1088/1475-7516/2007/12/010
\bibitem{DFCQ} H. Duan, G. M. Fuller, J. Carlson, and Y. Qian, \Journal{\PRD}{77}{085016}{2008} and references therein. doi:10.1103/PhysRevD.77.085016
\bibitem{Pretel}
A. Esteban-Pretel, S. Pastor, R. Tomas, and G. G. Raffelt, \Journal{\PRD}{77}{065024}{2008}. doi:10.1103/PhysRevD.77.065024 
\bibitem{BY}  
A. B. Balantekin and H. Yuksel, \Journal{\NJP}{7}{51}{2005}. doi:10.1088/1367-2630/7/1/051
\bibitem{PBKY}
Y. Pehlivan, A. B. Balantekin, T. Kajino, and T. Yoshida, \Journal{\PRD}{84}{065008}{2012}. doi:10.1103/PhysRevD.84.065008
\bibitem{Malkus} A. Malkus, J. P. Kneller, G. C. McLaughlin, and R. Surman, \Journal{\PRD}{86}{085015}{2012}. doi:10.1103/PhysRevD.86.085015
\bibitem{Duan2006} H. Duan, G. M. Fuller, and Y. Z. Qian, \Journal{\PRD}{74}{123004}{2006}. doi:10.1103/PhysRevD.74.123004
\bibitem{DFQ} H. Duan, G. M. Fuller, and Y. Z. Qian, \Journal{\ARNPS}{60}{569}{2010}.  doi:10.1146/annurev.nucl.012809.104524 
\bibitem{Wu2015} M.-R. Wu, Y.-Z. Qian, G. Mart$\acute{\mbox{i}}$nez-Pinedo, T. Fischer, and L. Huther, \Journal{\PRD}{91}{065016}{2015}. doi:10.1103/PhysRevD.91.065016
\bibitem{Ko2020} H. Ko, M.-K. Cheoun, E. Ha et al., \Journal{\ApJL}{891}{L24}{2020}. doi:10.3847/2041-8213/ab775b
%\bibitem{Pinedo-EPJA} G. Martinez-Pinedo, B. Ziebarth, T. Fischer, and K. Langanke, \Journal{\EPJ}{47}{98}{2011}. doi:10.1140/epja/i2011-11098-y
%\bibitem{Dasgupta} B. Dasgupta, A. Dighe, G. G. Raffelt, and A. Y. Smirnov, \Journal{\PRL}{103}{051105}{2009}. doi:10.1103/PhysRevLett.103.051105    
\bibitem{Sasaki} H. Sasaki, T. Kajino, T. Takiwaki, T. Hayakawa, A. B. Balantekin, and Y. Pehlivan, \Journal{\PRD}{96}{043013}{2017}. doi:10.1103/PhysRevD.96.043013 
\bibitem{Raffelt2011} G. G. Raffelt, S. Sarikas, and D. de Sousa Seixas, \Journal{\PRL}{111}{091101}{2011}. doi:10.1103/PhysRevLett.111.091101
\bibitem{Chakraborty2014} S. Chakraborty and A. Mirizzi, \Journal{\PRD}{90}{033004}{2014}. doi:10.1103/PhysRevD.90.033004
\bibitem{Zaizen} M. Zaizen, T. Yoshida, K. Sumiyoshi, and H. Umeda, \Journal{\PRD}{98}{103020}{2018}. doi:10.1103/PhysRevD.98.103020
\bibitem{Wu2014} M.-R. Wu, T. Fischer, L. Huther, G. Mart\'inez-Pinedo, and Y.-Z. Qian, \Journal{\PRD}{89}{061303(R)}{2014}. doi:10.1103/PhysRevD.89.061303
\bibitem{Pllumbi2015} E. Pllumbi, I. Tamborra, S. Wanajo, H.-T. Janka and L. H$\ddot{\mbox{u}}$depohl, \Journal{ApJ}{808}{188}{2015}. doi:10.1088/0004-637X/808/2/188
\bibitem{Xiong2019} Z. Xiong, M.-R. Wu, and Y.-Z. Qian, \Journal{\ApJ}{880}{81}{2019}. doi:10.3847/1538-4357/ab2870 
\bibitem{Sawyer2009} R. F. Sawyer, \Journal{\PRD}{79}{105003}{2009}. doi:10.1103/PhysRevD.79.105003
\bibitem{Dasgupta2017} B. Dasgupta, A. Mirizzi and M. Sen, \Journal{\JCAP}{1702}{019}{2017}. doi:10.1088/1475-7516/2017/02/019
\bibitem{Xiong2020} Z. Xiong, A. Siverding, M. Sen, and Y.-Z. Qian, \Journal{\ApJ}{900}{144}{2020}. doi:10.3847/1538-4357/abac5e
\bibitem{SBK} T. Suzuki, A. B. Balantekin, and T. Kajino, \Journal{\PRC}{86}{015502}{2012}. doi:10.1103/PhysRevC.86.015502 
\bibitem{SBKC} T. Suzuki, A. B. Balantekin, T. Kajino, and S. Chiba, \Journal{\JPG}{46}{075103}{2019}. doi:10.1088/1361-6471/ab1c11
\bibitem{Balant} A. B. Balantekin, \Journal{\EPJ}{52}{341}{2016}. doi:10.1140/epja/i2016-16341-5
\bibitem{Berryman} J. M. Berryman, V. Brdar, and P. Huber, \Journal{\PRD}{99}{055045}{2019}. doi:10.1103/PhysRevD.99.055045
\bibitem{Patton} K. Patton, J. Engel, G. MacLaughlin, and N. Schunck, \Journal{\PRC}{86}{024612}{2012}. doi:10.1103/PhysRevC.86.024612 
\bibitem{Freedman} D. Z. Freedman, \Journal{\PRD}{9}{1389}{1974}. doi:10.1103/PhysRevD.9.1389 
\bibitem{FST} D. Z. Freedman, D. N. Schramm, and D. L. Tubbs, \Journal{\ARNPS}{27}{167}{1977}. doi:10.1146/annurev.ns.27.120177.001123 
\bibitem{Drukier} A. Drukier and L. Stodolsky, \Journal{\PRD}{30}{2295}{1984}. doi:10.1103/PhysRevD.30.2295  
\bibitem{Akimov2017}  D. Akimov et al. (COHERENT Collaboration), \Journal{\SCI}{357}{1123}{2017}. doi:10.1126/science.aao0990
\bibitem{SO2008} T. Suzuki and T. Otsuka, \Journal{\PRC}{78}{061301}{2008}. doi:10.1103/PhysRevC.78.061301      
\bibitem{Suzuki-NP} T. Suzuki, \Journal{\NPA}{687}{119}{2001}. doi:10.1016/S0375-9474(01)00610-8
\bibitem{Langanke2002} E. Kolbe, K. Langanke, and P. Vogel, \Journal{\PRD}{66}{013007}{2002}. doi:10.1103/PhysRevD.66.013007
\bibitem{Langanke1996} K. Langanke, P. Vogel, and E. Kolbe, \Journal{\PRL}{76}{2629}{1996}. doi:10.1103/PhysRevLett.76.2629
\bibitem{Nakazato2018} K. Nakazato, T. Suzuki, and M. Sakuda, \Journal{\PTEP}{2018}{123E02}{2018}. doi:10.1093/ptep/pty134 
\bibitem{Icarus} ICARUS Collaboration, INFN Report No. INFN/AE-85/7, 1985 (unpublished); ICARUS-II. A Second-Generation Proton Decay Experiment and Neutrino Observatory at the Gran Sasso Laboratory, Proposal No. LNGS 95-10 (1995), http://www.cern.ch/icarus
\bibitem{Cavanna} F. Cavanna, The 8th Inyernational Workshop on Neutrino Nucleus Interactions in the Gew-GeV Region (NuInt12), Rio de Janeiro, 2012.   
\bibitem{SHAr40} T. Suzuki and M. Honma, \Journal{\PRC}{87}{014607}{2013}. doi:10.1103/PhysRevC.87.014607
\bibitem{VMU} T. Otsuka, T. Suzuki, M. Honma, Y. Utsuno, N. Tsunoda, K. Tsukiyama, and M. Hjorth-Jensen, \Journal{\PRL}{104}{012501}{2010}. doi:10.1103/PhysRevLett.104.012501
\bibitem{OSFGA} T. Otsuka, T. Suzuki, R. Fujimoto, H. Grawe, and Y. Akasishi, \Journal{\PRL}{95}{232502}{2005}. doi:10.1103/PhysRevLett.95.232502
\bibitem{Ormand} W. E. Ormand, P. M. Pizzochero, P. F. Bortignon, and R. A. Broglia, \Journal{\PLB}{345}{343}{1995}. doi:10.1016/0370-2693(94)01605-C  
\bibitem{Bhattacha} M. Bhattacharya, C. D. Goodman, and A. Garcia, \Journal{\PRC}{80}{055501}{2009}.  doi:10.1103/PhysRevC.80.055501
\bibitem{Karakoc} M. Karakoc, R. G. T. Zegers, B. A. Brown et al., \Journal{\PRC}{89}{064313}{2014}. doi:10.1103/PhysRevC.89.064313  
\bibitem{Kolbe} E. Kolbe, K. Langanke, G. Mart\'inez-Pinedo, and P. Vogel, \Journal{\JPG}{29}{2569}{2003}. doi:10.1088/0954-3899/29/11/010;\\ 
I. Gil-Botella and A. Rubbia, \Journal{\JCAP}{10}{9}{2003}. doi:10.1088/1475-7516/2003/10/009
\bibitem{Gardiner2021} S. Gardiner, \Journal{\PRC}{103}{044604}{2021}. doi:10.1103/PhysRevC.103.044604
%\bibitem{argo0} R.B. Wiringa, R.A. Smith, and T.L. Ainsworth, \Journal{\PRC}
%29} {1207} {1984}
%\bibitem{urbv14} I.E. Lagaris and V.R. Pandharipande, \Journal{\NPA}{359} 
%{331} {1981}
\end{thebibliography}
\end{document}